\documentclass[useAMS,usenatbib]{mnras}
\usepackage{rotating}
\usepackage{lscape}
\usepackage{longtable}
\usepackage{lastpage}

\def\vsini{$v$sin$i$~}
\def\teff{$T_{\rm{eff}}$~}
\def\logg{log~g}
\def\ms{m~s$^{-1}$}
\def\cms{cm~s$^{-1}$}
\def\kms{km~s$^{-1}$}
\def\mj{M$_{\rm{J}}$}
\def\me{M$_{\rm{\oplus}}$}
\def\rhk{log$R'_{\rm{HK}}$}
\def\msun{M$_{\odot}$~}

\def\msini{m~sin~$i$}

\title[New Planetary Systems from the CHEPS]{New Planetary Systems from the Calan-Hertfordshire Extrasolar Planet Search}
\author[J.S. Jenkins et al.]{J.S. Jenkins$^{1}$\thanks{E-mail: jjenkins@das.uchile.cl}, H.R.A. Jones$^2$, M. Tuomi$^2$, M. D\'iaz$^1$, J.P. Cordero$^1$, A. Aguayo$^1$, \newauthor
B. Pantoja$^1$, P. Arriagada$^3$, R. Mahu$^4$, R. Brahm$^5$, P. Rojo$^1$, M.G. Soto$^1$, O. Ivanyuk$^6$, \newauthor 
N. Becerra Yoma$^4$,  A.C. Day-Jones$^1$, M.T. Ruiz$^1$, Y.V. Pavlenko$^{2,6}$, J.R. Barnes$^7$, \newauthor
F. Murgas$^8$, D.J. Pinfield$^2$, M.I. Jones$^9$, M. L\'opez-Morales$^{10}$, S. Shectman$^{11}$, \newauthor
R.P. Butler$^3$, D. Minniti$^{12}$\\
$^1$Departamento de Astronomia, Universidad de Chile, Camino el Observatorio 1515, Las Condes, Santiago, Chile, Casilla 36-D\\
$^2$Center for Astrophysics, University of Hertfordshire, College Lane Campus, Hatfield, Hertfordshire, UK, AL10 9AB\\
$^3$Carnegie Institution of Washington, Dept. of Terrestrial Magnetism, 5241 Broad Branch Rd. NW, 20015, Washington D.C., USA\\
$^4$Dept. of Electrical Engineering, Universidad de Chile, Av. Tupper 2007, PO Box 412-3, Santiago, Chile\\
$^5$Instituto de Astrof\'isica, Facultad de F\'isica, Pontificia Universidad Cat\'olica de Chile, Av. Vicu\~na Mackenna 4860, Santiago, Chile\\
$^6$Main Astronomical Observatory, Academy of Sciences of Ukraine, Golosiiv Woods, Kyiv-127, 03680 Ukraine\\
$^7$Department of Physical Sciences, The Open University, Walton Hall, Milton Keynes, MK7 6AA, UK\\
$^8$Instituto de Astrof\'isica de Canarias, Via Lactea, E38205, La Laguna, Tenerife, Spain\\
$^9$Center of Astro-Engineering UC, Pontificia Universidad Cat\'olica de Chile, Av. Vicu\~na Mackenna 4860, Santiago, Chile\\
$^{10}$Harvard-Smithsonian Center for Astrophysics, 60 Garden Street, Cambridge, MA 02138\\
$^{11}$Carnegie Institution of Washington, The Observatories, 813 Santa Barbara Street, Pasadena, CA 91101-1292, USA\\
$^{12}$Departamento de Ciencias Fisicas, Universidad Andres Bello, Republica 220, Santiago, Chile}
\begin{document}

\date{Submitted January 2011}

\pagerange{\pageref{firstpage}--\pageref{lastpage}} \pubyear{2002}

\maketitle

\label{firstpage}

\begin{abstract}

We report the discovery of eight new giant planets, and updated orbits for four known planets, orbiting dwarf 
and subgiant stars using the CORALIE, HARPS, and MIKE instruments as part of the Calan-Hertfordshire Extrasolar Planet Search. 
The planets have masses in the range 1.1-5.4\mj's, orbital 
periods from 40-2900~days, and eccentricities from 0.0-0.6.  They include a double-planet system orbiting the most 
massive star in our sample (HD147873), two eccentric giant planets (HD128356$b$ and HD154672$b$), and a rare 14~Herculis analogue (HD224538$b$). 
We highlight some population correlations from the sample of radial velocity detected planets orbiting nearby stars, including the mass function 
exponential distribution, confirmation of the growing body of evidence that low-mass planets tend to be found orbiting more metal-poor stars 
than giant planets, and a possible period-metallicity correlation for planets with masses $>$0.1~\mj, based on a metallicity difference of 0.16~dex 
between the population of planets with orbital periods less than 100 days and those with orbital periods greater than 100 days.

\end{abstract}


\section{Introduction}

After the discovery and confirmation of the hot Jupiter Dimidium (aka. Helvetios~$b$ or 51~Pegasi~$b$) in 1995, our view of giant 
planets was changed forever.  Giant planets orbiting dwarf stars like the Sun have been found to inhabit many regions of the 
parameter space.  The first of these were found orbiting their 
stars with periods much shorter than those of Jupiter in our solar system (e.g. \citealp{mayor95}; 
\citealp{marcy96}).  More recently as the number of giant planets has grown, a new 
population of \emph{eccentric gas giants} has been shown to exist (e.g. \citealp{jones06}; \citealp{tamuz08}; \citealp{arriagada10}), 
making up a large fraction of the known systems, at least around main sequence stars since the same high fraction does not 
appear to be present around giant stars (\citealp{jones14}).  These planets seem to span the full 
range of masses from the sub-Jupiter range all the way up to 
the mass boundary between giant planets and brown dwarfs.  

It has been thoroughly demonstrated that there is a clear bias to metal-rich stars hosting giant planets across the currently sampled range of orbital 
separations (\citealp{fischer05}; \citealp{sousa11}), a bias that seems to hold for stars beyond the main 
sequence (e.g. \citealp{reffert15}; \citealp{soto15}; \citealp{jones16}).  In comparison, recent work appears to 
show that stars that host lower-mass planets have a metallicity distribution that is indistinguishable 
from those that host no known planets at all (\citealp{udry07}; \citealp{jenkins13a}; \citealp{sousa11}; 
\citealp{buchhave15}).  Furthermore, \citeauthor{jenkins13a} suggests there exists a boundary that delimits a planet 
desert for the lowest mass planets in the metal-rich regime.  Such results show there 
is a fundamental relationship between the proto-planetary disc metallicity and the mass of planets that 
form in these discs, indicating further study of the planetary mass function and its relationship with 
metallicity is warranted. 

Since metallicity plays a key role in the formation of the observed planetary systems, explained well by 
nature's merging of core accretion and planet migration (e.g. \citealp{ida04}; \citealp{mordasini12}), and 
now we are reaching a population size where less obvious correlations 
can reveal themselves, studying the current population of known planets can provide a window into the 
fundamentals of planet formation and evolution for the stars nearest to the Sun.

In 2007 we started a radial velocity planet search program on the HARPS instrument at La Silla Chile, with the 
goal of finding more gas giant planets orbiting super metal-rich stars to increase the statistics, whilst also following up any discoveries 
to search for transit events.  The first results from this work and the target sample were discussed in \citet{jenkins09} 
and since then our program has expanded to make use of the CORALIE 
spectrograph \citep{jenkins11a,jenkins11b,jenkins13b}.  This current paper announces the first planets from this survey detected 
using CORALIE and goes on to study the planet mass-metallicity plane to search for correlations between these 
two parameters.

In this work we describe the latest efforts from our Calan-Hertfordshire Extrasolar Planet Search, which builds on previous 
work by this group.  We include the new giant planets we have detected in this program with the large sample of gas giants 
detected by radial velocity measurements that already exist in the literature, in order to continue the search for emerging 
correlations that allow us a more stringent insight into the nature of these objects.  In $\S~\ref{observations}$ we describe 
the measurements used in this work.  In $\S~\ref{star}$ we discuss the sample selection, introduce the new giant planet 
detections from this survey, and discuss some characteristics of their host stars.  In $\S~\ref{results}$ we perform tests of 
the mass function and its relationship to stellar metallicity, along with searching for correlations between these parameters 
and planetary orbital period and we briefly discuss the impact of these results.  Finally, in $\S~\ref{conclusions}$ we 
summerise the main points of the present work.

\section{Observations and Reduction}\label{observations}

The radial velocity datasets for these stars were observed using the precision radial velocity 
spectrographs CORALIE, HARPS, and MIKE.  Both the CORALIE and HARPS spectrographs are physically located at the ESO La 
Silla Observatory in Chile, where CORALIE is mounted on the Swiss Euler telescope and HARPS 
is fed by light from the ESO 3.6m telescope.  The MIKE spectrograph is located at the Las Campanas Observatory 
and is mounted on the Magellan Clay 6.5m telescope.

In this work, 570 radial velocities are reported, with a fairly even split between CORALIE 
and HARPS observations, in comparison to the smaller fraction of MIKE data.  The baseline of observations for 
the CORALIE data run from 2009 November 25$^{\rm{th}}$ until 2015 October 
23$^{\rm{rd}}$ (BJD 2455160.53623 - 2457318.85147), whereas the HARPS data runs from 2007 May 
28$^{\rm{th}}$ until 2013 September 28$^{\rm{th}}$ (BJD 2454248.60231 - 2456563.90982) showing 
that the HARPS data covers a longer baseline but the CORALIE data has better sampling coverage in general 
for these targets.  The MIKE velocities run from 2003 August 13$^{\rm{th}}$ until 2009 July 7$^{\rm{th}}$ 
(BJD 2452864.57934 - 2455019.6938), covering a baseline of six years that overlaps with the 
HARPS baseline but not the CORALIE data.

\subsection{CORALIE}

The analysis of CORALIE data involves the normal echelle reduction steps, such as debiasing 
the images using CCD bias frames, order location and tracing using polynomial fitting 
methods and aperture order filtering, pixel-to-pixel sensitivity correction (flatfielding) by building a 
normalised master flatfield image and dividing out the master flatfield from the other images, scattered-light 
removal by measuring the contribution to the total light profile in the inter-order regions, order extraction 
using a profile fitting method (\citealp{marsh89}), and finally building a precise 2D wavelength solution that is good 
to $\sim$2.5~\ms overall precision (\citealp{jordan14}; \citealp{brahm16}).  The final steps in the analysis include cross-correlating the individual 
spectra with a binary mask (either G2, K0, K5, or M2 depending on the spectral type of the star, see 
\citealp{baranne96} and \citealp{pepe02}) and fitting the cross-correlation function (CCF) with a gaussian to measure the radial velocity, 
along with the width of the CCF to generate realistic uncertainties.  The instrumental drift is then measured and removed from the 
overall velocity by performing a second cross-correlation between the simultaneously measured Thorium-Argon (ThAr) lamp 
(simultaneously referring to observations where the second optical fibre illuminates the spectrograph with ThAr light) and 
a previously measured double ThAr observation where both fibres have been fed by light from the ThAr calibration lamp.  These double 
ThAr measurements are generally taken every 1.5-2 hours throughout the night to continually reset the wavelength solution zero-point.  
These steps were discussed in \citet{jordan14}, however the overall performance in term of stability is shown in 
Appendix~\ref{appendix_stable}.

An important additional step in the calculation of the radial velocities from CORALIE is the characterisation of the offset between the data 
collected prior to the November 2014 upgrade of the instrument.  As part of this upgrade, the CORALIE circular fibres were replaced with octagonal 
fibres to increase illumination stability, the double-scrambler was reintroduced into the system, and a focal mirror that focuses light on the 
guiding camera was replaced.  Such instrumental upgrades are expected to introduce systematic offsets in the radial velocity measurements compared 
to pre-upgrade observations.  

In an attempt to account for the offset between the CORALIE reference velocities observed before and after the upgrade, we first determined this offset in 
our two radial velocity reference targets, HD72673 and HD157347. We denote the mean estimate for the offset $x_{0}$ based on these reference targets as 
$\mu_{0}$ and its standard error as $\sigma_{0}$. We then used these numbers to construct a prior probability for the offset $x_{1}$ in the first data set of 
our sample such that $\pi(x_{1}) = \mathcal{N}(\mu_{0}, \sigma_{0}^{2})$. This prior was used to calculate a posterior for the offset and we denote the mean and 
standard deviation of this obtained posterior with $\mu_{1}$ and $\sigma_{1}$, respectively. After that, we adopted the refined estimate of the offset of 
$\mu_{i} \pm \sigma_{i}$ to construct the prior for the offset $x_{i+1}$ in the next data set, such that $\pi(x_{i+1}) = \mathcal{N}(\mu_{i}, \sigma_{i}^{2})$. This process, 
called 'Bayesian updating' because the prior is updated into a posterior density that is in turn used as as the next prior, was repeated until all the data sets were 
analysed. This process helps to account for all the information regarding the offsets from all the data sets without having to analyse them simultaneously. As a 
result, we summarise the information regarding the offset as a probability distribution that is almost Gaussian in the sense that the third and fourth moments 
are very close to zero. This density has a mean of 19.2 ms$^{-1}$ an a standard deviation of 4.8 ms$^{-1}$ , which indicates that an offset in the reference velocity 
of roughly 20 ms$^{-1}$ is significantly present in the CORALIE data sets after the upgrade for the stars included in this work.

\subsection{HARPS}

For HARPS, the steps are similar to those mentioned above, but the data is automatically processed by the HARPS-DRS version 3.5 which 
is based in general on the procedure explained in \citet{baranne96}.  The nightly drift of the ThAr lines are found to be below 0.5~\ms 
and including the other sources of uncertainty such as centering and guiding ($<$30~\cms), a stability of less than 1~\ms 
is found for this spectrograph over the long term (see \citealp{locurto10}).

\subsection{MIKE}

For the radial velocities measured using the MIKE spectrograph, the reduction steps are similar but the analysis 
procedure is different.  MIKE uses a cell filled with molecular iodine (I2) that is placed directly in the beam of light from the 
target star before entering the spectrograph, and this is used to record the instrument point spread function (PSF) and provide 
a highly accurate wavelength fiducial.  The full analysis procedure is explained in \citet{butler}, but in short, the velocities 
are measured by comparing each star+I2 spectrum to that of a template measurement of the same star.  This template is 
observed without the I2 cell in place, such that one can deconvolve the instrument's PSF from the template observation, 
usually accomplished by observing a rapidly rotating B-star with the I2 cell in place before and after the template observation 
and extrapolating the PSF from these observations to that of the template.  The deconvolved template can then be used to 
forward model each observation by convolving it with a very high resolution and high S/N I2 spectrum and a modeled PSF.  
The final stability of MIKE is found to be between that of CORALIE and HARPS, around the 5~\ms level of velocity precision.

\section{The Stars and their Doppler Signals}\label{star}

\begin{table*}
\caption{Stellar parameters for the hosts.}
\tiny
\label{tab:values}
\begin{tabular}{lllllll}
\hline
\multicolumn{1}{l}{Parameter}& \multicolumn{1}{l}{HD9174}& \multicolumn{1}{l}{HD48265}& \multicolumn{1}{l}{HD68402}& \multicolumn{1}{l}{HD72892}&
\multicolumn{1}{l}{HD128356} & \multicolumn{1}{l}{HD143361}\\ \hline

Spectral Type$_{Hipp}$          & G8IV                       & G5IV/V                   & G5IV/V                   & G5V                         & K3V                       & G6V   \\
$B-V_{Hipp}$                         & 0.761$\pm$0.002  & 0.747$\pm$0.014  & 0.660$\pm$0.021 &  0.686$\pm$0.015  & 1.017$\pm$0.015$^*$ & 0.773$\pm$0.004 \\
$V$                                      & 8.40                        & 8.05                       & 9.11                      & 8.83                          & 8.29                      & 9.20   \\
$\pi$ (mas)                          & 12.67$\pm$0.62     & 11.71$\pm$0.58   & 12.82$\pm$0.61  & 13.74$\pm$0.83       & 38.41$\pm$0.77   & 15.23$\pm$1.18 \\
distance (pc)                        & 78.93$\pm$3.86     & 85.40$\pm$4.23   & 78.00$\pm$3.71  & 72.78$\pm$4.40      & 26.03$\pm$0.52   & 65.66$\pm$5.09  \\
log$R'$$_{\rm{HK}}$                & -5.23                     & -5.24                      & -4.95                    & -5.03                       & -5.07                     & -5.12 \\
Hipparcos $N$$_{\rm{obs}}$     & 92                       &  98                           & 107                        & 92                            & 72                          & 96   \\
Hipparcos $\sigma$              & 0.014                  & 0.009                       & 0.017                     & 0.016                       & 0.013                     & 0.016  \\
$\Delta$$M_{V}$                    & 1.478                 & 1.914                        &  0.108                    &   0.405                     &  0.553                   &   0.349    \\
$T$$_{\rm{eff}}$ (K)                 & 5577$\pm$100   & 5650$\pm$100    & 5950$\pm$100   & 5688$\pm$100     & 4875$\pm$100  & 5505$\pm$100  \\
$L_{\rm{\star}}$/$L_{\odot}$       & 2.41$\pm$0.18  & 3.84$\pm$0.19  & 1.17$\pm$0.06     & 1.40$\pm$0.09      & 0.36$\pm$0.01 & 0.81$\pm$0.06 \\
$M_{\rm{\star}}$/$M_{\odot}$     & 1.03$\pm$0.05   & 1.28$\pm$0.05     & 1.12$\pm$0.05 & 1.02$\pm$0.05      & 0.65$\pm$0.05 & 0.95$\pm$0.05  \\
$R_{\rm{\star}}$/$R_{\odot}$       & 1.67$\pm$0.07   & 2.05$\pm$0.05    & 1.02$\pm$0.05 & 1.22$\pm$0.06      & 0.85$\pm$0.02  & 0.99$\pm$0.08 \\
$[$Fe/H$]$                           & 0.39$\pm$0.10  & 0.40$\pm$0.10     & 0.29$\pm$0.10    & 0.25$\pm$0.10    & 0.17$\pm$0.10 & 0.22$\pm$0.10 \\
\logg                                    & 4.03$\pm$0.05  & 3.92$\pm$0.03     & 4.47$\pm$0.05 & 4.27$\pm$0.05       & 4.52$\pm$0.06   & 4.42$\pm$0.08 \\
U,V,W (km/s)                         &  22.2,-56.5,-29 & -14.2,-24.0,4.5  & -37.5,-16.3,-17.3 & 72.2,-2.0,-15.8  & 30.8,-28.7,8.8     & -24.4,-49.5,3.7 \\
\vsini (km/s)                         & 2.1$\pm$0.2     & 3.1$\pm$0.3        & 2.9$\pm$0.2          &  2.7$\pm$0.2      & 1.3$\pm$0.1       & 1.5$\pm$0.1    \\
Age (Gyrs)                              & 9$\pm$3           &  5$\pm$3                & 2$\pm$3           &   8$\pm$3             & 10$\pm$5           & 5$\pm$5          \\
\hline
\multicolumn{1}{l}{Parameter}& \multicolumn{1}{l}{HD147873}& \multicolumn{1}{l}{HD152079}& \multicolumn{1}{l}{HD154672} & 
\multicolumn{1}{l}{HD165155}& \multicolumn{1}{l}{HD224538}& \multicolumn{1}{l}{}\\ \hline

Spectral Type$_{Hip}$         & G1V                          & G6V                       & G3IV                     & G8V                        &   F9IV/V &   \\
$B-V_{Hip}$                        & 0.575$\pm$0.012   & 0.711$\pm$0.025 & 0.713$\pm$0.013 & 1.018$\pm$0.095 &  0.581$\pm$0.006  &  \\
$V$                                       & 7.96                        & 9.18                        & 8.21                       & 9.36                      &  8.06  &   \\
$\pi$ (mas)                       & 9.53$\pm$0.99       & 12.00$\pm$1.52   & 15.44$\pm$0.84    & 15.39$\pm$1.72  &  12.86$\pm$0.73  &  \\
distance (pc)                       & 104.93$\pm$10.90 & 83.33$\pm$10.56  & 64.77$\pm$3.52 & 64.98$\pm$7.26     &  77.76$\pm$4.41  &  \\
log$R'$$_{\rm{HK}}$                  & -5.27                       & -4.99                     & -5.12                   & -5.18                      &  -4.99  &  \\
Hipparcos $N$$_{\rm{obs}}$          & 88                            & 84                         & 120                        & 67                         &    163  &   \\
Hipparcos $\sigma$                 & 0.011                       & 0.014                    & 0.014                     & 0.019                     &   0.017  &    \\
$\Delta$$M_{V}$                     &  1.337                      & 0.508                    & 0.943                     &  1.373                    &   0.627  &   \\
$T$$_{\rm{eff}}$ (K)                & 5972$\pm$100   & 5726$\pm$100 & 5655$\pm$100      & 5426$\pm$100 &  6097$\pm$100  &      \\
$L_{\rm{\star}}$/$L_{\odot}$     & 5.99$\pm$0.62   &1.28$\pm$0.16 & 1.91$\pm$0.10     & 0.70$\pm$0.08 &   2.95$\pm$0.17  &   \\
$M_{\rm{\star}}$/$M_{\odot}$   & 1.38$\pm$0.05   & 1.10$\pm$0.05 & 1.08$\pm$0.05     & 1.02$\pm$0.05 &  1.34$\pm$0.05  &     \\
$R_{\rm{\star}}$/$R_{\odot}$      & 2.29$\pm$0.10   &1.15$\pm$0.13 & 1.44$\pm$0.05     & 0.95$\pm$0.11 &   1.54$\pm$0.06  &     \\
$[$Fe/H$]$                           & -0.03$\pm$0.10 & 0.16$\pm$0.10 & 0.11$\pm$0.10    & 0.09$\pm$0.10 &  0.27$\pm$0.10  &    \\
\logg                                   & 3.86$\pm$0.05   & 4.36$\pm$0.10 & 4.15$\pm$0.05    & 4.49$\pm$0.11 &  4.19$\pm$0.04  &    \\
U,V,W (km/s)                       & 18.7,-18.2,3.2    & -39.6,-46.2,10.2 & -20.0,-18.7,-29.1 & 13.3,9.8,-20.5  &  -29.1,-15.0,+7.2  &   \\
\vsini (km/s)                        & 5.9$\pm$0.5       & 1.8$\pm$0.1        & 2.2$\pm$0.2         & 1.5$\pm$0.1 &  3.9$\pm$0.3  &   \\
Age (Gyrs)                             & 4$\pm$3             & 3$\pm$3              & 8$\pm$3            & 11$\pm$4        &    2$\pm$3   &   \\

\hline
\end{tabular}

* - colour calculated from magnitudes drawn from the Tycho-2 catalogue (\citealp{hog00}). \\
We assign a standard $\pm$100K uncertainty to the \teff measurements and $\pm$0.10~dex to the metallicities. \\
$\pi$ values come from \citet{vanleeuwen07} and all other Hipparcos parameters are taken from \citet{perryman97}. \\
Evolutionary bulk properties and spectrally measured indices were either computed in this work or taken from \citet{jenkins08}, \citet{jenkins11a}, and \citet{murgas13}. \\
$[$Fe/H$]$ abundances, spectroscopic \logg ~values, and rotational velocities were calculated using the procedures in \citet{pavlenko12} 
and taken from \citet{ivanyuk16}.

\end{table*}

\begin{figure*}
\vspace{7.0cm}
\hspace{-4.0cm}
\includegraphics{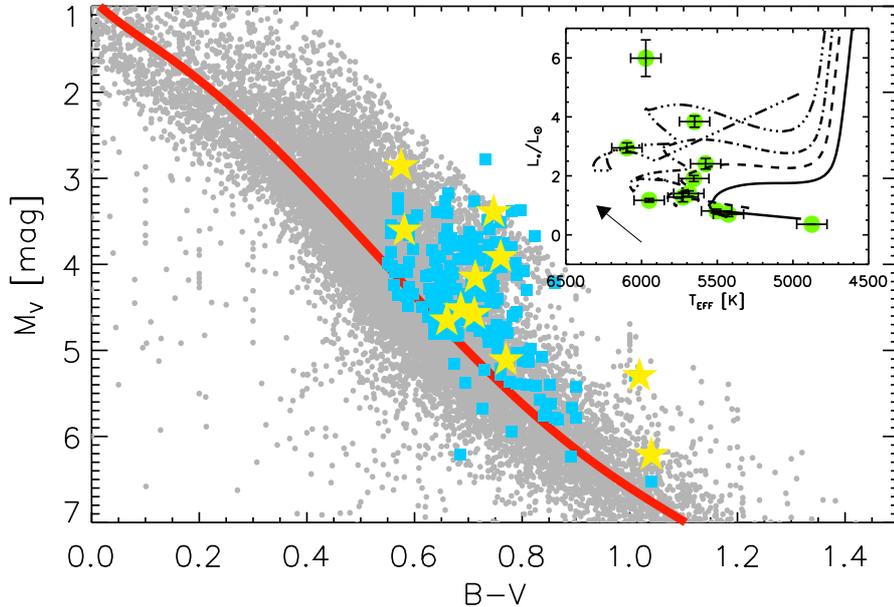}
\vspace{1.0cm}
\caption{Positions of these stars on a colour magnitude diagram.  The grey filled circles are nearby stars drawn from the Hipparcos catalogue, the filled squares are 
the full sample of CHEPS targets, the filled stars are the positions of the stars discussed in this work, and the solid curve marks the position of the main sequence.  
In the inset we show an HR-diagram with the positions of the 
stars discussed here, where the Y2 evolutionary models for masses of 1.0 (solid), 1.1 (dashed), 1.2 (dot-dashed), and 1.3 (dot-dot-dashed) and metallicities of 
+0.2~dex are shown increasing towards higher luminosities and temperatures.  The arrow shows the approximate direction of increasing mass.}
\label{hr_plot}
\end{figure*}

The selection of the Calan-Hertfordshire Extrasolar Planet Search (CHEPS) target sample is discussed in \citet{jenkins09} but we outline the main selection criteria here.  
The stars are generally selected to be late-F to early-K stars, with a $B-V$ range between 0.5-0.9,  and with a small number of redder stars included to allow the 
study of activity correlations and timescales between activity indicators and the measured Doppler velocities.  The positions of the CHEPS 
sample on a colour-magnitude diagram are shown in Fig.~\ref{hr_plot} (squares) compared to a general selection of nearby stars from the Hipparcos catalogue (circles), 
along with the targets discussed in this work (stars).  The magnitude limits we 
set are between 7.5-9.5 in the $V$ optical band and after preliminary screening with the FEROS spectrograph (\citealp{jenkins08}), 
we predominantly selected chromospherically quiet stars, and those that are metal-rich (primary sample having log$R'_{\rm{HK}}$ indices 
$\le$ -4.9~dex and [Fe/H] $\ge$ +0.1~dex, with some stars outside of these selection limits to use as comparisons).  The positions 
of the stars reported here on an HR-diagram are shown in the inset plot in Fig.~\ref{hr_plot}.  The Y2 isomass tracks are also shown 
(\citealp{demarque04}) for masses of 1.0-1.3~\msun, increasing in mass towards higher luminosities, and with a fixed metallicity of +0.2~dex.  
The characteristics of the targets we discuss in this work are shown in Table~\ref{tab:values}.

Since the planets reported here are gas giants, the radial velocity signals of these stars are fairly large, and hence 
one might expect that we can detect frequencies for all of them using standard periodogram analyses to 
hunt for power peaks in the Fourier power spectrum, or minimum mean square error (MMSE) spectrum (e.g. \citealp{dawson10}; 
\citealp{jenkins14a}).  However, the search for the best solutions can be complicated, particularly for combined data sets from 
independent spectrographs because the sets have different baselines, data samplings, and uncertainties that all have to be 
accounted for in a search for periodic signals.  The signals we discuss in this work are a mixture of well, and moderately well sampled data,  
such that some signals are more problematic to detect than others.  Moreover, the 
inclusion of a linear trend that could be present in the data of a given target due to a long period companion, will cause 
considerable correlations to the probability densities of the model parameters.  

In Appendix~\ref{appendix0} we show the MMSE power spectra for our sample of stars and it is clear that a number of deep troughs are 
present in the data, yet some sets show no significant power troughs over the frequency space searched.  We thus applied the delayed-rejection adaptive-Metropolis 
algorithm (DRAM; \citealp{haario06}) together with tempered Markov chains when searching for solutions to our Keplerian models 
(see \citealp{tuomi14a,tuomi14b}) and the simpler adaptive-Metropolis algorithm (\citealp{haario01}) when obtaining estimates for the 
model parameters.  This method has previously been applied in \citet{tuomi13a}, \citet{jenkins13a}, \citet{jenkins13c}, and \citet{jenkins14b}, for example.  
The vertical dashed lines in the MMSE periodograms mark the positions of these detected signals.

The DRAM algorithm works by using a sequence of proposal densities that are each narrower than the last. Here "narrower" is to be understood 
as a multivariate Gaussian proposal density, on which the adaptive-Metropolis algorithm is based, that has a smaller 
variance for at least one of the parameters in the parameter vector. In short, if a value proposed by drawing it from an initial proposal density is 
rejected, we continue by modifying the proposal density with respect to the period parameter(s) by multiplying it with a factor of 0.1 such that 
another value is proposed from a narrower area surrounding the current state of the chain. This enables us to visit the areas of high probability 
in the period space repeatedly and reliably and to see which periods correspond to the highest values of the posterior probability density.

We applied a statistical model that contains reference velocities of each instrument, a linear trend, Keplerian signals, and excess white noise. By 
simplifying the statistical model in \citet{tuomi14b} by removing the intrinsic correlations that we do not expect to play a significant role due 
to the low number of measurements and the (relatively) high amplitude variations in the data, the model can be written as

\begin{equation}
\label{eq:rv_model}
m_{i,l} = \gamma_{l} + \dot{\gamma}t_{i} + f_{k}(t_{i}) + \epsilon_{i,l} 
\end{equation}

where $m_{i,l}$ is the $i$th measurement made by the $l$th telescope/instrument, $\gamma_{l}$ is the systemic velocity offset for each instrument, 
$\dot{\gamma}$ represents the linear trend, $f_{k}$ is the 
superposition of $k$ Keplerian signals at time $t_{i}$, and $\epsilon_{i,l}$ is a Gaussian random variable with zero mean and a variance of 
$\sigma_{l}^{2} + \sigma_{i}^{2}$, where $\sigma_{l}$ is a free parameter in our analyses that 
represents the stellar jitter noise and $\sigma_{i}$ is the measurement uncertainty.  We use prior 
probability densities as described in \citet{tuomi12} but apply an informative prior density for the orbital eccentricities such 
that $\pi(e) \propto \mathcal{N}(0, 0.3^{2})$ that penalises, but does not exclude, eccentricities close to unity (\citealp{anglada-escude13}; 
\citealp{tuomi13b}).  Therefore, given the previous tests performed using this model, in particular using high-precision HARPS data, 
we expect this model to represent a sufficiently accurate description of the velocities analysed in the current work.  The good agreement between 
the parameters published for the stars we discuss that overlap with previous work, is testament to the model's applicability.

Posterior densities for the periods, semi-amplitudes, and eccentricities for all signals are shown in 
Appendices~\ref{appendix1},~\ref{appendix2}, and \ref{appendix3}, and below we discuss each of the systems independently.  
Note that all the signal fits are shown phase folded to highlight the phase coverage we have observed.  We did not show the posteriors for 
the residuals to any of the systems (more than a single planet model fit) unless we detected a unique solution that relates to a second Doppler 
signal.

\subsection[signals]{HD9174}

\begin{figure}
\vspace{5.0cm}
\hspace{-4.0cm}
\includegraphics{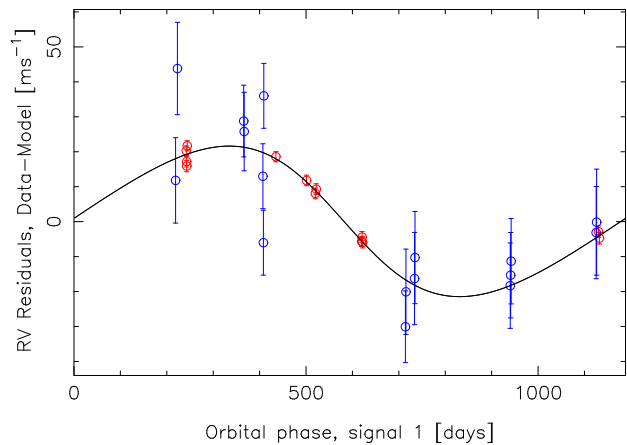}
\vspace{0.5cm}
\caption{Phase folded CORALIE (blue) and HARPS (red) velocities for HD9174.  The solid curve is the best fit Keplerian model.}
\label{phased1}
\end{figure}

HD9174 is classed as a G8 subgiant star in the Hipparcos catalogue (\citealp{perryman97}) since it has a distance from the 
main sequence ($\Delta M_V$) of 1.5 and a $B-V$ colour of 0.76.  The star is also extremely metal-rich, with an [Fe/H] of nearly +0.4~dex, 
indicating a high probability of hosting a giant planet.  A signal with a semi-amplitude of 21~\ms has been detected with a period of 
nearly 1200~days.  The star is very chromospherically quiet and exhibits low rotational velocity, indicating it is an ideal subgiant 
to search for orbiting exoplanets using the radial velocity method, and hence we do not believe the signal is induced by 
activity features, and therefore conclude that HD9174 hosts an orbiting planet.  

In order to understand the significance of the detected signals we present here, we calculated log-Bayesian evidence ratios for each of 
the signals presented here based on the MCMC samples drawn from a mixture of prior and posterior densities, as discussed in \citet{newton94} (see Eqs. 15 $\&$ 16). 
This is a version of importance sampling that uses a mixed distribution to obtain a sampling distribution that has "heavier" tails than 
the posterior (see \citealp{nelson16} for further discussion of this method).  We note that all of these signals are so strong that they 
would pass essentially any meaningful significance test, from likelihood-ratio tests to tests applying any other information criteria.  

The log-Bayesian evidence ratio for the signal in the data of HD9174 was found to be 42.1 for the 1-planet model.  
The signal is also readily apparent in the MMSE periodogram.  No secondary signal was detected.  Considering we set the threshold boundary for a statistically 
significant signal to be at the level of $10^4$, or 9.21 in log-Bayesian units, it is clear that this signal is very significant.  Our 
best fit to the data yields a non-circular eccentricity, only at the level of around 0.12, and with a stellar mass commensurate with 
that of the Sun, the orbiting planet has a minimum mass of 1.11~\mj.  The phase folded velocities and model fit are shown in Fig.~\ref{phased1}

\subsection{HD48265}

\begin{figure}
\vspace{5.0cm}
\hspace{-4.0cm}
\includegraphics{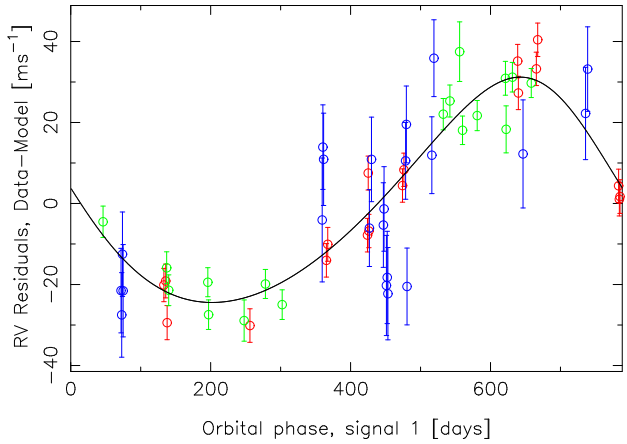}
\vspace{0.5cm}
\caption{Phase folded CORALIE (blue), HARPS (red), and MIKE (green) velocities for the long period signal detected in the HD48265 velocities.  
The solid curve represents the best fit Keplerian model to the data.}
\label{phased2}
\end{figure}

HD48265 is classed in the Hipparcos catalogue as a G5IV/V star, however we find a $\Delta M_V$ of 1.9, 
therefore this star is a subgiant.  Again, it is extremely metal-rich, towards the top end of the metallicity scale, since it's 
[Fe/H] is found to be +0.4~dex.  A planetary companion to this star was previously published in \citet{minniti09} and an update to the planet's 
orbital parameters was published in \citet{jenkins09}.  Here we report our latest orbital solution for this system, including more 
HARPS velocities and with the addition of CORALIE and MIKE data.  

We found a signal with a period of 780~days and with a semi-amplitude of 28~\ms (Fig.~\ref{phased2}) in the radial velocities of HD48265.  
The log-Bayesian evidence for this signal is 103.1, well above the significance threshold.  Although there is a trough in the MMSE close to this 
period when compared to the surrounding parameter space, it is significantly lower than the noise floor at shorter periods, in particular around 
a period of 20~days (deepest trough), therefore the MMSE periodogram can not be used to confirm the existence of this signal.  We studied this 
20~day trough to look for evidence of a signal here but nothing was clear.  In fact, this trough was found to be dependent on the noise properties 
of the data, since when the jitter parameter was considered in the periodogram analysis the evidence pointed towards a period of $\sim$40~days, 
close to twice the period, indicating it is an artifact of the interference pattern from the window function beating with the real signal in the data.

The evolved nature of the star ensures that it is an inactive and 
slowly rotating star and hence it is unlikely that stellar activity is the source of these signals.  The minimum mass of the planet is found 
to be 1.47~\mj\ and a non-zero eccentricity was found for the signal.  
We note that a significant linear trend was also found that possibly indicates more companions that await discovery in this system, 
particularly at longer orbital periods.  The method also indicated that a shorter period signal may be present, with a period 
close to 60~days, yet the current data does not allow us to confirm this as a genuine second planet since the signal disappears when we 
introduce the linear activity correlation parameters in the analysis.  Also, this signal could be an alias of the detected planetary signal. 

The planetary parameters we find here are generally in good agreement with those published in \citet{minniti09} and 
\citet{jenkins09}, however the period we quote is significantly larger than that found by Jenkins et al. by $\sim$80~days.  
The differences in the periods are attributed to the lower number of high precision HARPS data in that work, which was causing 
the fitting algorithm to weight heavily towards those data points, even though they were much fewer compared to the MIKE data.  
This had the effect of significantly increasing the precision of the fit compared to Minniti et al., but in the presence of a linear 
trend, it also skewed the orbital period to lower values.  All other parameters are in excellent agreement within the uncertainties.

\subsection{HD68402}

\begin{figure}
\vspace{5.0cm}
\hspace{-4.0cm}
\includegraphics{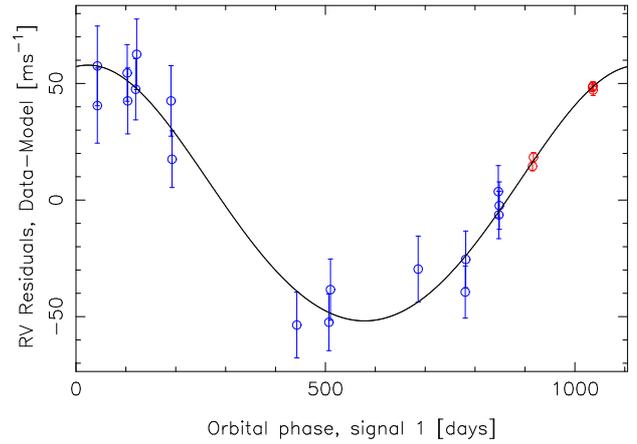}
\vspace{0.5cm}
\caption{Phase folded CORALIE (blue) and HARPS (red) velocities for HD68402.  The solid curve is the best fit Keplerian model.}
\label{phased3}
\end{figure}

The star HD68402 is classed as G5IV/V, though since we find a $\Delta M_V$ of around 0.1 magnitudes, 
we believe the star to be a dwarf.  The metallicity of HD68402 is found to be +0.29~dex.  A signal with a semi-amplitude of over 
50~\ms was found at a period ($\sim$1100~days) similar to that in the HD9174 data.  The MMSE shows two long period troughs 
for this data, with the detected signal found to be the second trough behind a slightly stronger trough close to 3000~days.  
The signal was found to have an eccentricity of 0.03, but to within the limits it can be considered as circular.  

Given the mass of HD68402 is 
1.12~\msun, we compute a minimum mass of just over 3~\mj\ for the planet.  The phase folded velocities for the planetary signal are shown in 
Fig.~\ref{phased3}. Since the log-Bayesian evidence is 49.4, strongly confirming the existence 
of a signal, and it was also found in the MMSE analysis, and by a manual fitting approach, the solution is significantly well constrained, even 
though there is a small gap in the phase folded curve.  

\subsection{HD72892}

\begin{figure}
\vspace{5.0cm}
\hspace{-4.0cm}
\includegraphics{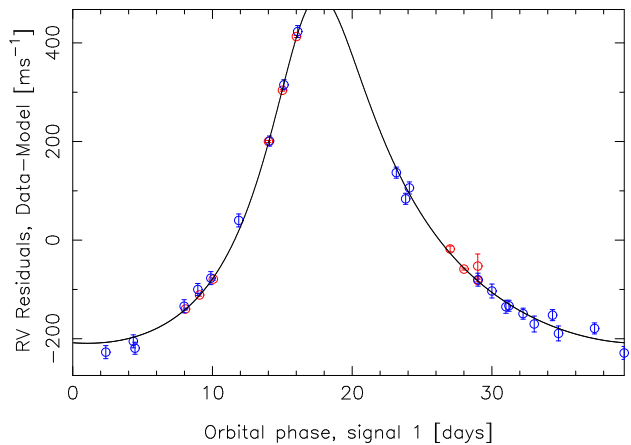}
\vspace{0.5cm}
\caption{Phase folded CORALIE (blue) and HARPS (red) velocities for HD72892.  The solid curve is the best fit Keplerian model.}
\label{phased4}
\end{figure}

This G5V star is located at a distance of 73~pc from the Sun and has a metallicity of +0.25~dex.  The star is also very inactive 
(log$R'_{\rm{HK}}$ = -5.02) and a slow rotator (\vsini = 2.5~\kms) representing an excellent target to search for planets.  We have detected a signal 
with a period of nearly 40~days and semi-amplitude of 320~\ms.  The signal has a log Bayesian evidence value of 903.0, 
confirming the signal at very high significance.  Given the S/N ratio between the signal amplitude and the HARPS and 
CORALIE uncertainties, we fully expected a large evidence ratio to confirm the nature of the signal.  The signal is also clearly apparent in 
the MMSE periodogram as the strongest trough, adding to its reality.  

We find a mass for the 
star of 1.02~\msun, leading to a planetary minimum mass of 5.5~\mj.  This super-Jupiter also has appreciable eccentricity, at the level of 0.4, 
and the final Keplerian model to the phase folded velocities are shown in Fig.~\ref{phased4}.  We note that the solution gives rise to 
a transit probability for this star of 1.6\%, a relatively high likelihood of transit for such a planet-star separation.

\subsection{HD128356}

\begin{figure}
\vspace{5.0cm}
\hspace{-4.0cm}
\includegraphics{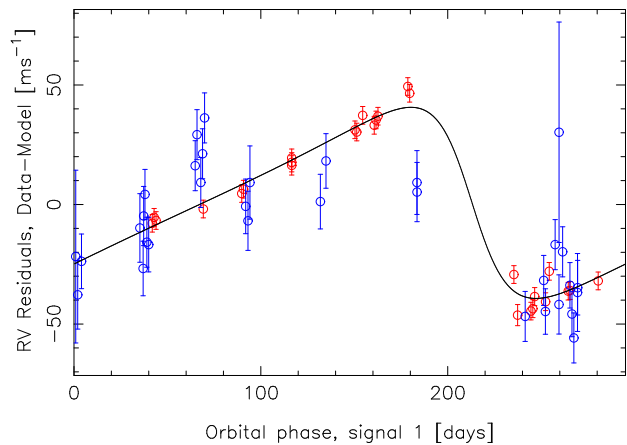}
\vspace{0.5cm}
\caption{Phase folded CORALIE (blue) and HARPS (red) velocities for HD128356.  The solid curve is the best fit Keplerian model.}
\label{phased5}
\end{figure}

HD128356 is the coolest star we have included in the CHEPS sample and is classed as a main sequence star by Hipparcos (K3V).  
We find the star to have a [Fe/H] of almost +0.2~dex, and with a $\Delta M_V$ of 0.55, it may be a subgiant star, or at least in the 
process of evolving onto the subgiant branch. 

Using our Bayesian approach we detected a signal with a semi-amplitude of 37~\ms and period approaching 300~days that had 
a log-Bayesian evidence ratio of 144.5.  The star is found to be a very slow rotator, having a \vsini of only 1.3$\pm$0.2~\kms, 
and with a low chromospheric activity of log$R'_{\rm{HK}}$ of -4.8.  We note that this activity level is significantly lower than the 
one reported in \citet{jenkins11a}, due to an updated $B-V$ colour used here.  The colour used in Jenkins et al. was drawn from the 
Hipparcos catalogue ($B-V =$ 0.685), yet the Tycho-2 catalogue magnitudes (\citealp{hog00}) give a colour $>1$, agreeing with 
what is expected for a mid-K star.  Even if the star was moderately active, we would still expect the jitter to be low since mid-K type stars are not as 
$Doppler-noisy$ as earlier type stars for a given activity level (e.g. \citealp{isaacson10}).  This, combined with the very slow 
rotation of the star, indicates that it is likely a signal with a peak-to-peak amplitude of nearly 80~\ms is not caused by modulated 
activity effects, especially with a period close to 300~days and eccentricity of 0.8.  

The signal detected in our MCMC analysis is also clearly apparent in the MMSE, being the deepest trough, despite the high eccentricity.  
However, an additional trough at a much longer period is also approaching a similar level of significance.  There is also no apparent 
correlations with the activity indicators, as discussed in the next section, nor are there any detected periods in the activity measurements.  
Given the multi-method signal detection and lack of activity correlations, we can confidently conclude that the signal we detected has a 
Doppler origin, and since the mass of HD128356 was found to be around 0.65~\msun, the measured minimum mass for the planet is 0.9~\mj.

\subsection{HD143361}

\begin{figure}
\vspace{5.0cm}
\hspace{-4.0cm}
\includegraphics{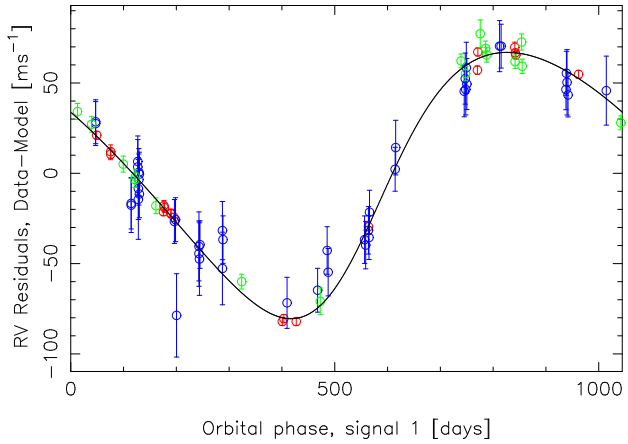}
\vspace{0.5cm}
\caption{Phase folded CORALIE (blue), HARPS (red), and MIKE (green) velocities for HD143361.  The solid curve is the best fit Keplerian model.}
\label{phased6}
\end{figure}

The star HD143361 was previously shown to have a planet with a period of 1057~days (\citealp{minniti09}; \citealp{jenkins09}) and 
we have been conducting further reconnaissance to search for additional companions and to better constrain the orbital 
characteristics of the previously detected planet.  The host star is a chromospherically quiet (log$R'_{\rm{HK}}$ = -5.12 dex) and 
metal-rich ([Fe/H] = +0.22~dex) G6V star, located at a distance of 66~pc.  

Our Bayesian search found a signal with a period of 1046~days with a Bayesian evidence of 491.9, relating to a planet 
orbiting the star with a minimum mass of 3.5~\mj\ (Fig.~\ref{phased6}).  From the 
MMSE periodogram the signal is clearly detected, being one of the most significant periodogram detection's in our sample.  
The final parameters are in good agreement with those published in Minniti et al. and Jenkins et al.  The orbital period 
found here is lower by 40~days ($\sim$4\%) compared to that published in Minniti et al. but only lower by 11~days ($\sim$1\%) to that published in 
Jenkins et al., and both are in agreement within the quoted uncertainties, which are a factor of 28.1 and 6.3 lower here than 
in those previous two works, respectively.  Although in agreement within the quoted uncertainties, our semi-amplitude is 
higher than those published in the previous two works, by 9.1 and 7.1~\ms, respectively.  
No strong evidence for a second companion was found in this system with the current data set.

\subsection{HD147873}

\begin{figure}
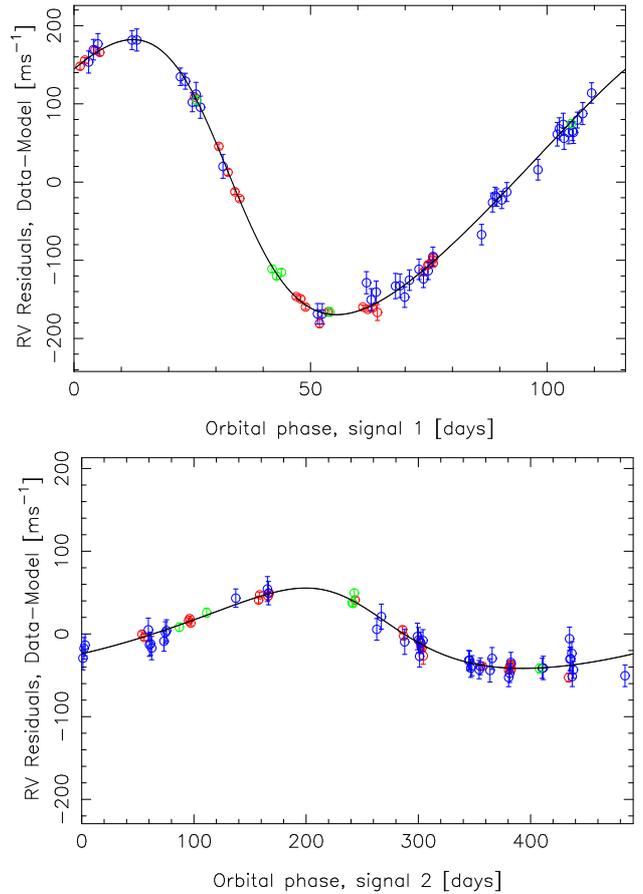

\vspace{5.0cm}
\hspace{-4.0cm}
\includegraphics{Fig10.ps}
\includegraphics{Fig11.ps}
\vspace{6.5cm}
\caption{Phase folded CORALIE (blue), HARPS (red), and MIKE (green) velocities for HD147873 for the short (upper panel) and long (lower panel) period signals.  
The solid curves represent the best fit Keplerian models to the data.}
\label{phased7}
\end{figure}

The star HD147873 is the earliest type star in this sample of planet-hosts and is reported as a G1V star in the Hipparcos 
catalogue.  Given its distance of 105~pc, the star is a little brighter than 8~magnitudes in $V$.  The star also appears to have 
a solar metallicity ([Fe/H] = -0.03~dex), is extremely inactive (log$R'_{\rm{HK}}$ = -5.27~dex), and rotates at the level of nearly 6~\kms.  
We find a Y2 evolutionary track mass for HD147873 of 1.38~\msun.

The Bayesian search for signals in the Doppler data for this star detected two strong periodic signals with semi-amplitudes of 
170 and 50~\ms for HD147873$b$ and $c$ respectively.  The periods of the signals were found to be at 117~days for the 
stronger signal and 492~days for the weaker of the two signals.  The log-Bayesian evidences we found for these signals were 
1131.5 and 145.2.  The MMSE periodogram also detected both these signals rather easily; the second becoming detectable in 
the residuals of the data once the first signal was removed, as shown in Appendix~\ref{appendix0}.   
We find planetary minimum masses of 5.1 and 2.3~\mj\ for the short and longer period planets, respectively.  Both 
Keplerian fits to the data are shown in the upper and middle plots in Fig.~\ref{phased7}.  The inner planet is also only one of 
two in this sample that has a transit probability of over 2\%, a value that encourages the search for transits from intermediate 
period planets.

Given we have discovered two giant planets with a semimajor axis difference of only 0.84~AU between them, we decided to 
test if the system architecture was dynamically stable.  We ran Gragg-Burlich-Stoer integrations in the Systemic Console 
(\citealp{meschiari09}) over a period of 10~Myrs to study the evolution of the orbits of both planets.  We find the eccentricity 
of the orbits librate with a period of around 12000~years but the system itself remains dynamically stable across this timespan.  
Systems with multiple giant planets are interesting laboratories for dynamical studies and this system may warrant further detailed 
dynamical study, especially if more massive companions are discovered with the addition of more data.

\subsection{HD152079}

\begin{figure}
\vspace{5.0cm}
\hspace{-4.0cm}
\includegraphics{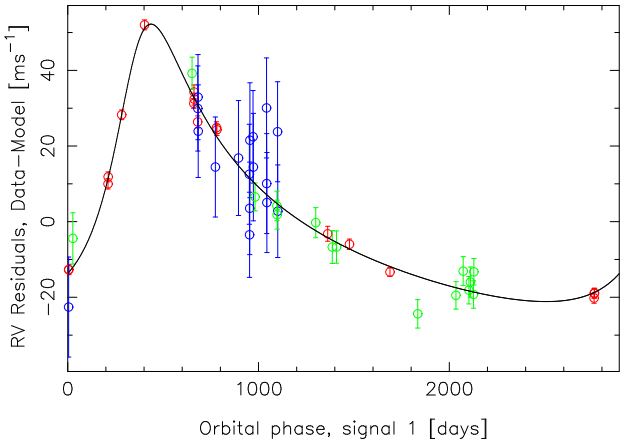}
\vspace{0.5cm}
\caption{Phase folded CORALIE (blue), HARPS (red), and MIKE (green) velocities for HD152079.  The solid curve is the best fit Keplerian model.}
\label{phased8}
\end{figure}

HD152079 is classed as a G6 main sequence star in the Hipparcos catalogue and our previous work found it to be 
inactive (\rhk=-4.99~dex) and metal-rich (+0.16~dex in [Fe/H]), which may explain the 0.5 magnitude $\Delta M_V$.  
We found the mass of the star to be 1.1~\msun.  This is also one of the stars in this sample with a previously announced planet 
candidate detected in orbit (\citealp{arriagada10}).

A signal with a period of 2900~days and semi-amplitude of 31~\ms was detected in the Doppler data of HD152079.  The 
log-Bayesian evidence ratio was found to be 99.1, highly significant, and the signal was found to have an eccentricity 
over 0.5.  It is likely for this data set that the moderate eccentricity of the signal is hampering its detection in the MMSE periodogram.  In 
addition, there is the presence of a linear trend in the data that indicates there is a long period secondary companion to this star, and 
since linear trends are not considered in the MMSE model, the interference here could also be confusing the algorithm. Yet 
there is a fairly strong trough showing at a period of $\sim$1400 days, which is close to half the Bayesian detected signal, and could 
be related to the Doppler signal, or an additional companion that is at a 2:1 resonance site, which could also explain the eccentric 
shape of the one planet signal (\citealp{marcy01}; \citealp{anglada-escude10}).  In any case, we found the minimum mass of HD152079$b$ to be 
2.2~\mj\ and the Keplerian model fit is shown in Fig.~\ref{phased8}.  These values are in good agreement with those presented 
in Arriagada et al., except the precision quoted here is much higher.  For instance, the period of 2097$\pm$930~days quoted in 
their work has been constrained to 2899$\pm$52~days here, a factor of 18 increase in precision and pushing the planet's orbit upwards 
by nearly 900~days.   This precision increase is also mirrored directly in the semi-amplitude precision, lowering it from 
58$\pm$18~\ms to 31.3$\pm$1.1~\ms.

\subsection{HD154672}

\begin{figure}
\vspace{5.0cm}
\hspace{-4.0cm}
\includegraphics{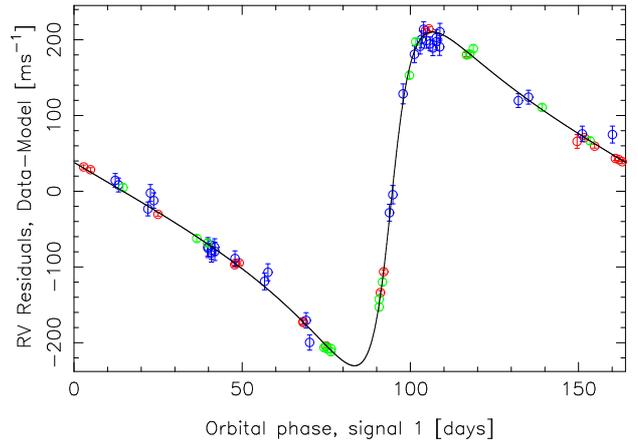}
\vspace{0.5cm}
\caption{Phase folded CORALIE (blue), HARPS (red), and MIKE (green) velocities for HD154672.  The solid curve is the best fit Keplerian model.}
\label{phased9}
\end{figure}

This star has a Hipparcos classification of G3IV, confirmed by our measurement of 0.94~magnitudes from the 
main sequence.  Part of the elevation from the main sequence can also be explained by the metallicity enrichment of 0.11~dex.  
The star is also a slow rotator, having a \vsini of 2.2~\kms and was found to be very chromospherically inactive (\rhk = -5.12~dex).  
The position on the HR-diagram gives rise to a mass of 1.08~\msun.

A signal with a period of 164~days and semi-amplitude of 176~\ms was found in the Doppler timeseries of HD154672, with a 
log-Bayesian evidence ratio of 1709.2, the most significant signal in this data set.  The signal is also clearly apparent in the 
MMSE periodogram, the period trough being significantly 
stronger than any other periods across the parameter space.  The eccentricity of the signal was found to be 0.6, giving rise to a planet with a 
minimum mass of nearly 5~\mj.  The values we find are in good agreement with the values previously published for this 
planet in \citet{lopez-morales08}, with the period agreeing to within one hour and the minimum mass being slightly lower 
here by only 0.23~\mj, but well within the 1$\sigma$ uncertainties.  Given the inclusion of higher quality data in this analysis, 
we find the jitter for this star to be ~2~\ms, a factor two lower than that quoted in Lopez-Morales et al., demonstrating that a significant fraction 
of their jitter was instrumental noise.  The model fit is shown in Fig.~\ref{phased9}.  

Our search for additional planets in the 
combined data sets did not yield any positive results, therefore no firm evidence exists for any additional companions in this system.
If the eccentricity from 
this planet is genuine and not due to the super-position of mixed signals from other giant planets in resonant orbits (see 
\citealp{anglada-escude10}; \citealp{wittenmyer12}), then the transit probability for this object is found to be the highest in 
the current sample of intermediate and long period planets, at 2.5\%.  

\subsection{HD165155}

\begin{figure}
\vspace{5.0cm}
\hspace{-4.0cm}
\includegraphics{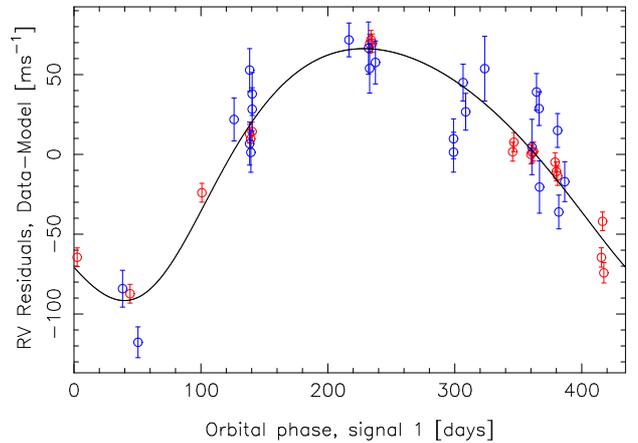}
\vspace{0.5cm}
\caption{Phase folded CORALIE (blue) and HARPS (red) velocities for HD165155.  The solid curve is the best fit Keplerian model.}
\label{phased10}
\end{figure}

The Hipparcos catalogue classifies HD165155 as a G8 main sequence star, however with a elevation above the main sequence of 
1.4, this star can be considered as a subgiant.  The star is located at a distance of 65~pc, and from spectroscopy we have found a 
\rhk\ activity index of -5.18~dex, a rotational velocity of 1.5~\kms, and a [Fe/H] metallicity index of 0.09~dex.  Comparison 
to Y2 evolutionary models yield a mass for the star of 1.02~\msun.

A signal has been detected in the radial velocity data for HD165155 with a period of 435~days and a semi-amplitude of 
76~\ms.  The log-Bayesian evidence for the signal was found to be 168.9, securely above the significance threshold.  
The eccentricity was found to be 0.20 and therefore the final minimum mass of the companion is calculated as 2.9~\mj\ 
(Fig.~\ref{phased10}).  A two-planet model search produced statistically significant 
evidence for a second signal in the data, however given the limited number of measurements we could not confirm a unique 
secondary signal at this time.  No statistically significant troughs were detected in the MMSE periodogram for this star, 
which may be due to the presence of a secondary signal that is interfering with the primary signal.  Indeed, the inclusion of a 
strong linear trend was necessary to constrain this signal, and since linear trends are not included in the MMSE modeling 
approach, this trend is likely the reason why the MMSE approach failed to detect this signal.

\subsection{HD224538}

\begin{figure}
\vspace{5.0cm}
\hspace{-4.0cm}
\includegraphics{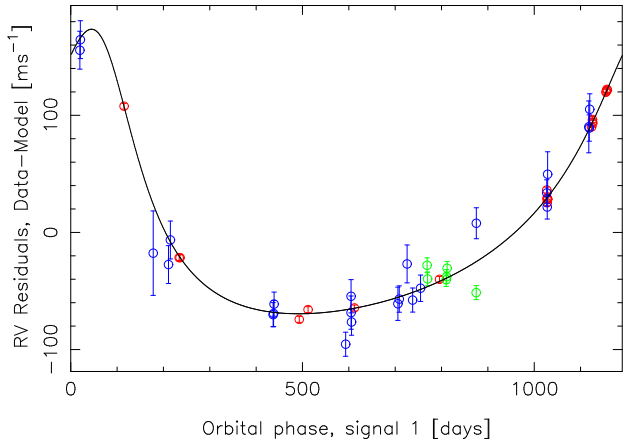}
\vspace{0.5cm}
\caption{Phase folded CORALIE (blue), HARPS (red), and MIKE (green) velocities for HD224538.  The solid curve is the best fit Keplerian model.}
\label{phased12}
\end{figure}

The main Hipparcos catalogue lists HD224538 as a F9 dwarf or subgiant located at a distance of 78~pc.   With a calculated $\Delta M_V$ 
of 0.63 and a high overabundance of metals in the star ([Fe/H]=+0.27~dex), the possibility remains that this star is either on the 
main sequence or crossing into the subgiant branch.  The star is both a slow rotator (\vsini = 3.9~\kms) and chromospherically inactive (\rhk = -4.99~dex).  
From comparisons to Y2 isomass tracks on a HR-diagram we found a mass of 1.34~\msun for HD224538.

Our Bayesian algorithm found a signal with a period of 1189~days, a semi-amplitude of 107~\ms, and an eccentricity of 0.46, shown in 
Fig.~\ref{phased12}.  The log-Bayesian evidence ratio for the signal was found to be highly significant at 391.0.  The MMSE periodogram also 
clearly detected this signal.  Therefore a planet 
with a minimum mass of 6.0~\mj\ is found to be orbiting this star.  This is reminiscent of the gas giant planet 14 Her~$b$ that has a broadly similar 
mass, period, and eccentricity (\citealp{butler03}) and such planets appear to be rare.  Even though 14~Her is an early 
K-dwarf star it does have a super solar metallicity (+0.43$\pm$0.08~dex) similar to HD224538, likely necessary to form such high-mass 
planets.  No additional statistically significant signals were found in the current data set.

\subsection[act_tests]{Line Modulation Tests}

\begin{figure}
\vspace{5.5cm}
\hspace{-4.0cm}
\includegraphics{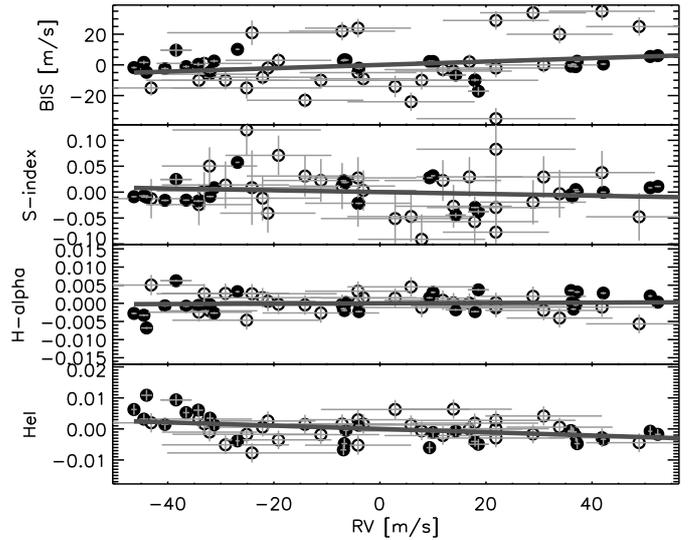}
\vspace{1.5cm}
\caption{The four plots from top to bottom show the linear correlations between the radial velocities and the bisector span velocities, 
the $S$-indices, the H$\alpha$ indices, and the He~I indices for HD128356, respectively, where CORALIE data is represented by open rings 
and HARPS data by filled circles.  The solid lines are the best fit linear trends to the data.}
\label{activity}
\end{figure}

Although these stars are very inactive and slowly rotating and the Doppler signals we have detected are generally very large 
compared to the uncertainties (most are significantly larger than 20~\ms), it is useful to rule out line modulations 
that could originate from stellar activity as the source of the variations.  The activity parameters employed are the calcium\sc ii\rm HK 
line doublet, the bisector span (BIS), the CCF FWHM, the $H\alpha$ line, and the He~I D3 line.  These indices were selected since they 
have previously been shown to be good tracers of stellar magnetic activity, and/or spectral line modulations (e.g. \citealp{queloz01}; \citealp{robertson14}; 
\citealp{santos10}).  In Fig.~\ref{activity} we show four of 
the tests we have carried out to rule out these modulations as the source of the detected radial velocity shifts for the star HD128356, originally 
believed to be the most active star in the sample due to the erroneous $B-V$ colour.  We note 
that we do not show the CCF FWHM test for this star since there is large variations with a few outliers, but no correlation exists.

In the upper plot of Fig.~\ref{activity} we show how the BIS values vary as a function of the radial velocity datasets.  The 
BIS values for HARPS were taken from the HARPS-DRS and measured following the method explained in \citet{queloz01}.  The CORALIE BIS 
values were calculated using a similar procedure.  No significant correlation between the 
radial velocities and the BIS measurements are found and we highlight this by showing the best fit linear trend to the data.  The unweighted 
Pearson rank correlation coefficient has a value of 0.23, signifying a weak correlation, however when the correlation is weighted by 
the measurement uncertainties on the radial velocity and BIS values, the coefficient drops to 0.11, or no evidence at all for any correlation.   

We also searched for a correlation between the chromospheric activity $S$-indices 
and the velocities as a second useful discriminant that activity is not the source of the observed variations.  The measurement of these $S$-indices for HARPS was 
briefly discussed in \citet{jenkins13a} and therefore here we only discuss the CORALIE 
activity measurement method in Appendix~\ref{appendix_act}.  In any case the method for both is similar, except for HARPS we use the extracted 1D 
order-merged spectrum, whereas for CORALIE spectra we perform the calculations using the extracted 2D order-per-order spectrum, similar to the method 
discussed in \citet{jenkins06}.

The second plot in Fig.~\ref{activity} shows these chromospheric activity $S$-indices as a function of the radial velocity measurements and no apparent 
correlation is found.  The best unweighted linear fit is shown by the solid line and confirms the lack of any correlation between the two parameters.  
The correlation coefficient also confirms this since an unweighted $r$ coefficient of -0.09, similar to the weighted BIS, is not statistically significant, dropping 
even lower when considering the weights.  

The lower two plots in the figure show the linear correlations against the measured H$\alpha$ and He~I D3 activity indicators, respectively.  These 
indices were calculated following the methods discussed in \citet{santos10}.  For both of these indices, no significant correlation is found when 
combining the CORALIE and HARPS data.  Some moderate correlation between the He~I index and the velocities is seen, with an unweighted $r$ correlation 
coefficient of -0.64 for the HARPS only measurements, decreasing to -0.39 when the CORALIE measurements are added.  Judging by the lower panel in the 
figure, no striking correlation is apparent, given what would be expected for this level of correlation, and once the measurement uncertainties are included to 
weight the correlation coefficient, the value drops significantly to be in agreement with zero correlation.  In fact, we can see that the majority of the data are 
uncorrelated, from radial velocities between -25 - +55~\ms, with only 
a few offset data points clustered around -40~\ms driving the correlation.  

As an aside, if we apply the relationships in \citet{saar} and \citet{hatzes02} to calculate the spot coverage expected for a star with the rotational period 
of HD128356, in order to produce a radial velocity amplitude in agreement with that observed here, then $\sim$5\% of disk spot coverage is required, 
which would likely exhibit as photometric variations that are not observed (see below).  

Although the activity indicators for the other stars reported in this work show no evidence for any strong linear correlations, measured by the 
Pearson Rank correlation coefficient, against the radial velocities, we report the moderately correlated data sets ($|0.5| \ge r \ge |0.75|$).  
For HD48265 the HARPS BIS values correlate with the radial velocities with a $r$ value of 0.59$\pm$0.25, indicating some moderate correlation between 
the two parameters.  We also note that both the H$\alpha$ and He~I indices have values of 0.47 and -0.41, respectively, yet there are large parts of 
parameter space that are under-sampled by including only the HARPS data alone.  When adding 
in the CORALIE measurements we find these values decrease to 0.22, 0.12, and -0.33 for the three quantities respectively, with uncertainties of $\pm$0.16, 
indicating that these correlations are not the source of the velocity signal for this star. 

The star HD68402 shows HARPS velocity correlations with the BIS and He~I indices with $r$ values of 0.79 and 0.52, respectively.  We note that there 
are only five HARPS data points for this star so no result here can be deemed significant.  Furthermore, once the CORALIE values are included we find 
values of 0.01 and 0.12 for these parameters respectively.  
When including the CORALIE measurements we find a moderate correlation appears between the velocities and the CCF FWHM measurements ($r =$ 0.52$\pm$0.23).  
Again there appears no significant correlations in the analysis.

Another star with a limited number of HARPS spectra (eight measurements) that give rise to an apparent moderate correlation between the radial 
velocities and activity indicators is HD72892.  The HARPS CCF FWHM and He~I measurements have Pearson rank correlation coefficients of 0.58 and 0.73, 
respectively, with uncertainties of $\pm$0.38.  When the CORALIE measurements are added to the HARPS data we find this correlation becomes insignificant, with 
values of only -0.47 and 0.10, respectively, and uncertainties of $\pm$0.19.  However, the $H\alpha$ indices now exhibit a moderate correlation with the velocities 
($r =$ 0.57).  The signal for this star has an amplitude of nearly 320~\ms and a period of $\sim$40~days.  Although 
40~days is a plausible rotational period for this type of very inactive G5 dwarf star, the fact that it is so inactive indicates that such a large signal 
would be difficult to produce through spot rotation.  In fact, 
if we calculate the spot coverage as above, then 20-25\% of disk spot coverage is required, which can be ruled out based 
on the photometric stability and the low \rhk\ measurement.

For HD152079 the 
HARPS $H\alpha$ index correlates with the velocities with a correlation coefficient $r$ of 0.66$\pm$0.24, although none of the the other indicators show 
any evidence for correlations, and when adding the CORALIE data, the correlation coefficient decreases to a value of only 0.47$\pm$0.17.  We note that the 
variations in the measurements are only changing at the few $\times$10$^{-3}$ level, from 0.203 to 0.208 in our HARPS $H\alpha$ index.

The star HD165155 shows a correlation between the velocities and the HARPS CCF FWHM with a value of -0.66$\pm$0.23, although with a large spread.  
Addition of the CORALIE CCF FWHM decreases the value to 0.12$\pm$0.17, rendering this result insignificant.  

HD224538 shows moderate correlation between the velocities and the $H\alpha$ indices at the level of 0.69$\pm$0.22 in the HARPS data, and at the level of 
0.27$\pm$0.15 when the CORALIE values are included.  Again the variation is only at the few $\times$10$^{-3}$ level, which is likely to be insignificant.  

The periodogram analysis for each of the five activity indicators did not reveal any significant periods that could explain the detected signals in any of 
the stars considered here (see Appendix~\ref{appendix5}), however a few features do appear.  For the star HD72892 there are emerging peaks in the 
periodograms of the BIS, FWHM, $S$ and He~I indices at periods between 11-13~days, and although this region is distinct from the detected planetary 
signal, these may be linked to the rotational period of the star, however there is a peak in the window function at 22.5~days that could be giving rise to 
a peak at the first harmonic in these indices.  For HD128356, the moderately active star, both the BIS and FWHM timeseries show 
peaks that agree with a period around 1250~days that has no counterpart in the window function, meaning this could be a magnetic cycle, 
but it is far from the detected Doppler signal period.  HD147873 
does show a unique peak close to the lower period signal in the radial velocities in the He~I indices with a period of around 120~days, 
yet the peak is not significant.  The H$\alpha$ indices show an emerging peak with a period of $\sim$70~days as the strongest signal for 
HD165155 that could point to the rotation period for this star but there is a window function peak emerging at 63.5~days, that is likely unrelated to this 
peak, but is worthy of note.  In any case, neither of these are related to the detected Doppler signal in the radial velocities.  
Finally, the star HD224538 shows evidence for two peaks in the periodograms of the BIS and H$\alpha$ indices that are in good agreement with 
signals with periods of 20~days, which again could be a good candidate for the rotational period for this star, however the fifth strongest peak in 
the window function is found to be at 19~days, meaning there is a non-insignificant probability that this peak is being boosted by the sampling.

\subsection[phot]{Photometric Analysis}

We decided to photometrically search for a secure rotational period for these stars by employing frequency analyses of the $V$-band All Sky Automated 
Survey (ASAS; \citealp{pojmanski97}) photometric data.  We have previously shown that 
such analyses can shed light on the rotational periods of planet-host stars and/or short period and long period magnetic cycles (e.g. \citealp{anglada-escude13}; 
\citealp{jenkins14b}).  We tend to focus on the best quality data, ASAS grade A or B using the smallest ASAS apertures that are best for point sources, 
and typical baselines cover $\sim$9~years at a sampling cadence of 
$\sim$3~days.  Out of all the 11 planet-hosts considered in this work, six show evidence for a significant rotational period or long period magnetic cycle in the 
photometry, and these are summarised below, with particular focus paid to HD165155.

The six stars showing evidence for a photometric signal in the ASAS timeseries are HD68402, HD147873, HD152079, HD154672, HD165155, and HD224538, and 
the Lomb-Scargle periodograms for these six stars are shown in Appendix~\ref{appendix4}.  We do not show the periodograms for the remaining six stars since 
they exhibit no significant frequency peaks.  The periodogram for HD68402 shows two strong peaks emerging at periods of 312~days and 2000~days.  Neither of 
these signals reside at periods close to the detected Doppler signal in the radial velocities, however their strength suggests there may be some long term 
magnetic cycle at play within this star.  

The stars HD147873, HD152079, HD154672, and HD224538 all show evidence for long period modulated spot activity, with periods at the extremities of the data 
timeseries and periodogram sampling $\sim$5000--10000~days.  These could be real long term spot cycles or they could be sampling features to due to the 
limited baselines of the data sets.  However, none of these features appear to coincide with the detected radial velocity signal periods, or harmonics there-of.  HD154672 
has a velocity signal detected at just over 160~days, far from any long period magnetic cycle, whereas the signal in the HD224538 velocities has a period of 
a few thousand days, which could agree with any potential photometric signal in the ASAS data that is not due to 
the limited data baseline.  However, our analysis never indicated any correlations were evident between the radial velocities and the activity indicators for 
this star, and its inactive nature, along with the strength of the radial velocity signal ($K =$ 110~\ms), would make an activity origin unlikely for a star of this type.  
The evidence from these analyses points to the origin of the detected signals as being due the gravitational influence of orbiting planets.

The star HD152079 has a detected signal in the velocities with a period of a few thousand days, which is approaching the regime where a long period magnetic cycle 
could be present due to the ASAS photometric periodogram, yet the amplitude of this signal is a little over 30~\ms for this very inactive star.  
The structure of the long period signals in the power spectrum of these stars are very similar, which argues that the frequencies emerge 
due to the sampling baseline.  A further secondary peak exists in the photometric periodogram 
for this star at $\sim$830~days.  Since this is too short to be associated with the signal in the radial velocities, it is not the origin of that signal but could 
be a possible sampled magnetic cycle.  In any case, the nature of the radial velocity signal is likely Doppler and from an orbiting planetary mass candidate.

We note that for HD147873 and HD224538, additional peaks arise in the periodograms at periods that could be in the range of rotational periods 
for stars with these types of rotational velocities and stellar radii, or could relate to additional magnetic cycles.  For HD147873, there is a strong peak at a 
period of 29.5~days, very close to the lunar cycle, and since this period was found to arise in other ASAS timeseries, it is likely this is a sampling alias and 
not the star's rotational period.  For HD224538, the next strongest peak is located at 385~days, with again the 29.5~day period being detected.  The 385~day 
peak is within a small cluster of peaks that surround the Earth's orbital period, therefore it is likely this is another sampling alias.  
Hence, it is unlikely that we have made a significant detection of the rotational period for any of these stars, with only tentative detections 
of long period magnetic cycles.

Finally, we discuss the photometric analysis for the star HD165155 independently, since there is an indication of a peak in the periodogram that is close to the 
period of the detected signal in the RV measurements.  Given that the signal is rather strong, it is unlikely that activity 
is the source of the signal at this type of period.  We also note that 
the orbital separation is too large to produce star-planet interactions that could cause any photometric signal.  From the ASAS periodogram 
in Appendix~\ref{appendix4} we can see that again a long period signal emerges, but after this signal, there are two fairly strong 
peaks with periods of 454 and 344~days.  The 454~day signal is the second strongest after the long period peak and it closely matches 
the period of the signal in the radial velocities at 452~days, indicating the signal could arise from activity or pulsations.  
As mentioned above, no similar periodicities were found in the activity indicators and 
since the Hipparcos photometry for this star only consists of 67 measurements, with a scatter in the data of 0.019 magnitudes, no significant periodicity was found in 
this data either.  We also note that by removing the long period trend with periods of 10000~days or more removes the signal at 454~days, 
signifying it is linked to this long period trend and therefore likely not a true magnetic cycle that could induce such a radial velocity signal as we observe.

There exists a small possibility that the detected radial velocity signal is not of Doppler origin and is due to line asymmetries from stellar activity on this 
subgiant star, even though the existence of this photometric period is difficult to causally connect to the origin of the radial velocity signal without corroborating 
periodicities in the spectral activity indicators.  Without garnering more data, and since the detected photometric peak in the periodogram could be an alias that is 
associated with a longer period signal or the window function of the data, we still consider the radial velocity signal as due to an orbiting companion.  If, on the other hand, 
the signal in the velocities is genuinely of astrophysical origin, this data set would serve as a warning when trying to understand 
the origin of signals in radial velocity timeseries of subgiant stars, even when the signal amplitude is relatively large, and there are no correlations or periodicities in the 
spectral activity indicators.  Thorough searches of existing photometric data should always be performed, where possible, to help to confirm the reality of proposed planetary 
systems.

\begin{table*}
\tiny
\caption{Orbital mechanics for all planetary systems described in this work.}
\label{tab:orb_params1}
\begin{tabular}{ccccccc}
\hline
\multicolumn{1}{c}{Parameter} & \multicolumn{1}{c}{HD9174$b$} & \multicolumn{1}{c}{HD48265$b$} & \multicolumn{1}{c}{HD68402$b$} & \multicolumn{1}{c}{HD72892$b$} & \multicolumn{1}{c}{HD128356$b$} & \multicolumn{1}{c}{HD143361$b$}  \\
\hline

Orbital period $P$ (days) & 1179$\pm$34 & 780.3$\pm$4.6 & 1103$\pm$33 & 39.475$\pm$0.004 & 298.2$\pm$1.6 & 1046.2$\pm$3.2  \\
Velocity amplitude $K$ (m/s) & 20.8$\pm$2.2 & 27.7$\pm$1.2 & 54.7$\pm$5.3 & 318.4$\pm$4.5 & 36.9$\pm$1.2 & 72.1$\pm$1.0  \\
Eccentricity $e$ & 0.12$\pm$0.05 & 0.08$\pm$0.05 & 0.03$\pm$0.06 & 0.423$\pm$0.006 & 0.57$\pm$0.08 & 0.193$\pm$0.015   \\
$\omega$ (rad) & 1.78$\pm$0.66 & 6.0$\pm$2.4 & 0.3$\pm$2.3 & 6.010$\pm$0.014 & 1.47$\pm$0.08 & 4.21$\pm$0.06  \\
M$_0$ (rad) & 3.5$\pm$1.3 & 4.9$\pm$1.4 & 6.0$\pm$2.2 & 2.714$\pm$0.010 & 3.1$\pm$0.7 & 3.21$\pm$0.14  \\

\msini (\mj) & 1.11$\pm$0.14 & 1.47$\pm$0.12 & 3.07$\pm$0.35 & 5.45$\pm$0.37 & 0.89$\pm$0.07 & 3.48$\pm$0.24  \\
Semimajor axis $a$ (AU) & 2.20$\pm$0.09 & 1.81$\pm$0.07 & 2.18$\pm$0.09 & 0.228$\pm$0.008 & 0.87$\pm$0.03 & 1.98$\pm$0.07  \\

$\gamma_{\rm{HARPS}}$ (m/s) & -7.2$\pm$1.4 & -1.5$\pm$1.6 & -34.2$\pm$8.2 & -37.8$\pm$1.7 & -0.1$\pm$1.9 & -1.2$\pm$0.8  \\
$\gamma_{\rm{CORALIE}}$ (m/s) & -1.6$\pm$3.0 &-4.3$\pm$2.6 & -10.6$\pm$4.6 & 48.7$\pm$3.2 & 9.4$\pm$2.7 & 3.4$\pm$2.2  \\
$\gamma_{\rm{MIKE}}$ (m/s) & -- & -3.5$\pm$1.4 & -- & -- & -- & -26.6$\pm$1.2  \\
$\sigma_{\rm{HARPS}}$ (m/s) & 1.8$\pm$0.6 & 6.0$\pm$0.6 & 1.7$\pm$0.9 & 2.2$\pm$0.7 & 3.9$\pm$0.7 & 2.3$\pm$0.6  \\
$\sigma_{\rm{CORALIE}}$ (m/s) & 2.2$\pm$1.0 & 2.7$\pm$1.1 & 2.0$\pm$1.0 & 2.0$\pm$1.0 & 2.1$\pm$1.0 & 1.8$\pm$0.9  \\
$\sigma_{\rm{MIKE}}$ (m/s) & -- & 2.8$\pm$0.8 & -- & -- & -- & 2.8$\pm$0.8  \\
$\dot{\gamma}$ [ms$^{-1}$year$^{-1}$] & -- & -- & -- & -- & -- & -- \\
$P_{T}$   &  0.3\% & 0.5\% & 0.2\% & 1.6\% & 0.4\% & 0.2\% \\
N$_{\rm{Obs}}$ & 29 & 57 & 20 & 32  & 60 & 80   \\
$\ln B(k,k-1)$ & 42.1 & 103.1 & 49.4 & 903.0 & 144.5 & 491.9  \\

\hline
\multicolumn{1}{c}{Parameter} &  \multicolumn{1}{c}{HD147873$b$} & \multicolumn{1}{c}{HD147873$c$} & \multicolumn{1}{c}{HD152079$b$} & \multicolumn{1}{c}{HD154672$b$} & \multicolumn{1}{c}{HD165155$b$} & \multicolumn{1}{c}{HD224538$b$}  \\\hline

Orbital period $P$ (days) & 116.596$\pm$0.023 & 491.54$\pm$0.79 & 2899$\pm$52 & 163.967$\pm$0.009 & 434.5$\pm$2.1 & 1189.1$\pm$5.1 \\
Velocity amplitude $K$ (m/s) & 171.5$\pm$1.2 & 47.9$\pm$1.7 & 31.3$\pm$1.1 & 176.3$\pm$0.7 & 75.8$\pm$3.0 & 107.0$\pm$2.4   \\
Eccentricity $e$ & 0.207$\pm$0.013 & 0.23$\pm$0.03 & 0.52$\pm$0.02 & 0.600$\pm$0.004 & 0.20$\pm$0.03 & 0.464$\pm$0.022   \\
$\omega$ (rad) & 1.40$\pm$0.05  & 0.73$\pm$0.20 &5.67$\pm$0.06 & 4.63$\pm$0.01 & 3.7$\pm$0.2 & 0.40$\pm$0.03 \\
M$_0$ (rad) & 1.65$\pm$0.07 & 3.09$\pm$0.20 & 0.8$\pm$0.8 & 3.60$\pm$0.02 & 0.9$\pm$0.8 & 0.3$\pm$0.3   \\

\msini (\mj) & 5.14$\pm$0.34  & 2.30$\pm$0.18 & 2.18$\pm$0.17 & 4.73$\pm$0.32 & 2.89$\pm$0.23 & 5.97$\pm$0.42   \\
Semimajor axis $a$ (AU) & 0.522$\pm$0.018  & 1.36$\pm$0.05 & 3.98$\pm$0.15 & 0.59$\pm$0.02 & 1.13$\pm$0.04 & 2.28$\pm$0.08   \\

$\gamma_{\rm{HARPS}}$ (m/s) & 59.0$\pm$1.2 & -- & -37.9$\pm$7.0 & 5.2$\pm$0.7 & -59.6$\pm$18.7 & -15.3$\pm$1.5  \\
$\gamma_{\rm{CORALIE}}$ (m/s) & 5.4$\pm$2.2 & -- & -44.8$\pm$8.5 & -44.6$\pm$2.0 & -87.7$\pm$20.1 & 27.0$\pm$2.7   \\
$\gamma_{\rm{MIKE}}$ (m/s) & 37.6$\pm$3.1 & -- & -13.6$\pm$6.8 & 28.2$\pm$1.2 & -- & 55.9$\pm$4.3   \\
$\sigma_{\rm{HARPS}}$ (m/s) & 2.6$\pm$0.7 & -- & 1.5$\pm$0.6 & 2.1$\pm$0.5 & 5.8$\pm$0.6 & 2.9$\pm$0.6  \\
$\sigma_{\rm{CORALIE}}$ (m/s) & 1.9$\pm$0.9 & -- & 2.0$\pm$1.0 & 2.2$\pm$1.0 & 3.7$\pm$1.2 & 2.0$\pm$1.0   \\
$\sigma_{\rm{MIKE}}$ (m/s) & 2.4$\pm$1.0 & -- & 2.7$\pm$0.8 & 3.6$\pm$0.7 & -- & 5.2$\pm$0.7  \\
$\dot{\gamma}$ [ms$^{-1}$year$^{-1}$] & 2.94$\pm$0.68 & -- & 1.72$\pm$0.47 & -- & 4.00$\pm$1.19 & --  \\
$P_{T}$ & 2.3\% & 0.7\% & 0.1\% & 2.5\% & 0.4\% & 0.2\%  \\
N$_{\rm{Obs}}$ & 66 & -- &  50 & 72 & 38 & 50 \\
$\ln B(k,k-1)$ & 1131.5  & 145.2 & 99.1 & 1709.2 & 168.9 & 391.0  \\

\hline
\end{tabular}

The uncertainties on the \msini~ and semimajor axis consider the uncertainties on our stellar mass estimate of 10\%.  \\
The $\gamma$ offset is the value after subtracting off the mean of the data set.\\
The $\sigma$ terms parameterise the excess noise in our model fits, aka jitter.\\
The ln B(k,k-1) are generally the 1-planet models (e.g. B(1,0)) except for HD147873 which is a 2-planet model (B(2,1)). \\
$P_{T}$ is each planet's transit probability.\\
N$_{\rm{Obs}}$ are the total number of radial velocities per target star.\\

\end{table*}

\section{Planet Population Distributions}\label{results}

The high metallicity selection bias in our program means we are generally targeting gas giant planets.  The working hypothesis being that if such planets are 
formed through core accretion processes, then large cores can form quickly due to the enrichment of the proto-planetary disk, which gives the planetesimals 
sufficient time to accrete gaseous material to reach large masses after they cross the critical core mass limit of around 10~\me (\citealp{mizuno80}).  

\subsection{Mass Function}

\begin{figure*}
\vspace{6cm}
\hspace{-4.0cm}
\includegraphics{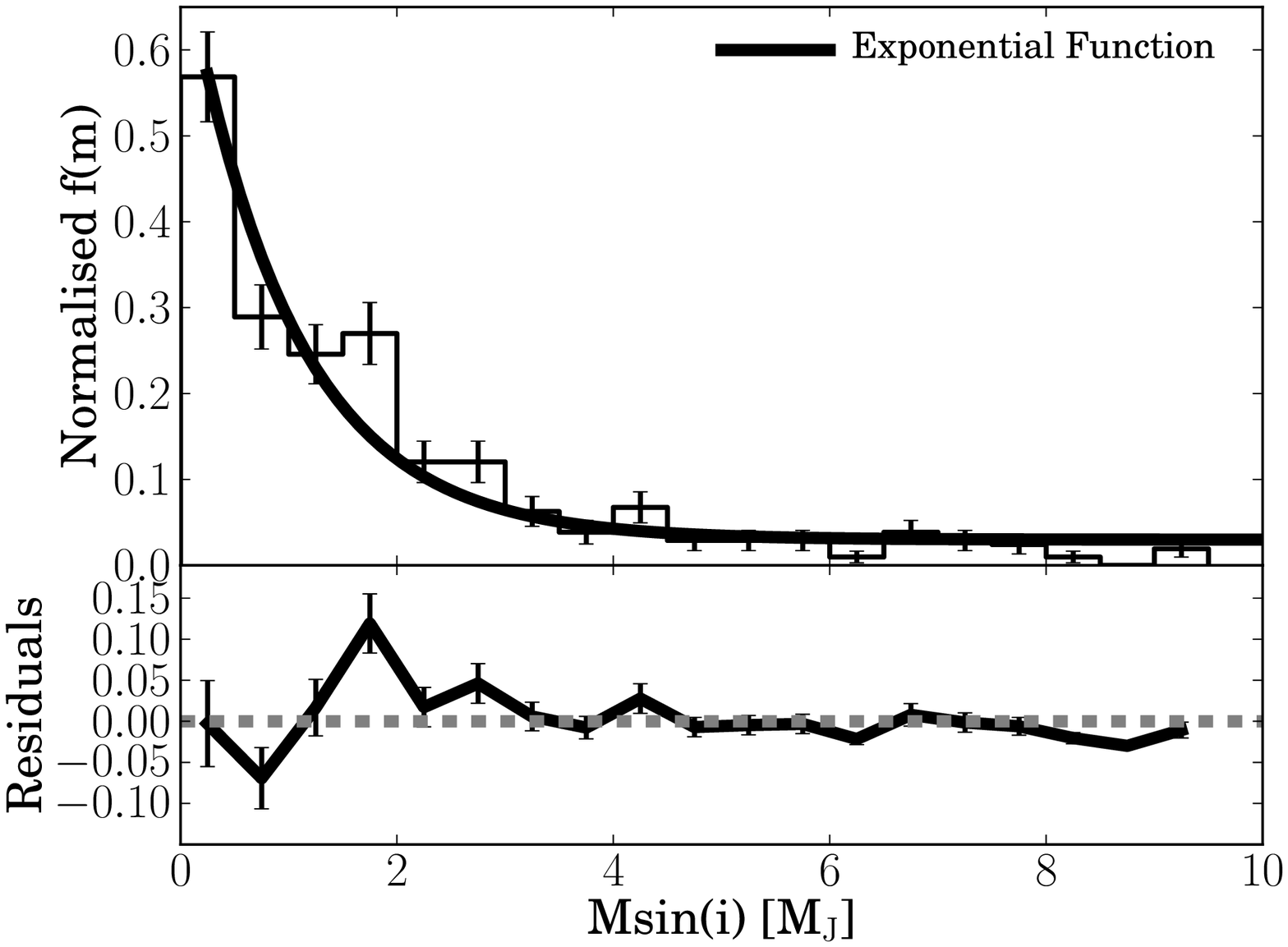}
\includegraphics{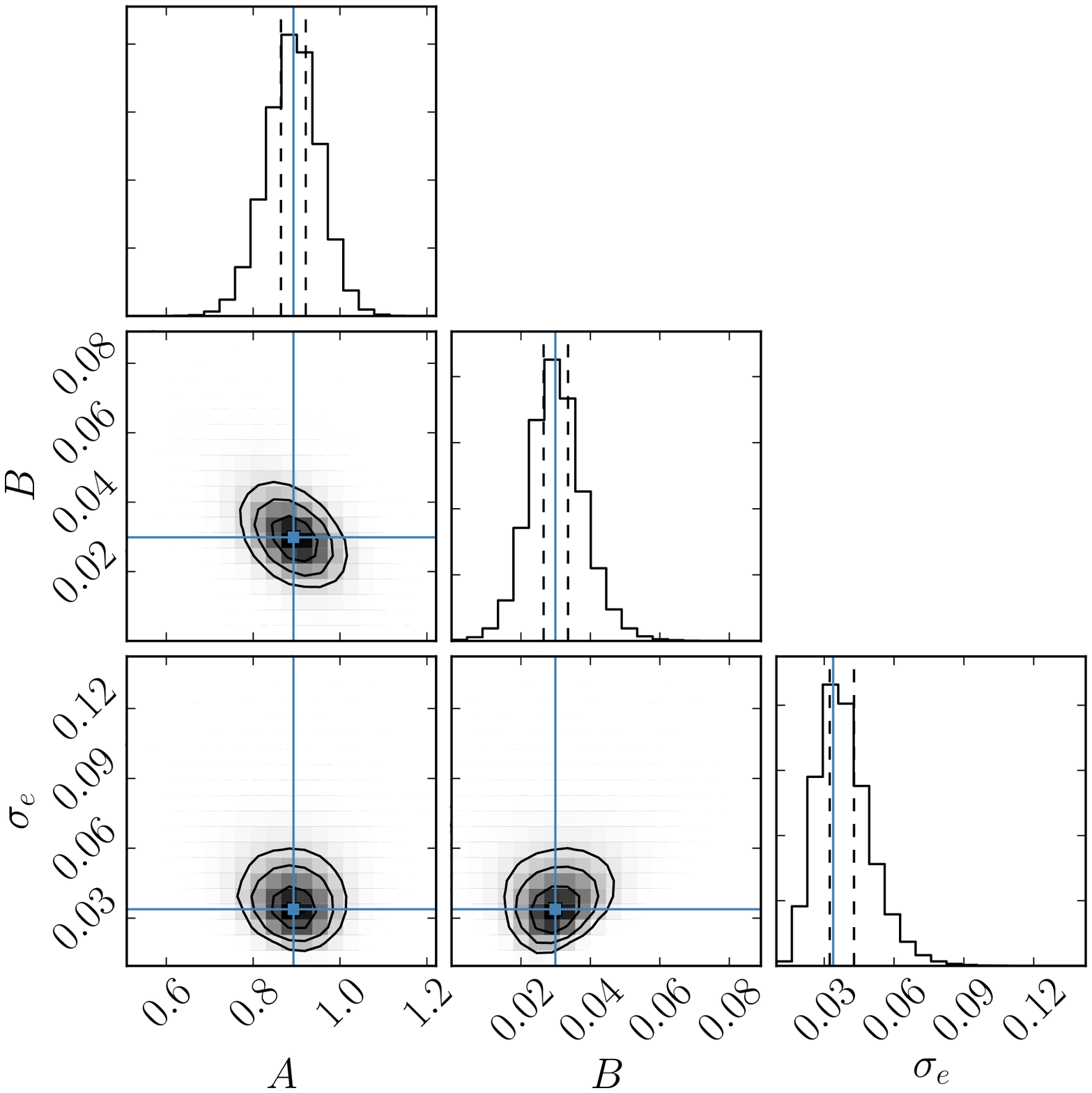}
\vspace{1.1cm}
\caption{Normalised mass function distribution showing an exponential fit to the data for 444 exoplanet candidates (left).  
The associated uncertainties have been calculated assuming Poisson statistics. The right plot shows the parameter space contours and histograms 
(aka. a corner plot; Dan Foreman-Mackey et al. (2016). corner.py: corner.py v1.0.2. Zenodo. 10.5281/zenodo.45906) constructed from the MCMC chains.  
The contours show the exponential scaling (A), the exponent ($\alpha$), 
and the excess noise for 1, 2, and 3$\sigma$ percentiles radiating outwards from the point of maximum probability of the distributions.  
The cross-hairs mark the values determined from the 
maximum likelihood best fit.  The right-edge plots show the histogrammed distributions collapsed in only the $x$ dimension, where the 
mean (solid) and 1$\sigma$ (dashed) ranges are highlighted.
}
\label{new_func}
\end{figure*}

The observed mass distribution is a key observational constraint for planet formation models, a constraint which has previously been fit by smooth 
power law trends with indices around -1 (e.g. \citealp{butler06}; \citealp{lopez12}).  In Fig.~\ref{new_func} we show the results of applying an exponential 
function to the data, which we found to be more suited to the current distribution of exoplanets that have been detected over a wide range in stellar mass 

\begin{equation}
f(m) = A \times e^{msin(i)} + B
\end{equation}

where $f(m)$ is the model function that we fit to the data and $A$ and $B$ are the scaling parameter and offset of the model that are left as free 
parameters to be found following a maximum likelihood procedure with the following Gaussian likelihood function:

\begin{equation}
\mathcal{L}(\Theta) = -0.5 \times \log(2\pi) - \sum_i \log(\sigma_{t,i}) - \frac{\sum_i (y_i-f(m)_i)^2}{\sigma_{t,i}^2} 
\end{equation}
\begin{equation}
\sigma_{t,i} = \sqrt{\sigma_{p,i}^2 + \sigma_{e,i}^2}
\end{equation}

Here $\mathcal{L}$ is the likelihood function for parameters $\Theta$, $y$ is the observed data (mass function histogram points) for all $i$ 
histogram points, and $\sigma_p$ and $\sigma_e$ are the 
Poisson uncertainties and any excess uncertainty for each of the values, respectively.  This procedure finds the following values for the modeled 
parameters 0.89$\pm$0.03, 0.030$^{+0.004}_{-0.003}$, and 0.034$^{+0.009}_{-0.002}$ for $A$, $B$, and $\sigma_e$, respectively.  The uncertainties on these 
parameters were determined using a Markov Chain Monte Carlo (MCMC) procedure in \sc Python\rm, employing the 
\sc emcee \rm numerical package (\citealp{foreman13}).  We used 100 walkers and ran chains of 10000 steps in length, with a 1000 step burn-in, which 
relates to a final chain length of 900,000 steps, with a final mean acceptance rate of 49\%.  The parameter values we measure are insensitive to small changes 
in the bin size used in the histogram, which we set to be 0.5\mj, a value that allows enough samples in most of the bins to reflect the smoothly varying 
distribution.

At the right of Fig.~\ref{new_func} we show the parameter extent probed by the chains, where we used uniform 
priors for the parameters except the excess uncertainty, where we employed a Jeffries prior where the probability is proportional to 1/$\sigma$.  
The distribution of the parameters are well confined to the region around the maximum likelihood value for each, showing the model we put forward 
is an acceptable representation of the current exoplanet mass function.  We note that the $A$ and $B$ parameters follow Gaussian distributions, 
whereas the excess noise parameter is more like a skewed Gaussian or Poisson distribution, where the lower 1$\sigma$ credibility limit is found 
to be close to the maximum likelihood value of 0.03.  In any case, it seems that the mass function appears to be fairly well described by an exponential 
function.

\subsection{Mass-Metallicity Functions}

\begin{figure}
\vspace{5.5cm}
\hspace{-4.0cm}
\includegraphics{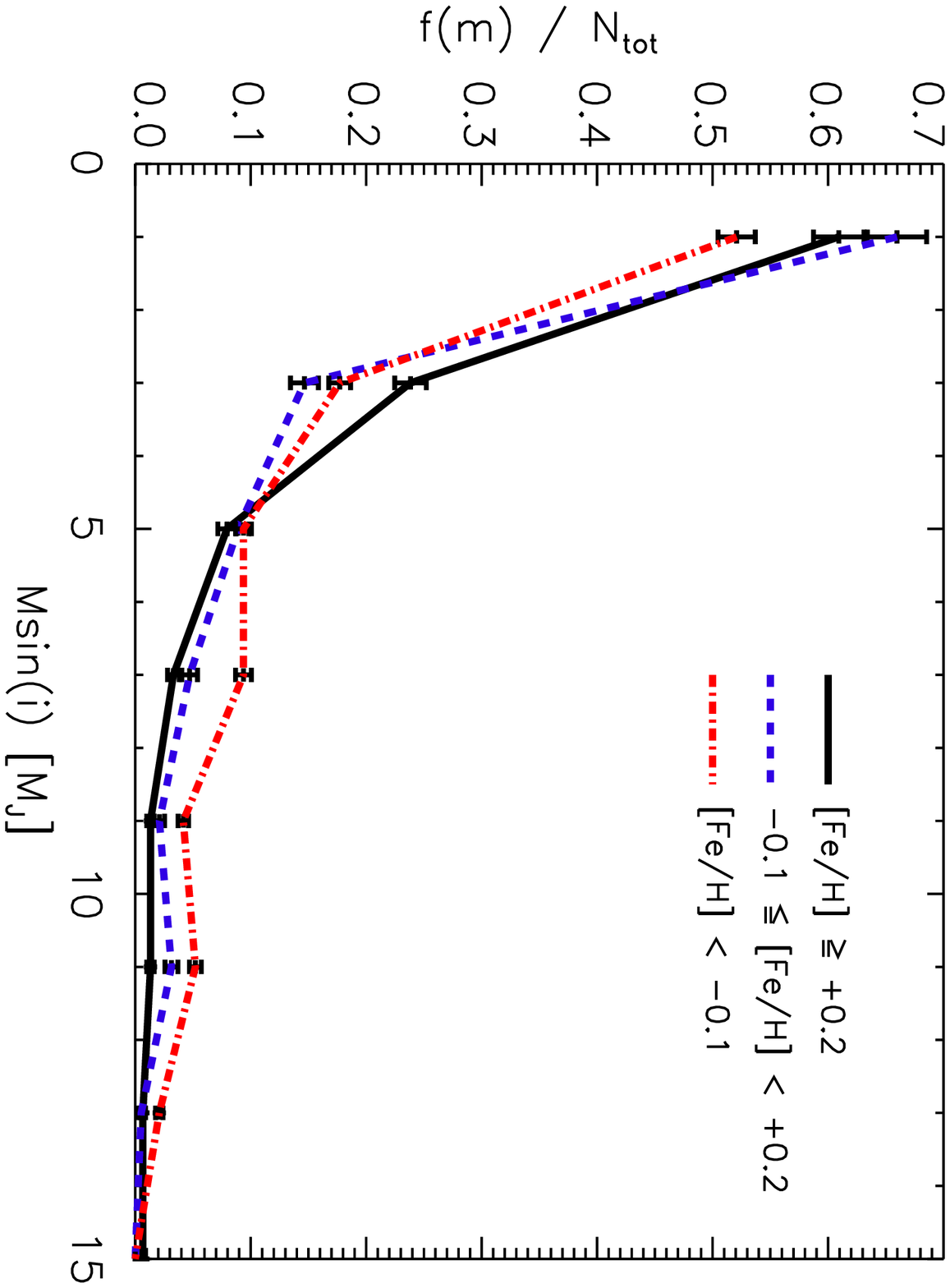}
\vspace{0cm}
\caption{The observed mass distribution split into three bins with different metallicity distributions and a binsize of 
2~\mj.  The solid black curve is for the most metal-rich planet-hosts, the dashed blue curve is for the intermediate metallicity stars, 
and the dot-dashed red curve is the distribution for all metal-poor stars.  The metallicity cuts are shown in the key.}
\label{met_mass}
\end{figure}

As Fig.~\ref{met_mass} shows, we tested if there was any metallicity dependence in the mass function.  In 
order to test this we split the sample into three metallicity bins, a high metallicity bin ([Fe/H]$\ge$+0.2~dex), an intermediate metallicity bin 
(-0.1$\le$[Fe/H]$<$+0.2~dex), and a low metallicity bin ([Fe/H]$<$-0.1~dex).  These bin sizes allowed a useful number of samples in each 
bin to statistically probe the distributions.  

Metallicity splitting gives us probabilities (D-statistics) from two-tailed KS-tests of 8\% (0.165) that the high metallicity 
bin and the low metallicity bin are drawn from the same parent population, and 6\% (0.161) that 
the intermediate metallicity planet-hosts and the low metallicity planet hosts are also drawn from the same population.  By combining the 
high metallicity bin and the low metallicity bin values and comparing those to the intermediate metallicity bin, the probability is essentially the same, 
only dropping the D-statistic by 0.01, with a probability of only 6\% that the two populations are statistically similar.  To perform 
this test we decided to remove the lowest mass planets from the metal-poor and intermediate-metallicity samples since \citet{jenkins13a} 
shows that there appears to be a correlation between the mass and metallicity in the low-mass regime.  Therefore, we only consider planets 
with minimum masses above 0.0184~\mj\ as this is the lowest mass planet in the high metallicity sample, neglecting the exceptional case 
of the planet orbiting Alpha Centauri~B (\citealp{dumusque12}) that \citet{hatzes13} and \citet{rajpaul15} claim may be attributed to other 
phenomena like stellar activity or sampling ghosts.

In order to firm up these statistics we also ran the samples through the Anderson-Darling (AD) test, which generally tends to be more sensitive 
than the standard KS test since it gives more statistical weight to the tails of the distribution.  From these tests we found p-values of 2\% 
and 5\% for the comparison between the high metallicity and low-metallicity samples, and between the intermediate and low-metallicity 
samples, respectively.  This is in good agreement with the KS test results, indicating that there is a correlation between mass and metallicity, 
whereby metal-rich stars produce many more Jupiter-mass planets compared to super-Jupiters, but metal-poorer stars produce a higher fraction 
of super-Jupiters than Jupiters compared to the metal-rich population.  However, the current sample of host star properties were 
not drawn from a homogeneous source and therefore the heterogeneous nature of the data could be influencing the results.  

To try to circumvent this problem, we decided to search for our sample of exoplanet-hosts in the SWEET-Cat catalogue (\citealp{santos13}).  
The SWEET-Cat is a project that plans to eventually contain all exoplanet host star properties like \teff and metallicity that have been measured 
using high resolution spectroscopy in a homogeneous fashion.  We were able to find 93\% of our sample in the SWEET-Cat, 
but some of these were not measured homogeneously.  From this sample we reran the KS tests and found 
probabilities of 9.1\% that the high and low metallicity bins are drawn from the same parent population and 13.8\% that the intermediate 
and low metallicity bins are drawn from the same distribution.

A further step that was taken was to remove even more information but improve the homogeneity of the sample.  We selected only those 
stars with a homogeneous flag of 1 in the SWEET-Cat, which means that the properties of these stars were measured using the same general 
methodology.  This selection resulted in a 20\% loss of information but still contained a total sample size of 358 planet-hosts, however 
the low metallicity bin only contained 64 stars, whereas the high and intermediate bins have sample sizes of 131 and 163 objects, respectively.  
The KS test probabilities are now significantly lower than the full sample, having values of 38.0\% and 52.0\% that the high and intermediate 
mass functions are statistically similar to that of the low metallicity bin.  These tests likely show that currently there are no statistically significant 
correlations between planetary mass and the metallicity of their host stars, as claimed by \citet{mortier12}, and the overabundance 
of Jupiter's is not due to the enhanced formation of such planets as a function of metallicity.  We did not run the SWEET-Cat samples through 
the AD test since the results were shown to be very similar to the KS tests for the full sample.

\subsection{Other Observational Properties}

Within the period-mass plane some features can be seen in the metallicities of exoplanet host stars.  An examination of the left plot in Fig.~\ref{properties} 
reveals that there is a broad mix of metallicities for the gas giant planets and the planets we publish here are located predominantly in the upper 
right quadrant of the plot, with only two having periods below 100~days.

\subsubsection{Host Star Metallicities}

\begin{figure*}
\vspace{7cm}
\hspace{-4.0cm}
\includegraphics{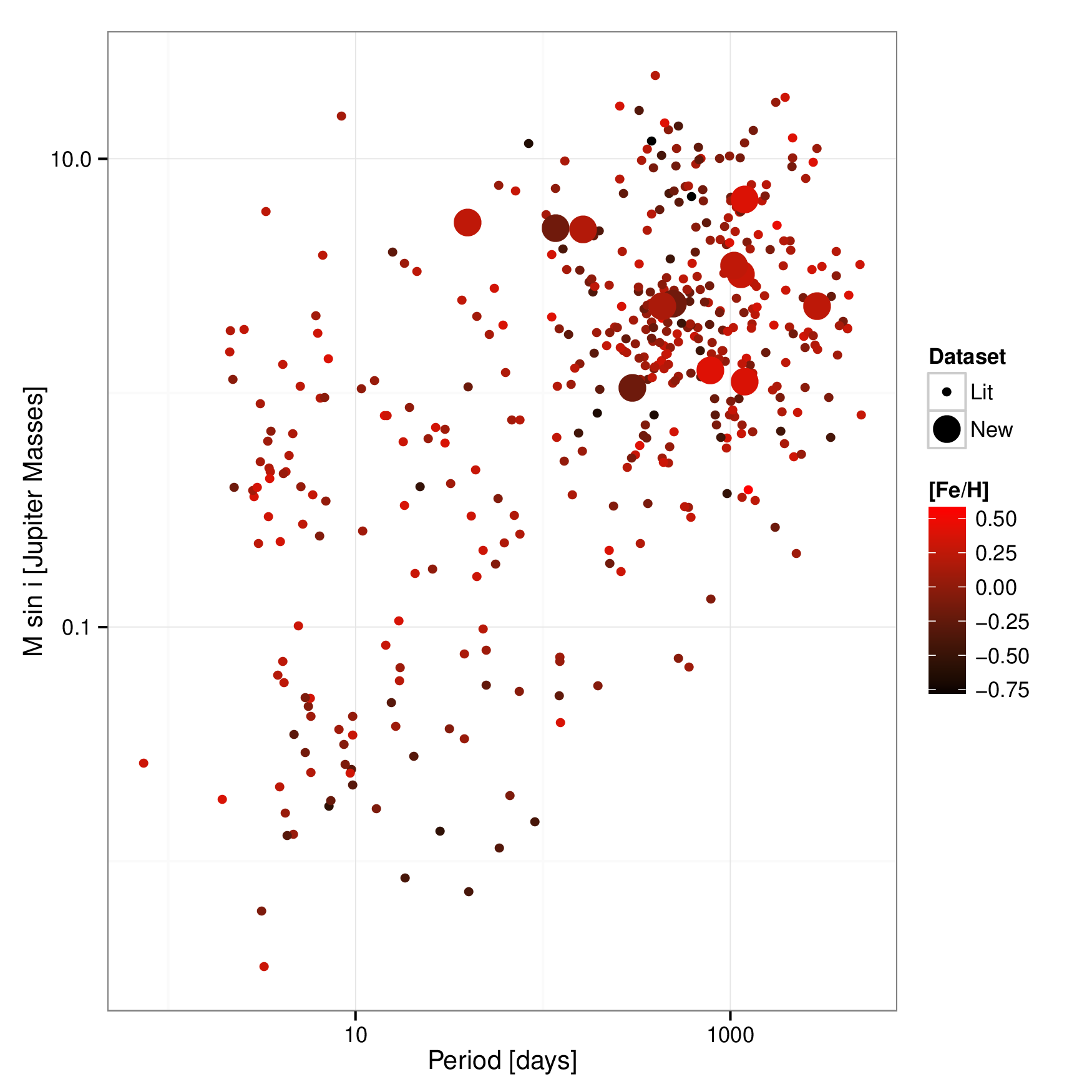}
\includegraphics{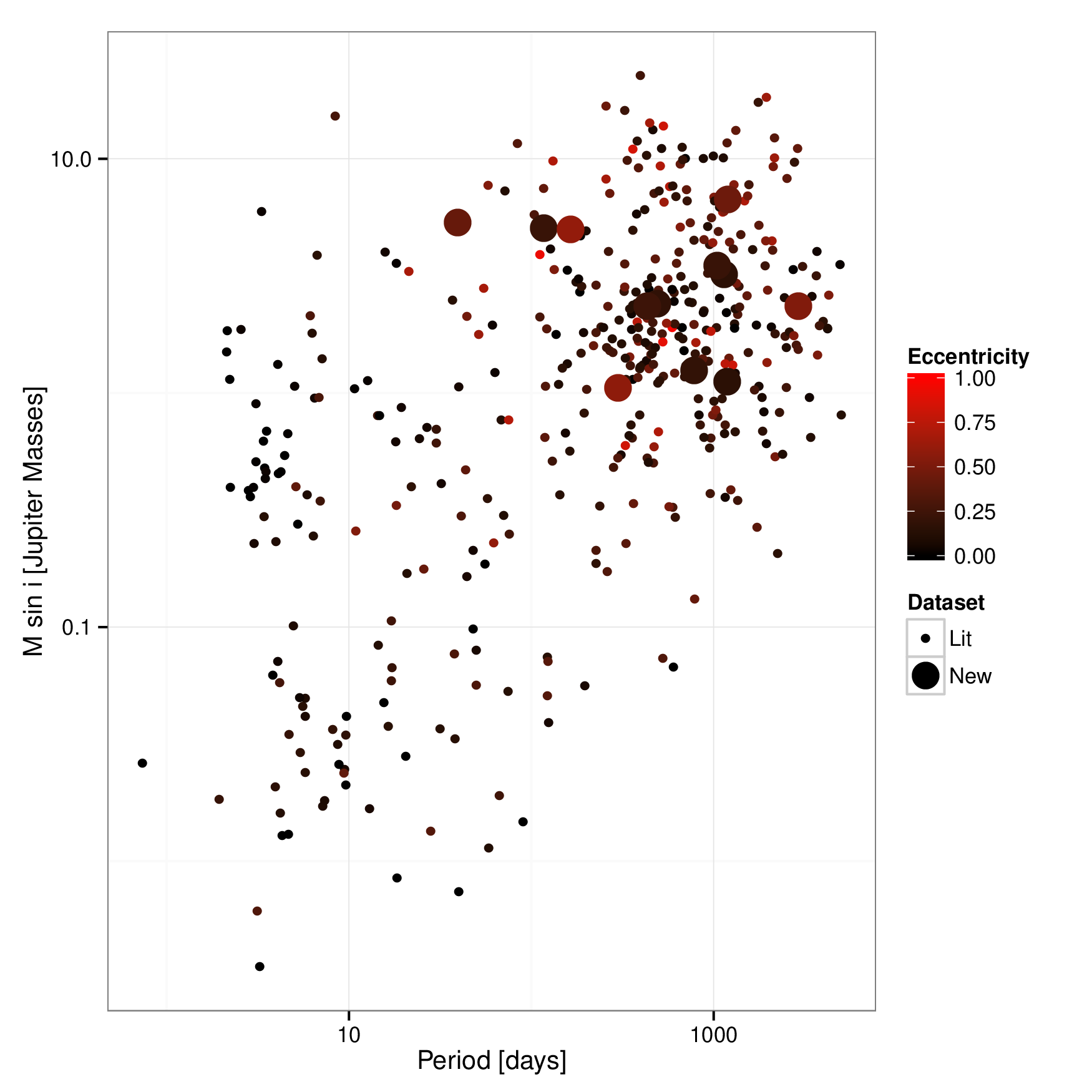}
\vspace{1cm}
\caption{Distribution of exoplanet metallicities (left) and eccentricities (right) within the period-mass plane.  The small circles are the 
literature values and the large circles are our targets from this work.  The colour scale to highlight the differing metallicities and 
eccentricities are shown at the right of both plots.}
\label{properties}
\end{figure*}

It appears that the lowest mass planets are found mostly on short period orbits, due to the inherent sensitivities of Doppler surveys, and they also appear to 
orbit metal-poor stars in general, hence the dominance of the black points towards the bottom left of the left hand panel in Fig.~\ref{properties}.  
This result was previously highlighted by \citet{jenkins13a}, revealing a 'planet desert' for the most metal-rich stars, and subsequent 
confirmation has also been discussed in \citet{marshall14}.  The nature of this desert could be explained by core accretion theory whereby the lower 
density discs have limited metals to form cores, whereas the high density discs can readily form high-mass cores that quickly grow to more massive objects, 
crossing the critical core mass limit and becoming gas giants.  In fact, the metallic properties of all planets with periods of less than 100~days appears to 
be different when we compare planets more massive or less massive than 0.1~\mj.  In the plot, this is shown by the significantly higher fraction of 
red data points above a mass of 0.1~\mj\ compared to below that limit. 

\begin{figure}
\vspace{5.5cm}
\hspace{-4.0cm}
\includegraphics{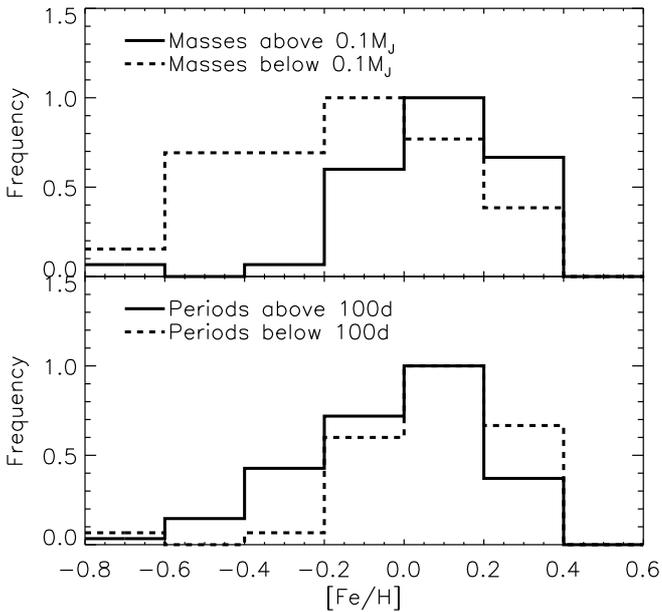}
\vspace{2.5cm}
\caption{The binned histogram of sub-100~day period exoplanet host star metallicities detected by radial velocity programs is shown in the top plot.  
The solid histogram represents the gas giant planets, minimum masses above 0.1~\mj, and the dashed histogram is for host stars with planets below this threshold.  
The lower plot shows the binned histogram of gas giant planets (minimum masses above 0.1~\mj) for orbital periods of 100~days or less (solid histogram) and those 
with periods above 100~days (dashed histogram).  All histograms have been normalised to the peak of the distribution to highlight their differences.}
\label{met_hist_sub100}
\end{figure}

More directly we can test the reality that the metallicity distribution for planets with periods of 100~days or less have a different metallicity 
distribution by again applying 
an AD test to the sample of known exoplanets.  We chose to apply the sample to the homogeneous samples that we previously 
cross-matched with the SWEET-Cat list.  This test reveals a T-statistic of 8.54 when adjusted for all non-unique values, revealing a probability 
of 2$\times10^{-4}$ that these samples are statistically similar.  A Kolmogorov-Smirnov test yields a similar probability value (4$\times10^{-4}$) 
with a D-statistic of 0.375.  The histograms of both populations are shown in the top plot of Fig.~\ref{met_hist_sub100}.  Although the 
two histograms appear to show similar forms, the host stars that contain lower-mass planets currently has a flatter shape than the host stars containing 
higher-mass planets in this period space.  Although the sample sizes are small, 48 objects in the low-mass population for example, 
it does appear that the lowest-mass planets are drawn from a different metallicity sample when compared with the most massive planets, 
within the limits of the current data set.  In the future with many more discoveries of very low-mass planets orbiting the nearest stars from Doppler 
surveys, since these represent the most precise metallicities that can be measured, trends such as those discussed here can be tested at a 
higher level of statistical significance.

Further to this, the high-mass planet sample may indicate there is a non-uniform metallicity distribution as a function of period.  To test this we split the high-mass 
planet sample into two bins with orbital periods less than or equal to 100~days and those beyond 100~days.  The binned histogram for both samples is shown in the 
lower plot of Fig.~\ref{met_hist_sub100}, and although the distributions appear less discrepant than the mass cut in the top plot, there is an indication of a functional 
change in the metal-poor regime.  The AD 
test of the metallicities from these samples returns a T-statistic of 7.62, leading to a probability of 4$\times10^{-4}$ that the samples are similar.  This time the 
Kolmogorov-Smirnov test reveals a slightly smaller probability of the null hypothesis, returning a value of 8$\times10^{-3}$ and a D-statistic of 0.223.  Therefore 
we find that, in general, short period giant planets have higher metallicities than those at longer periods, with a mean value of [Fe/H] of 0.16~dex for the 
sub-100~day planets and a value of 0.06~dex for the giant planets with orbital periods longer than 100~days, as suggested by \citet{sozzetti04} and \citet{pinotti05}.

\citet{mordasini12} constructed global population synthesis models of forming planets in a range of disc environments to search for expected correlations 
between planetary orbital parameters and bulk compositions against disc properties.  They found that planets tend to migrate more in low-metallicity 
discs compared to more metal-rich discs because the cores that form in the low-metallicity environment need to migrate 
more to undergo enough collisions to grow to the critical mass limit and transition from Type I migration to the slower Type II migration.  They suggest 
no clear correlation between semimajor axis, or orbital period, exists because the planets in low-metallicity discs also form further from the central star 
than in the high-metallicity discs, and so the increased efficiency of migration in the low-metallicity environment is compensated by the increased 
distance the planets need to travel inward towards the star.  

These modeling efforts tend to be at odds with the findings we have made unless certain conditions apply.  If giant planets in metal-poor discs migrate more 
then we would expect to see the opposite result, unless the planets start their journeys very far out in the disc before arriving at their current locations.  In 
addition, another scenario could be that the low-metallicity discs are dispersed faster than high-metallicity discs through photo-evaporation (\citealp{yasui09}; 
\citealp{ercolano10}).  This effect is thought to be due to the lower optical depth of the 
disc allowing the UV and X-ray flux to pass deeper into the disc, dispersing the inner regions faster, meaning there is no remaining gas and dust 
for the planet to interact with, essentially halting its migration earlier when compared to a planet migrating through a metal-rich disc.  Finally, in the 
Mordasini et al. model, they predominantly consider mostly inward migration of cores, however recent work has shown that disc structure is important 
in defining the dominant torques that drive planet migration and in the inner discs that are heated by the intensity of the young star's radiation field, 
corotation torques dominate over the differential Linblad torques, leading to outward migration of the cores (\citealp{kretke12}).  Further to this, 
random walk motion too can be important for migrating low-mass cores (\citealp{nelson04}; \citealp{laughlin04}; \citealp{nelson05}) and 
dead-zones in the disk can subsequently halt the migration of 
forming low-mass cores (\citealp{balmforth01}; \citealp{li09}; \citealp{yu10}).  All of these processes could lead to the preservation of a period-metallicity 
relationship that favours short period planets predominantly being found orbiting more metal-rich stars and longer-period planets being found 
in more metal-poor environments.

\subsection{Orbital Eccentricities}

In the right plot in Fig.~\ref{properties} we show the same period-minimum mass plane, yet this time the colour scaling highlights the eccentricity distribution.  
We can see that the majority of the short period planets ($P \le$10~days) are generally found to have circular orbits, a fact that can be attributed to the planets 
tidal interactions with the host star that tends to circularised their orbits.  We also see that the majority of the low-mass planets are found on circular orbits 
too (black points), in comparison to the high-mass planets where a significant fraction of them have moderate-to-high eccentricities (red points).  Note that there is also 
a selection bias towards the detection of higher eccentricities that depends on the quantity of radial velocity data points that describes a given signal 
(\citealp{otoole09}), whilst high eccentricity also elevates the amplitude of a given Doppler signal, which can sometimes make them easier to detect.

If we again split the planets up into two mass bins, 
where the low-mass planets have minimum masses of $\le$0.1~\mj\ and the high mass bin comprises all planets with minimum masses above this limit, the eccentricity 
means and standard deviations of the two populations are 0.13 and 0.12 for the low-mass planet population and 0.25 and 0.21 for the high-mass planet sample.
Taken at face value, the increased standard deviation for the higher mass sample tells us the spread in eccentricities in this mass regime is higher than for the 
lower mass planets.  There is a strong bias here where the low-mass planet sample has a significantly lower mean orbital period, with a much higher fraction 
of planets orbiting close enough to the star to be quickly circularised through tidal dissipation of the orbits.  Furthermore, there is a tendency to fix the eccentricity 
to zero when performing Keplerian fits to radial velocity data, in order to remove this additional degree of freedom and the degeneracy with other parameters being 
fit at the same time.

\section{Summary}\label{conclusions}

We have used the CORALIE, HARPS and MIKE spectrographs to discover eight new giant planets orbiting seven super metal-rich stars, and 
a star of solar metallicity, along with updated orbits for four previously published planets.  We include radial velocity data prior- and post-2014 
CORALIE upgrade and our Bayesian updating method returned a systematic offset of 19.2$\pm$4.8~\ms between the two velocity sets for our stars.    
The new planets cover a wide area of the 
giant planet parameter space, having a range of masses, periods, and eccentricities, including a double planet system that was 
found orbiting the most massive star in our list, and 
a 14~Herculis~$b$ analogue that has a minimum mass of $\sim$5.5\mj, an orbital period of nearly 1200~days, and significant 
eccentricity (e=0.46), adding another member to the sub-population of massive eccentric planets orbiting super metal-rich stars.

We introduced our method for measuring the chromospheric $S$-index that is a measure of the magnetic activity of Sun-like stars using 
CORALIE spectra.  These activities, along with bisector measurements, CCF FWHM's, H$\alpha$ indices, HeI indices, and Hipparcos and ASAS photometry, were used 
to rule out the origin of the planetary Doppler signal as being due to line modulations from rotationally influenced star spot migration or 
other activity phenomena like chromospheric plage or stellar pulsations.

We show that the mass function for planets is well described by an exponential function with a scaling parameter of 
0.89$\pm$0.03 and an offset of 0.030$^{+0.004}_{-0.003}$.  We confirm the lack of the lowest-mass planets orbiting metal-rich stars 
and we also find a period-metallicity correlation for giant planets.  The population of planets with masses $\ge$0.1~\mj\ and orbital 
periods less than 100~days is found to be more metal-rich than the same mass 
planets with orbital periods greater than 100~days.  The difference is significant at the 0.004\% level and the mean difference is found to be 
0.16~dex between the two populations.  This result could be describing the formation locations of planets in the early disks, with metal-rich 
disks forming planets in the inner regions and metal-poor disks forming planets further out in the disk, or that giant planets migrate more in 
metal-rich disks, due to a stronger torque interaction between the high surface density disk and the migrating planet.

\section*{Acknowledgments}

We thank the anonymous referee for providing a fair and helpful report.  We are also thankful for the useful discussions with Francois Menard and 
to Andres Jord{\'a}n for providing access to the CORALIE pipeline.  
JSJ acknowledges funding by Fondecyt through grants 1161218 and 3110004, partial support from CATA-Basal (PB06, Conicyt), the GEMINI-CONICYT FUND 
and from the Comit\'e Mixto ESO-GOBIERNO DE CHILE, and from the Conicyt PIA Anillo ACT1120.  MJ acknowledges financial support from Fondecyt project \#3140607.  
DM is supported by the BASAL CATA Center for Astrophysics and Associated Technologies through grant PFB-06, 
by the Ministry for the Economy, Development, and Tourism’s Programa Iniciativa Científica Milenio through grant IC120009, awarded to Millenium 
Institute of Astrophysics (MAS), and by FONDECYT Regular grant No. 1130196.  
This research has made use of the SIMBAD database and the VizieR catalogue access tool, operated at CDS, Strasbourg, France.

\bibliographystyle{mnras}
\bibliography{refs}

\appendix

\section{Stable Star Results}\label{appendix_stable}

To confirm the stability of the CORALIE pipeline reduction and analysis procedure we have observed a star known to 
be radial velocity stable at the $\sim$2~\ms level that should provide an ideal test candidate for our method.  HD72673 
is a bright ($V$=6.38), nearby (12.2~pc), and inactive (\rhk~=~-4.946 \citealp{isaacson10}) G9 dwarf star that has a metallicity of 
-0.38$\pm$0.04~dex (\citealp{marsakov88}; \citealp{santos04}; 
\citealp{valenti05}), and a mass and age of 0.814$\pm$0.032~\msun and 1.48$^{+5.44}_{-1.48}$~Gyrs (\citealp{takeda07}), respectively.  

\begin{figure}
\vspace{5.5cm}
\hspace{-4.0cm}
\includegraphics{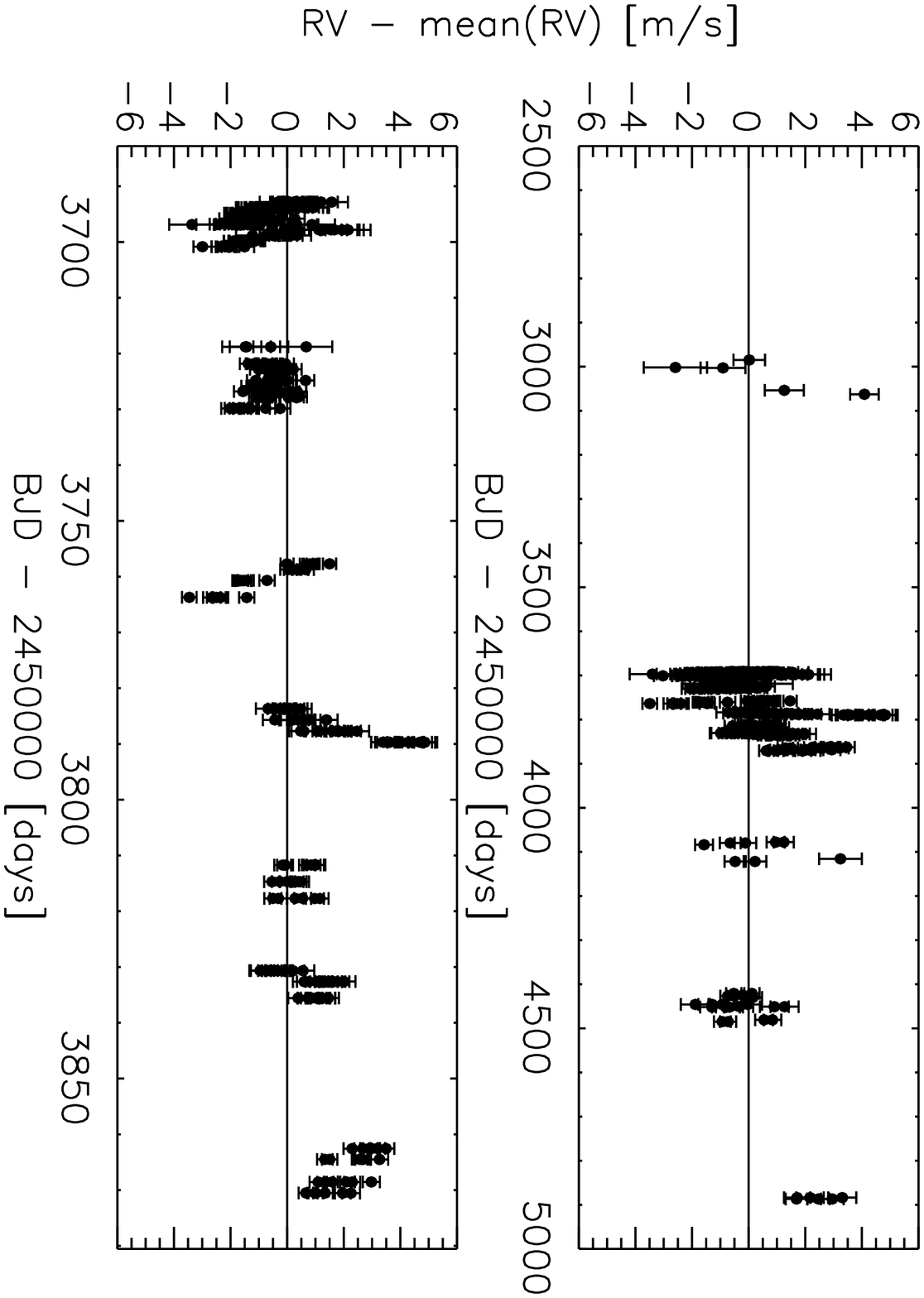}
\vspace{1cm}
\caption{HARPS radial velocity timeseries of HD72673 with the mean subtracted off the data.  The upper panel shows the full timeseries 
and the lower panel is a zoom in on the most densely observed epoch for this star.}
\label{hd72673_harps}
\end{figure}

In the upper panel of Fig.~\ref{hd72673_harps} we show the radial velocity timeseries for HD72673 observed with HARPS that was taken from the ESO 
Archive\footnote{Based on data obtained from the ESO Science Archive Facility under request number JJENKINS 50958.}. 
The data span a baseline of over 1900~days in total and comprise 363 individual radial velocity measurements where we have removed 5$\sigma$ outliers that corresponded to 
bad weather observations, and therefore had very low S/N data, like the point at BJD-2453724.79378.  After subtracting off the mean of the data, 
which we use as our standard flat noise model, we 
find a rms of 1.44~\ms.  The lower panel shows the same data except zoomed in on the most densely sampled observing epoch (BJD 2453500 - 
2454000).  Some structure is found in the radial velocities throughout this epoch and could be the first signatures of low-amplitude Doppler 
shifts induced by orbiting low-mass planets, or stellar activity signals affecting the velocities.  In any case the HARPS velocities agree that 
HD72673 does not show large radial velocity variations and is a useful star for testing the precision we can achieve with our CORALIE pipeline.

\subsection{CORALIE Observations}

Over the course of four years we have performed 108 observations of the star HD72673 with CORALIE, combining data from this project and 
also from the HAT-South (\citealp{bakos12}) CORALIE observations.  The sampling and time baseline provide an excellent diagnostic test of the 
long term stability that is currently attained with the CORALIE using the procedure described in \citet{jordan14}.  

\begin{figure}
\vspace{5.5cm}
\hspace{-4.0cm}
\includegraphics{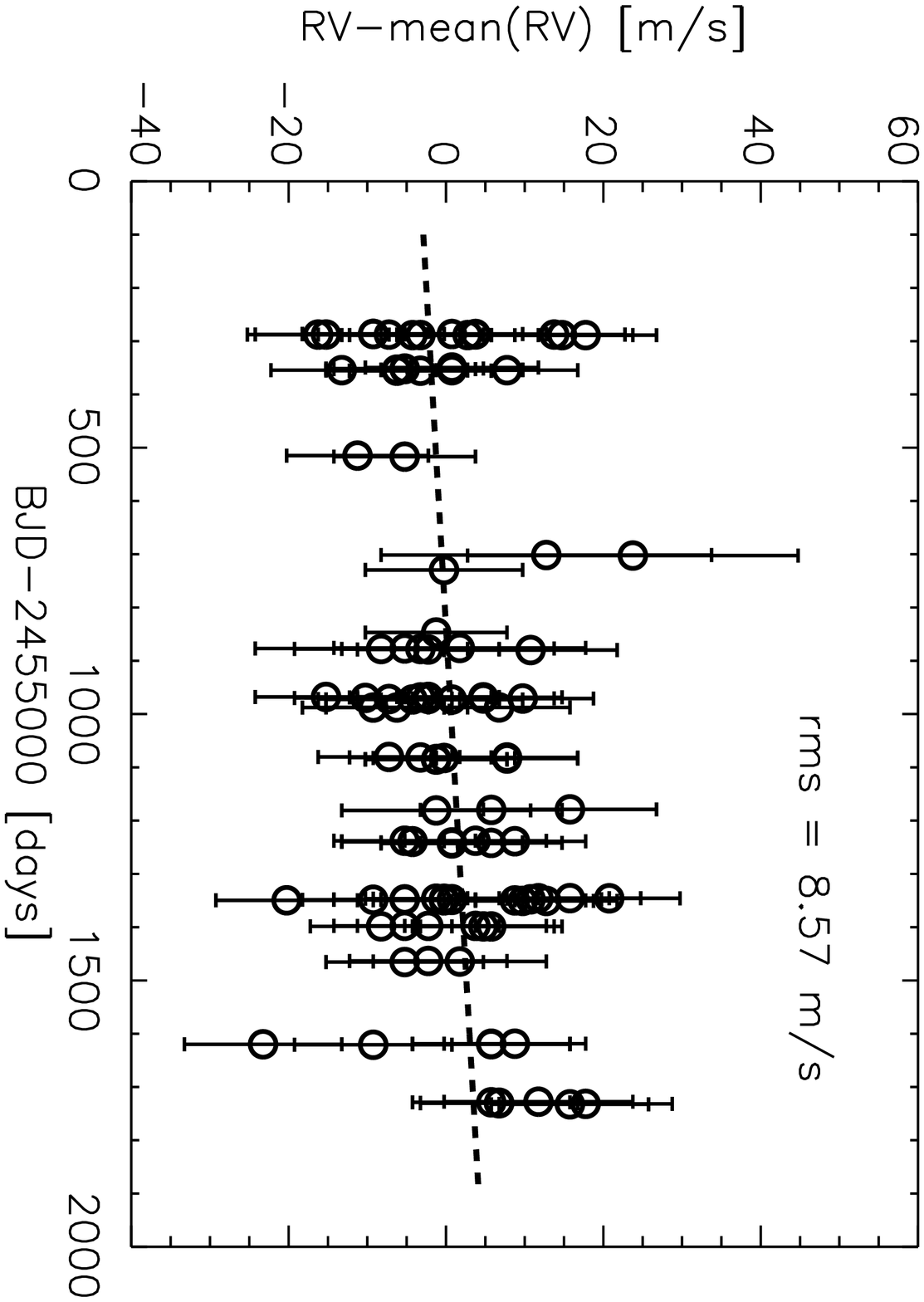}
\vspace{0.6cm}
\caption{Time-series and mean subtracted radial velocities for the star HD72673.  The dashed line represents the best straight line fit to the data.  
The rms scatter around the fit in \ms is shown in the plot.}
\label{hd72673_rv}
\end{figure}

For the observations of HD72673 we aimed to get a S/N of around 100 across the optical regime of interest, leading to typical integration times of $\sim$5~minutes.  
Fig.~\ref{hd72673_rv} shows the full radial velocity dataset 
as a function of time and clearly we see only a small linear trend over the full baseline of observations.  No large systematic trends are found in our 
dataset and the gradient of the best fit we show is 0.004~\ms/day, well below the intrinsic scatter of our procedure. 
The rms scatter for the full data is found to be 10.9~\ms, however after removing a 5-$\sigma$ outlier due to low S/N we arrive at a 
scatter of 8.7~\ms, or 8.6~\ms after subtraction of the linear trend shown in the figure.  Therefore, we consider the precision of the CORALIE observations 
to be 9~\ms, consistent with the precision reported by \citeauthor{jordan14}, but covering a longer time baseline.

\section{MMSE Periodograms}

The minimum mean square error periodograms from the radial velocity timeseries data discussed in this work.  The vertical dashed 
lines mark the signals detected by the MCMC search algorithm.

\begin{figure*}
\vspace{5.5cm}
\hspace{-4.0cm}
\includegraphics{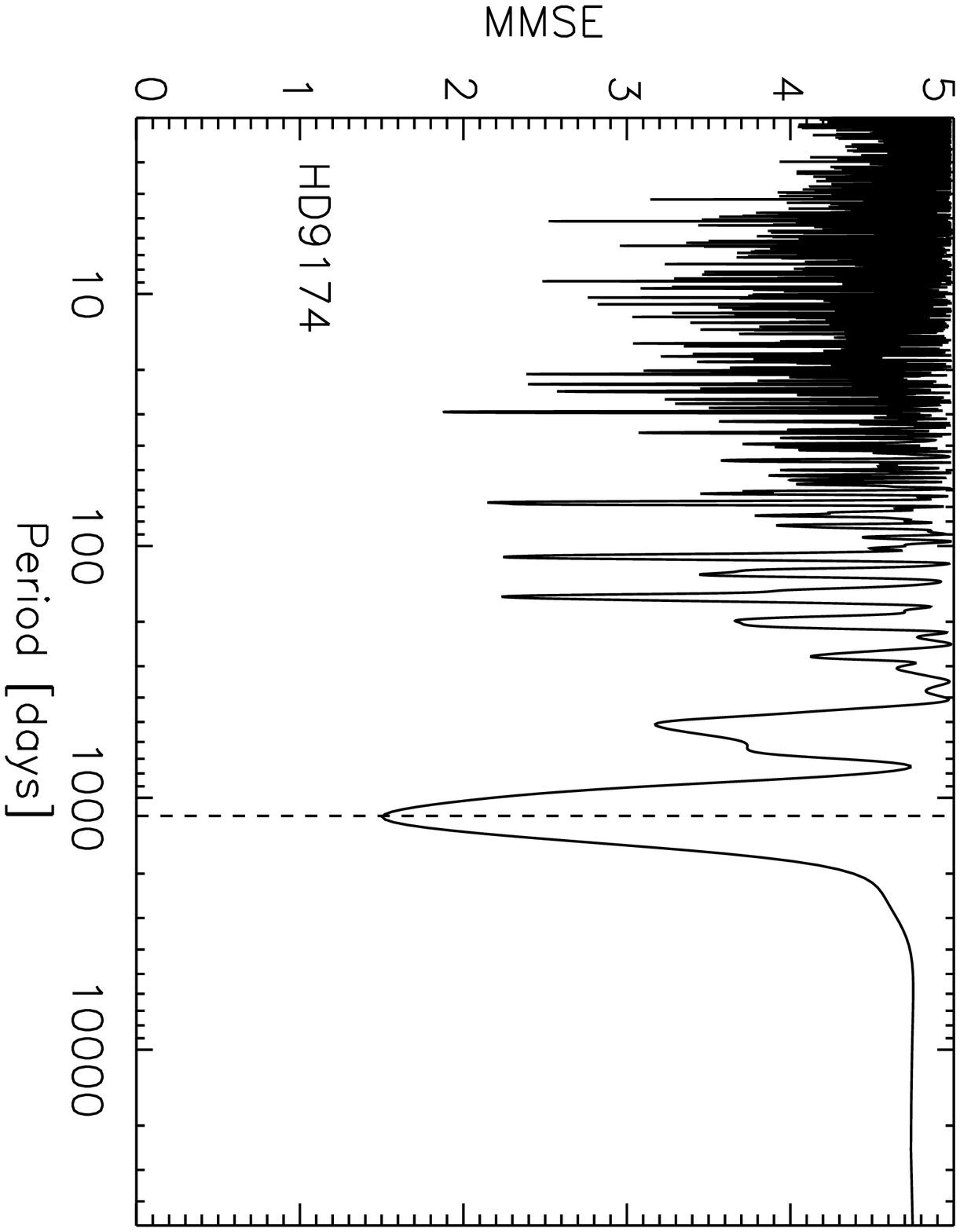}
\includegraphics{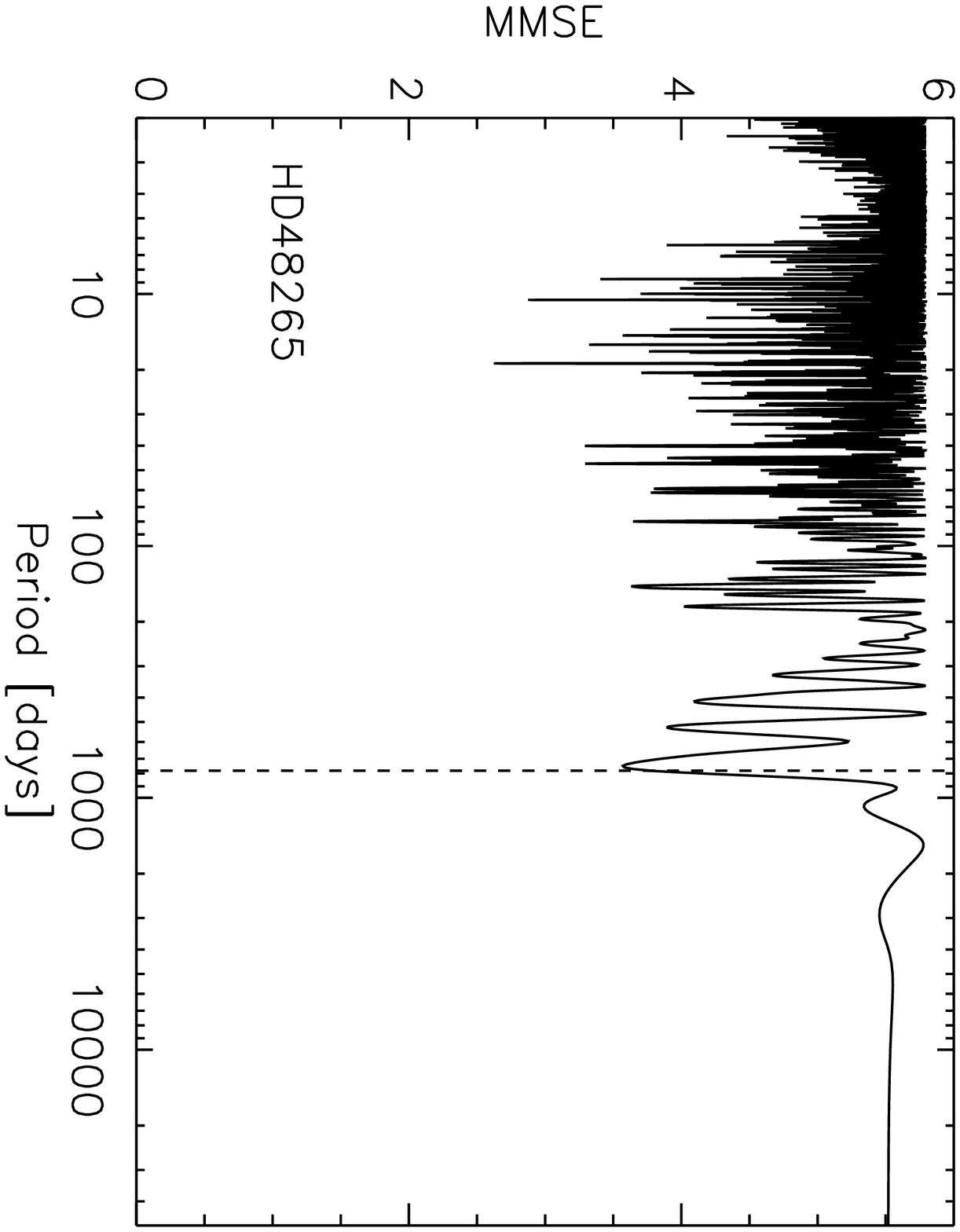}
\includegraphics{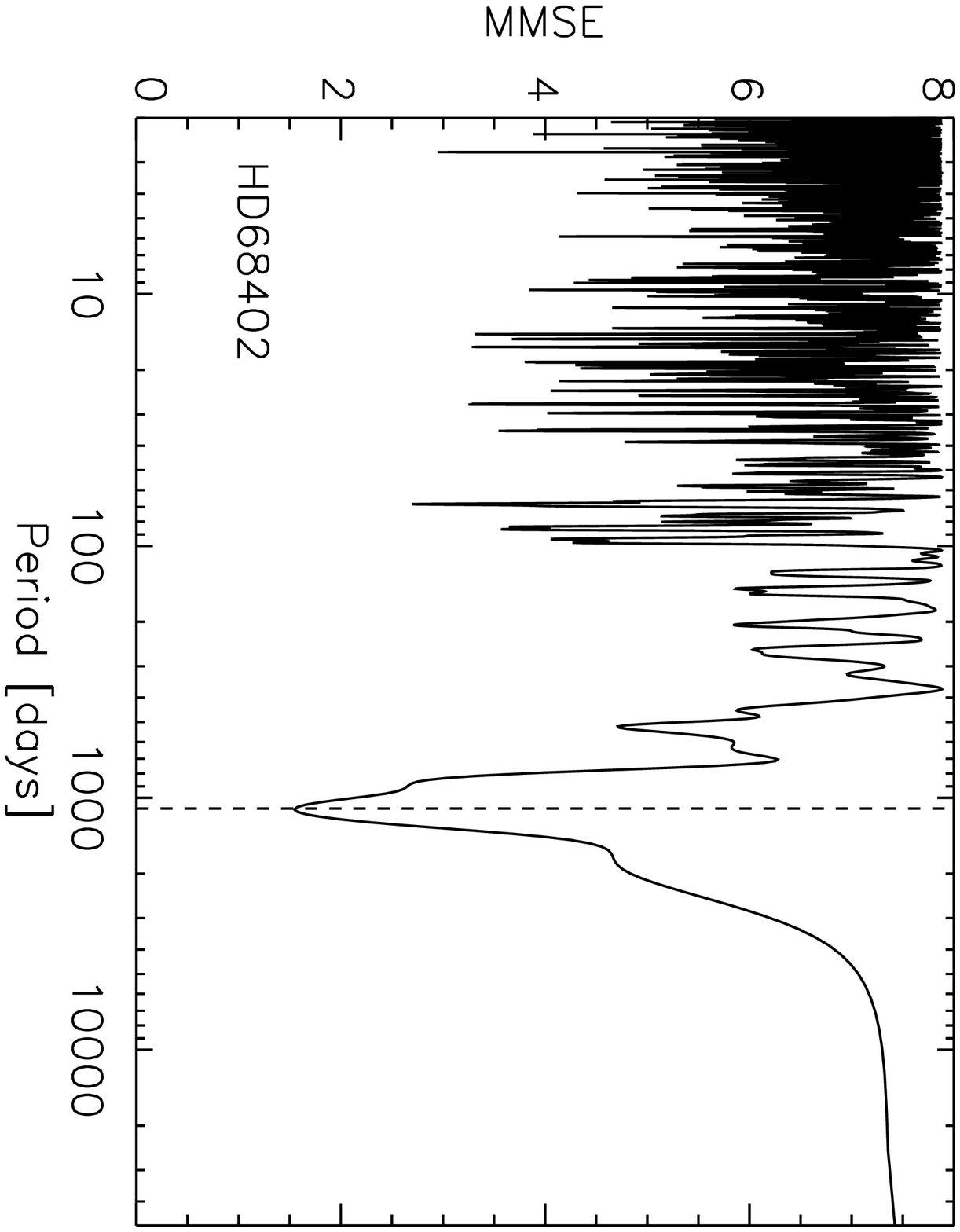}

\includegraphics{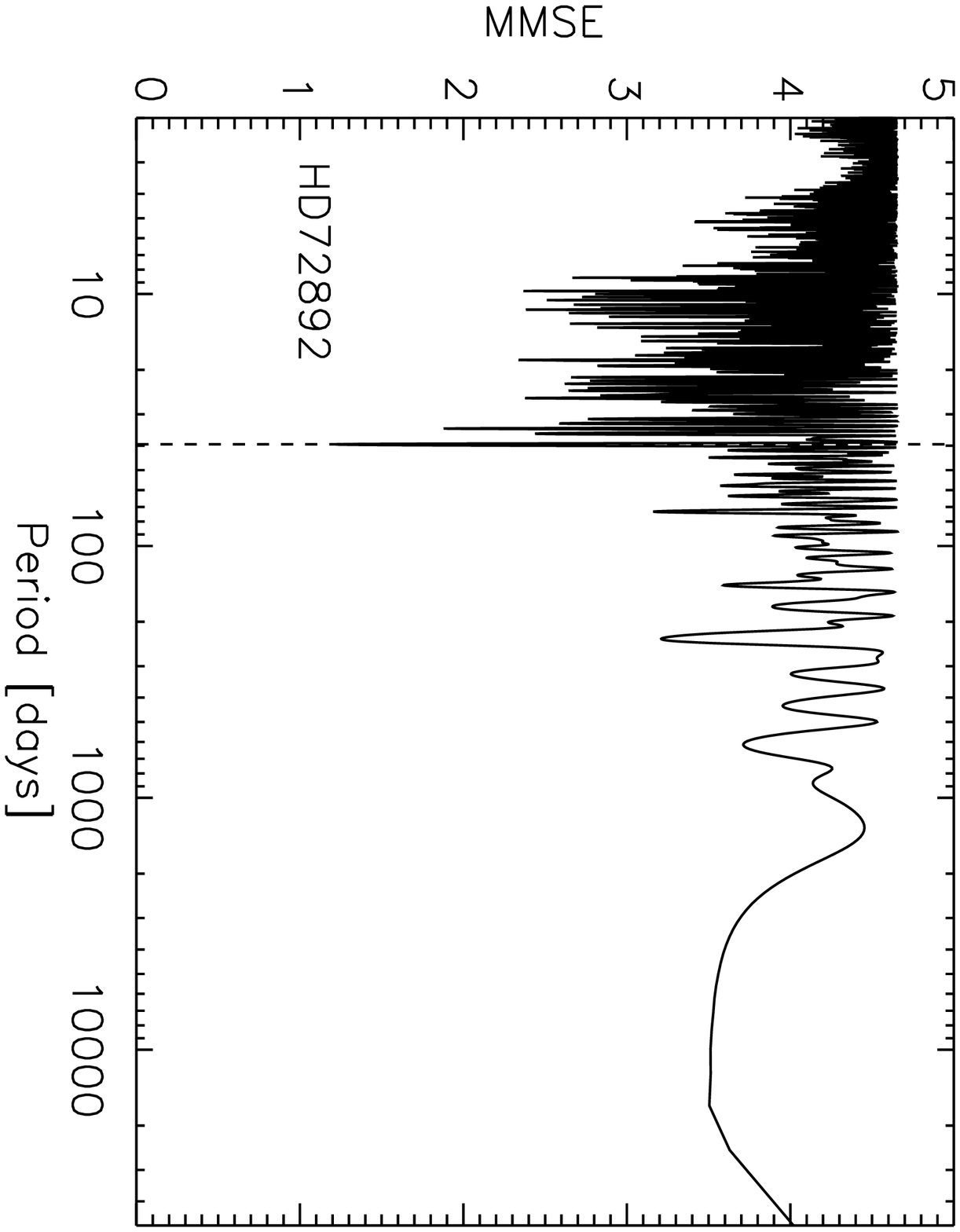}
\includegraphics{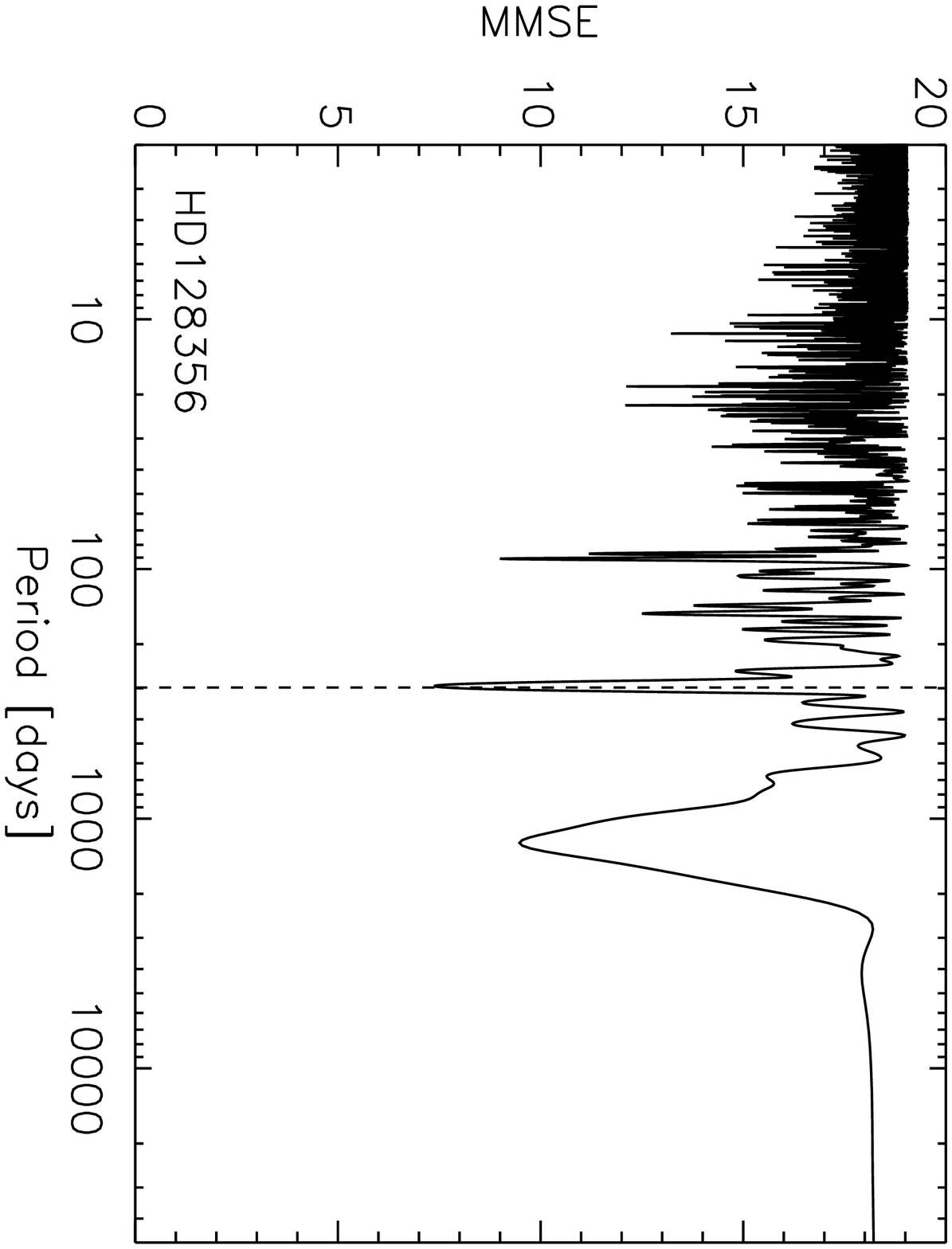}
\includegraphics{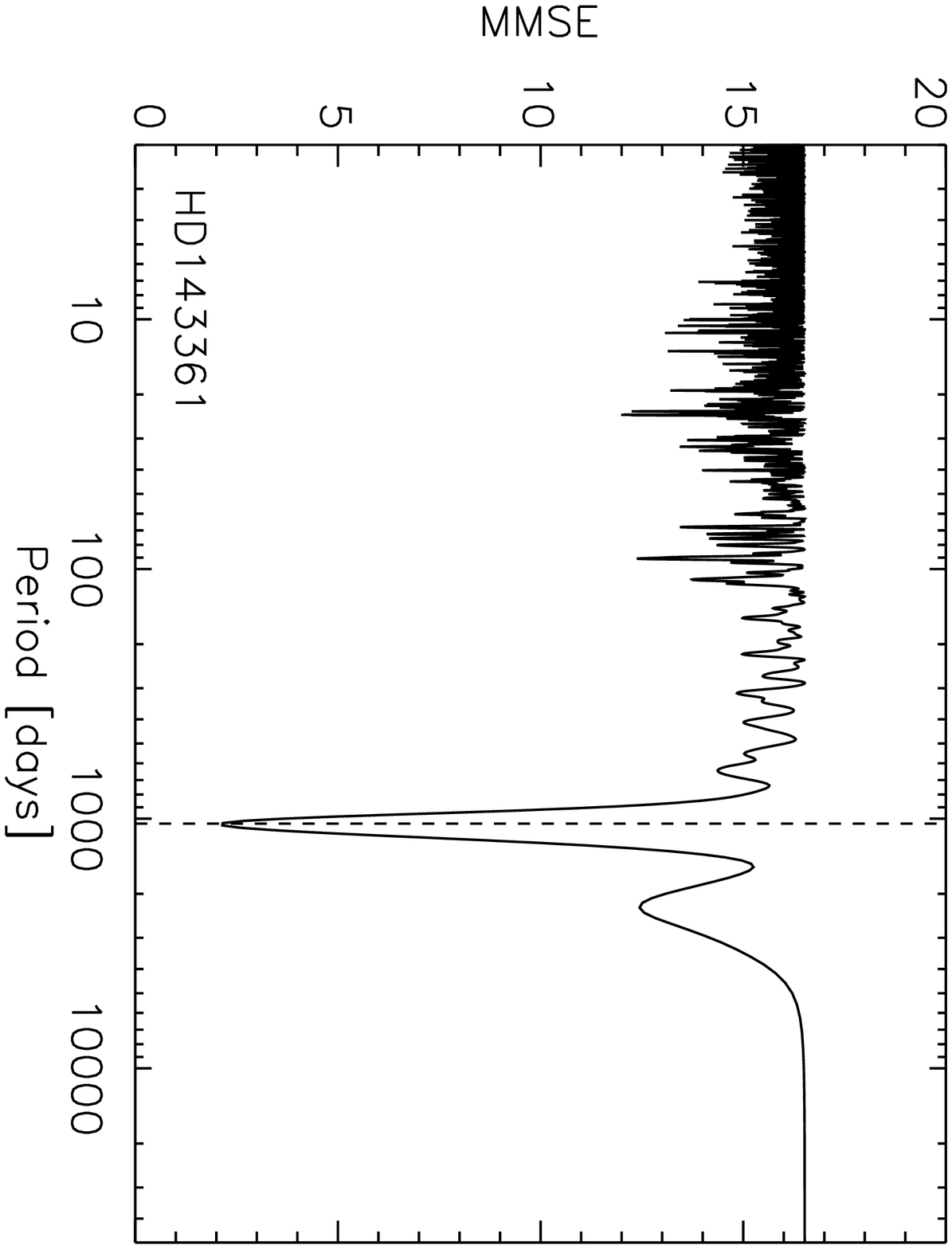}

\includegraphics{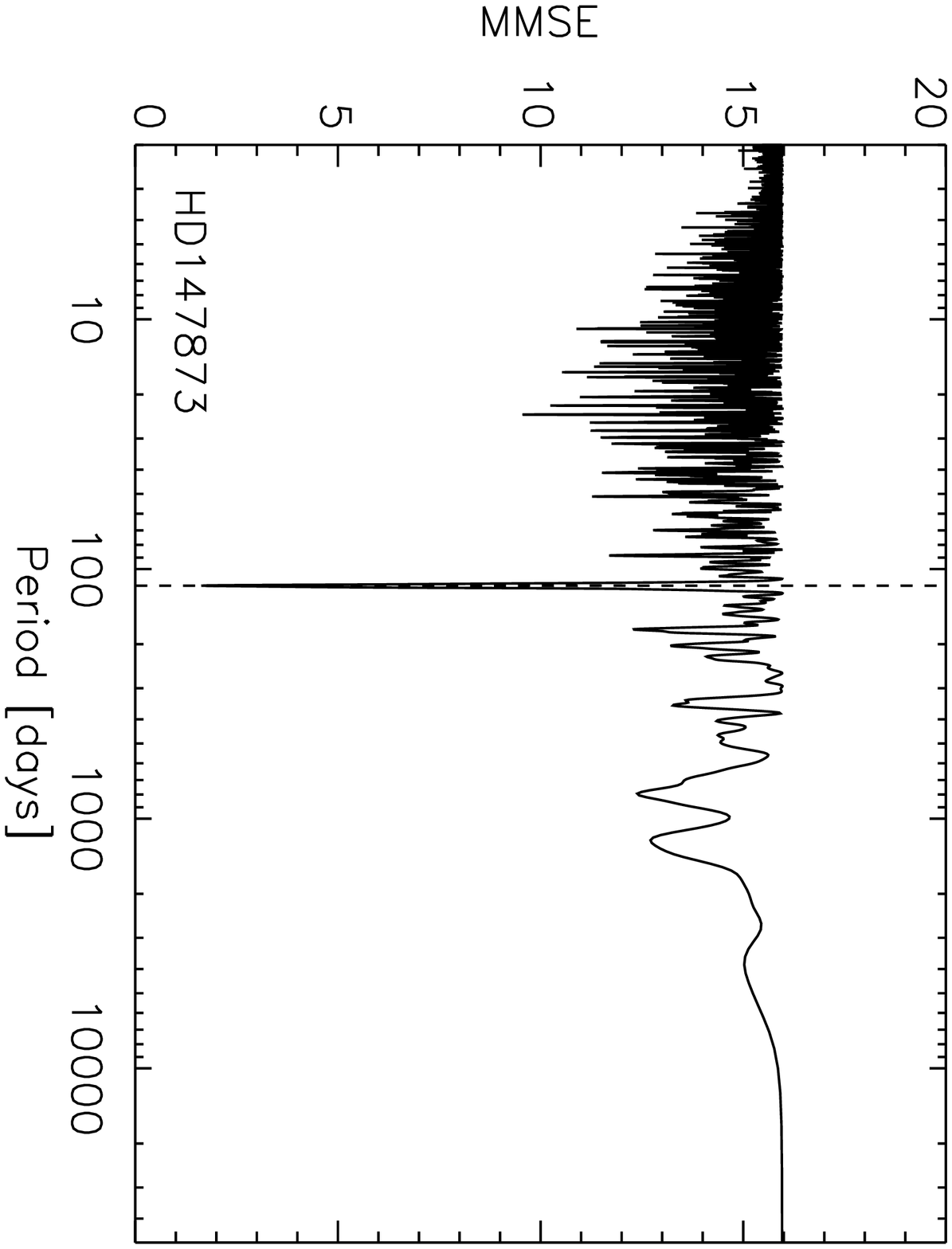}
\includegraphics{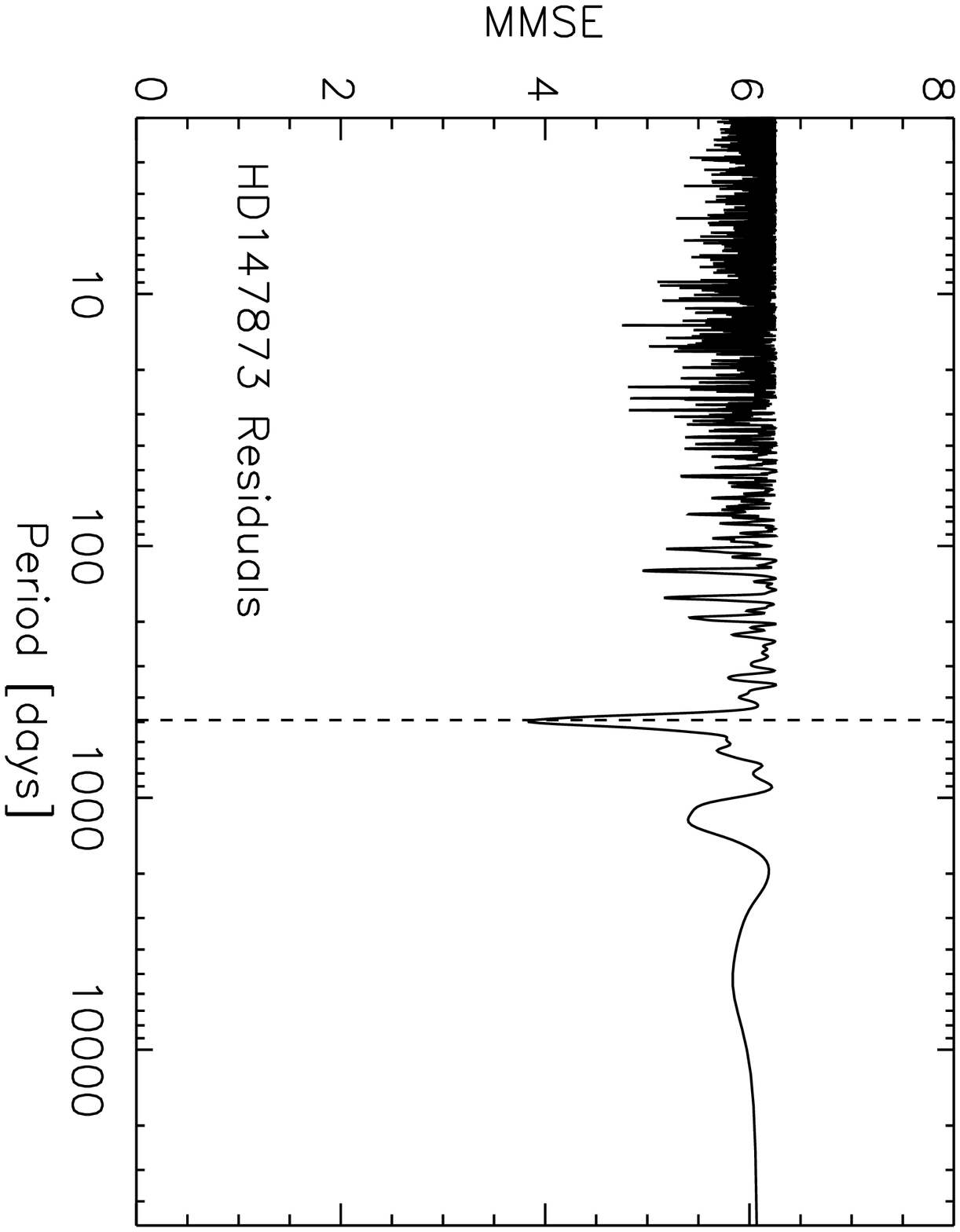}
\includegraphics{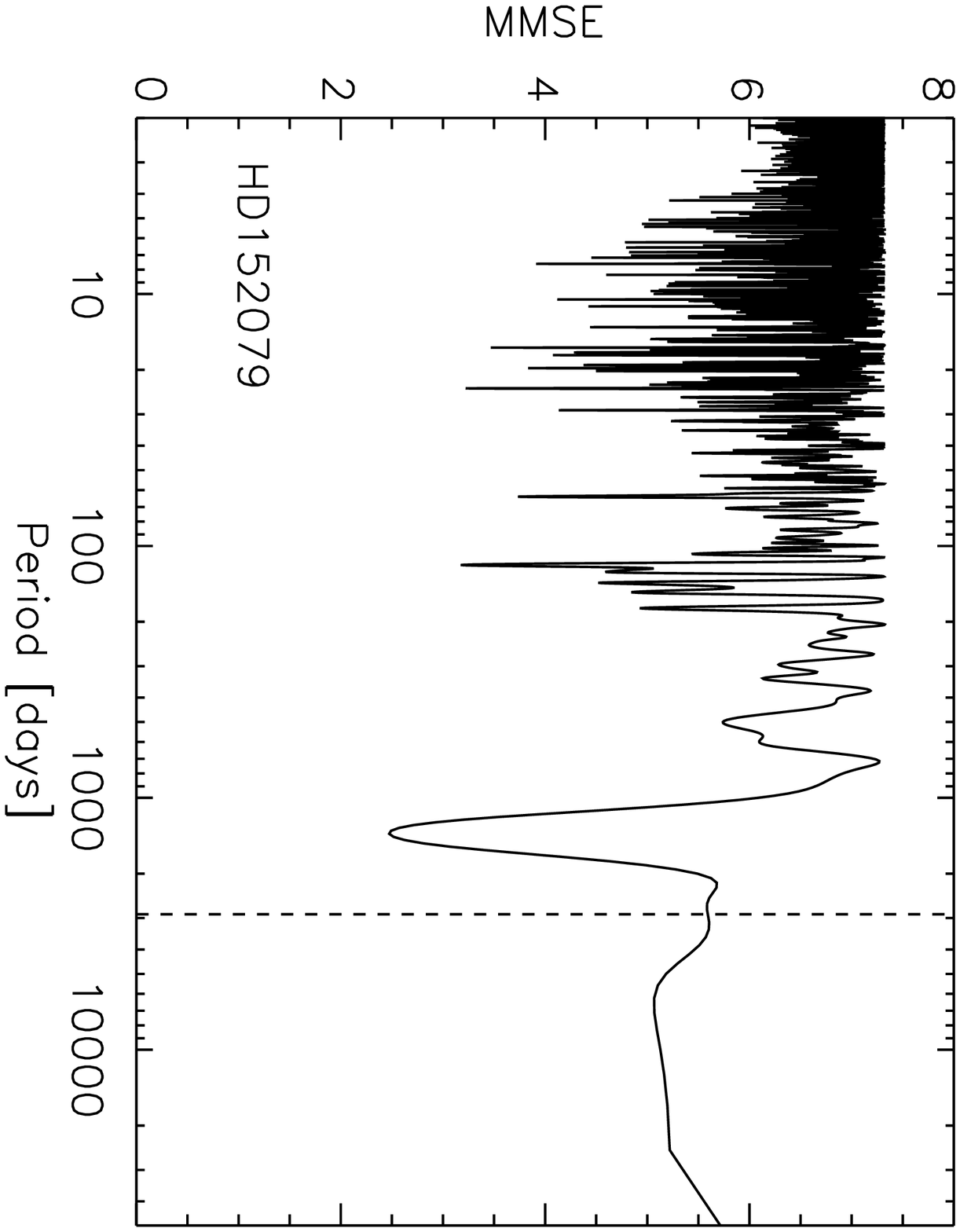}

\includegraphics{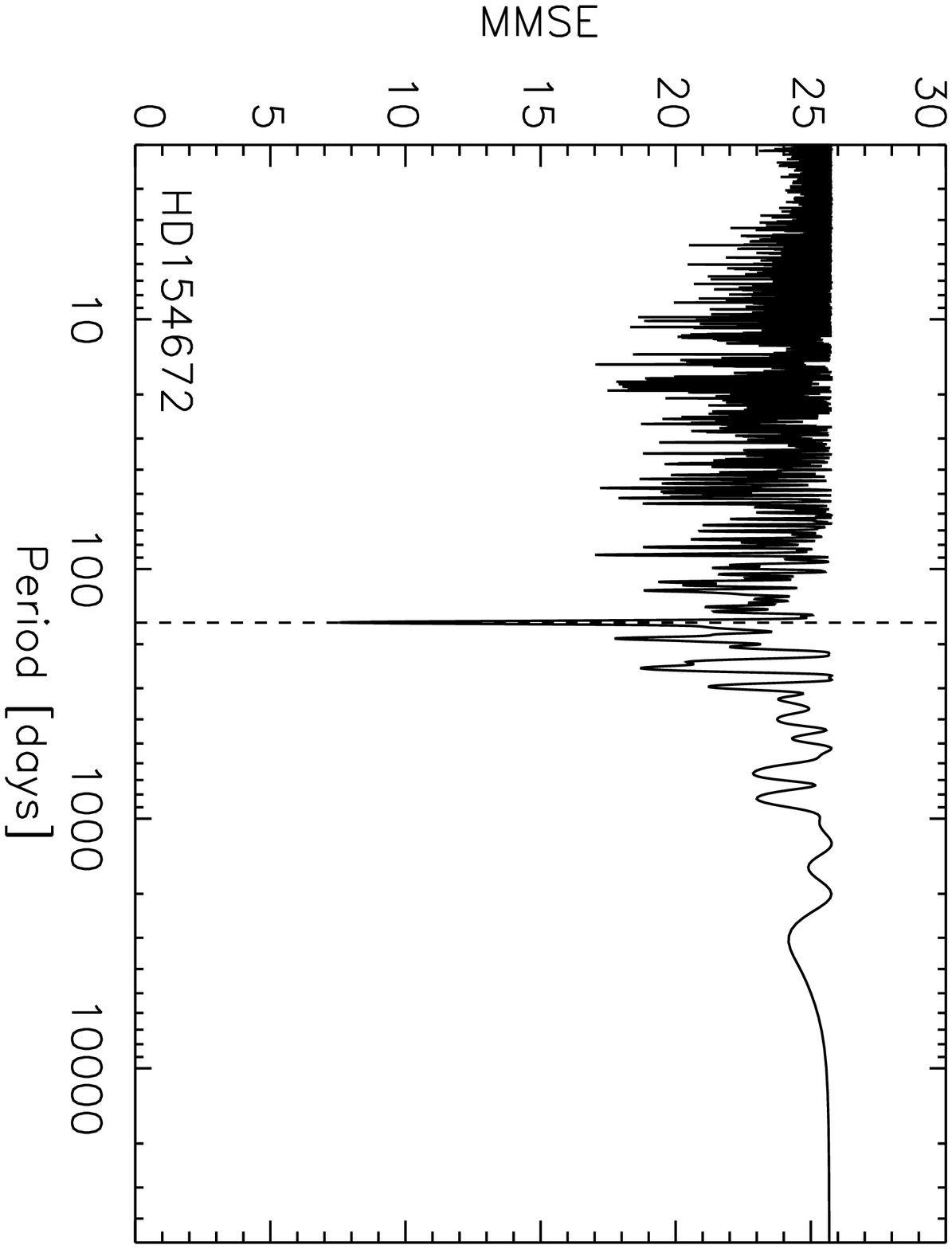}
\includegraphics{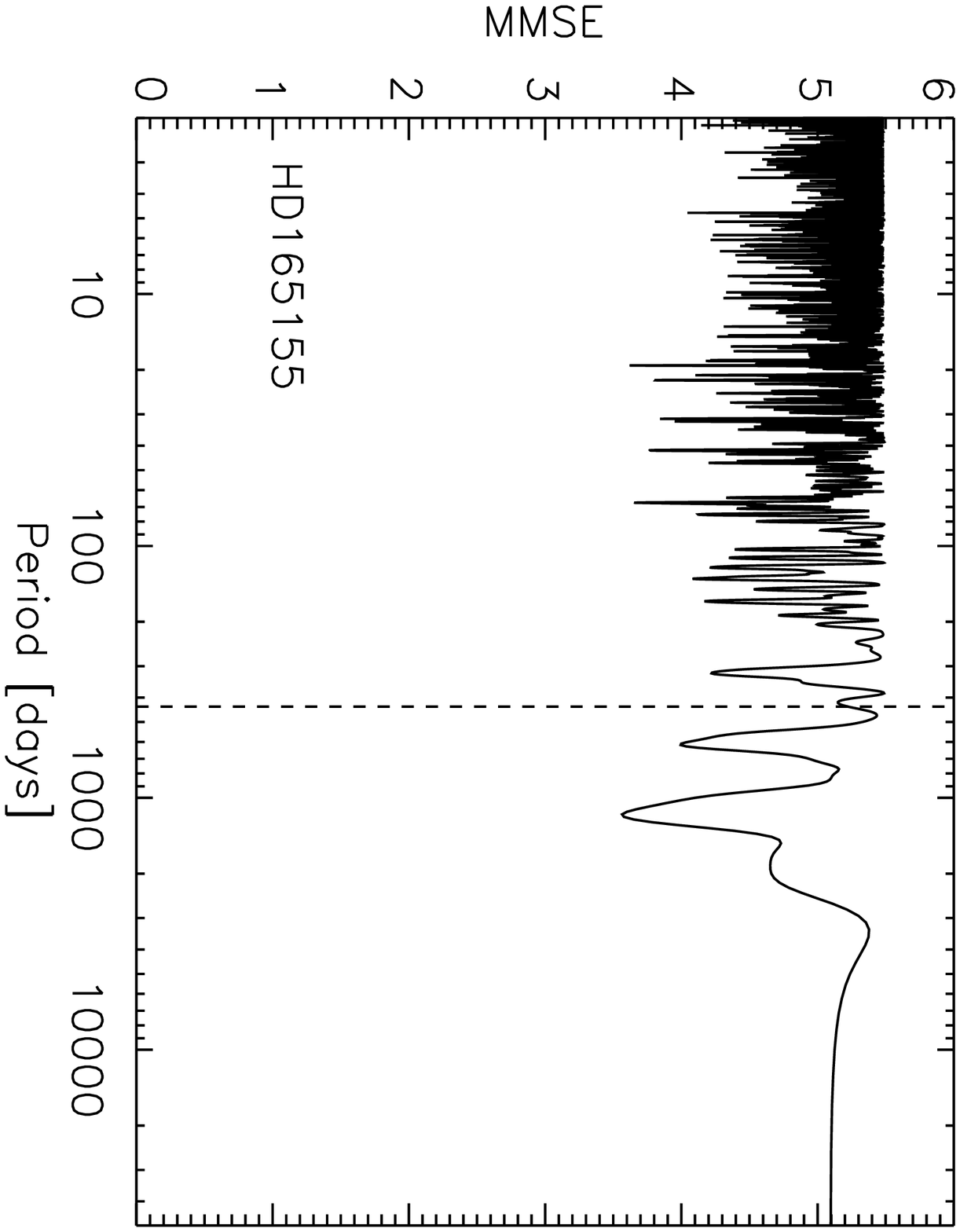}
\includegraphics{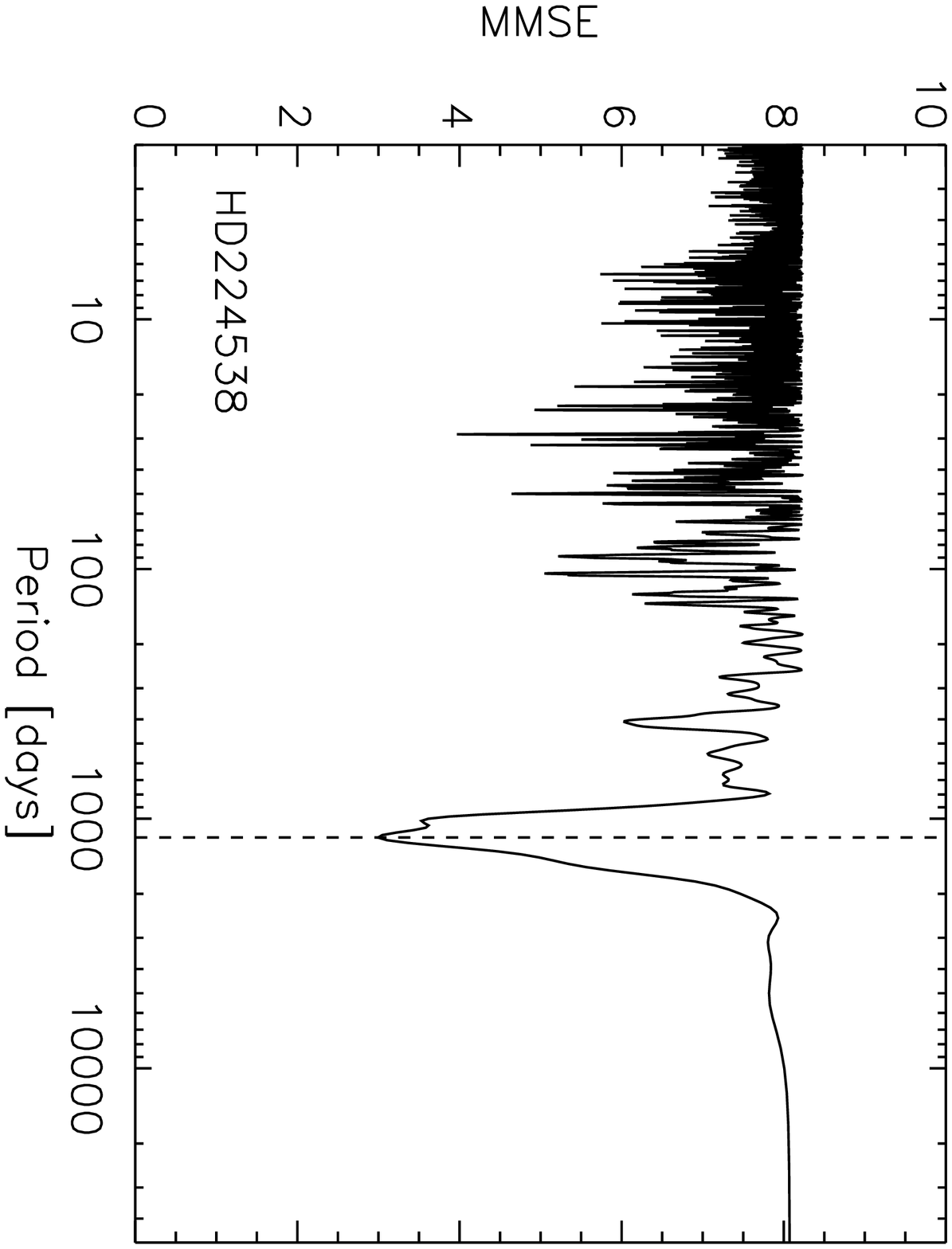}

\vspace{11.0cm}
\caption{Periodograms from top left to bottom right: HD9174, HD48265, HD68402, HD72892, HD128356, HD143361, HD147873, HD147873 residuals after fitting out the 
best fit Keplerian signal associated with the raw data primary spike, HD152079, HD154672, HD165155, and HD224538.  The dashed vertical lines represent the periods 
detected by the MCMC analysis.}
\label{appendix0}
\end{figure*}

\section{MCMC Posterior Densities}

Here we show the posterior densities from our MCMC search for signals in the radial velocities for all targets in this work.  We show the 
densities from the samplings for the periods, semiamplitudes, and eccentricities for all signals.

\begin{figure*}
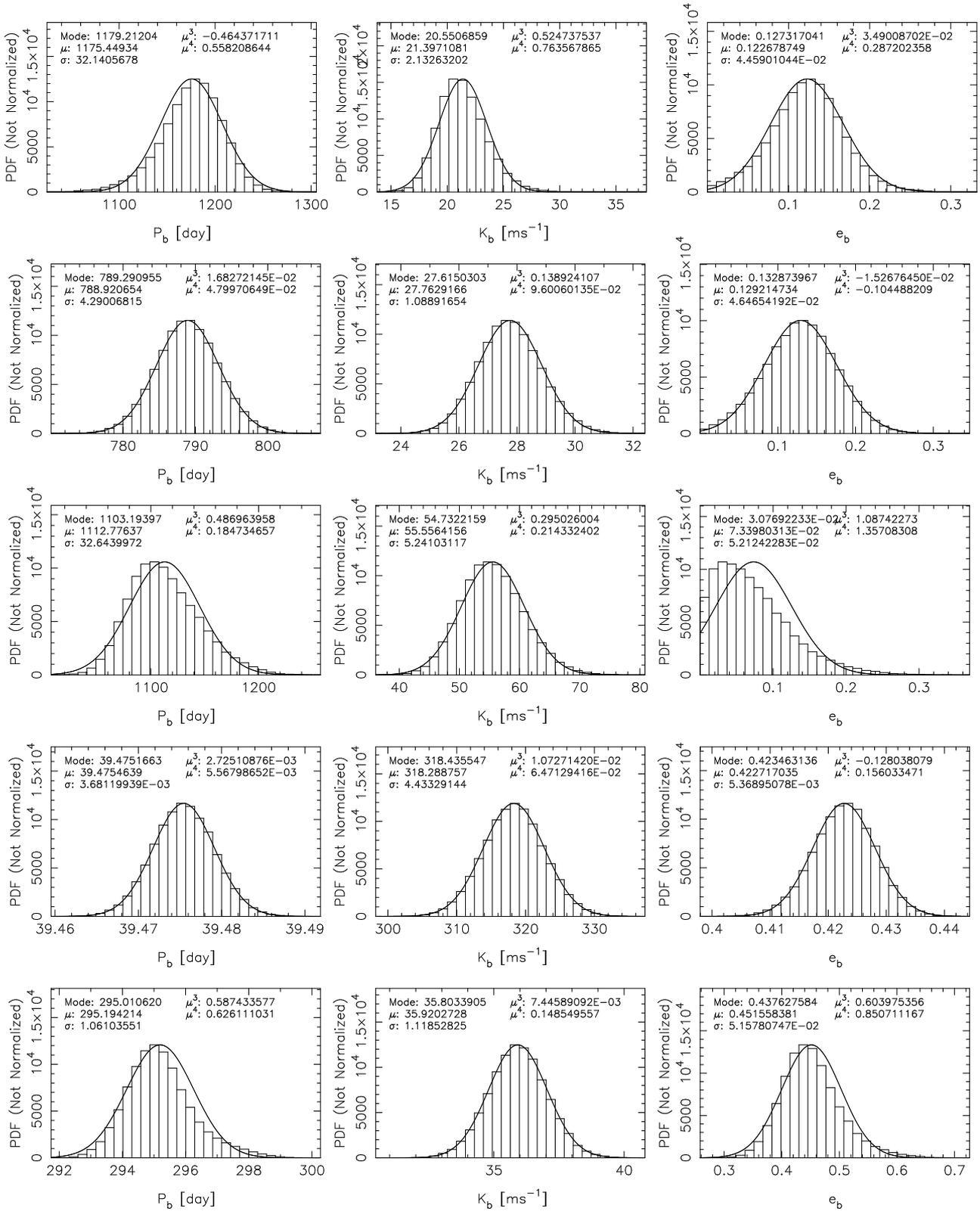

\vspace{5.5cm}
\hspace{-4.0cm}
\includegraphics{Fig40.ps}
\includegraphics{Fig41.ps}
\includegraphics{Fig42.ps}

\includegraphics{Fig43.ps}
\includegraphics{Fig44.ps}
\includegraphics{Fig45.ps}

\includegraphics{Fig46.ps}
\includegraphics{Fig47.ps}
\includegraphics{Fig48.ps}

\includegraphics{Fig49.ps}
\includegraphics{Fig50.ps}
\includegraphics{Fig51.ps}

\includegraphics{Fig52.ps}
\includegraphics{Fig53.ps}
\includegraphics{Fig54.ps}
\vspace{15.2cm}
\caption{Posterior densities for the signals HD9174$b$, HD48265$b$, HD68402$b$, HD72892$b$, and HD128356$b$.}
\label{appendix1}
\end{figure*}

\begin{figure*}
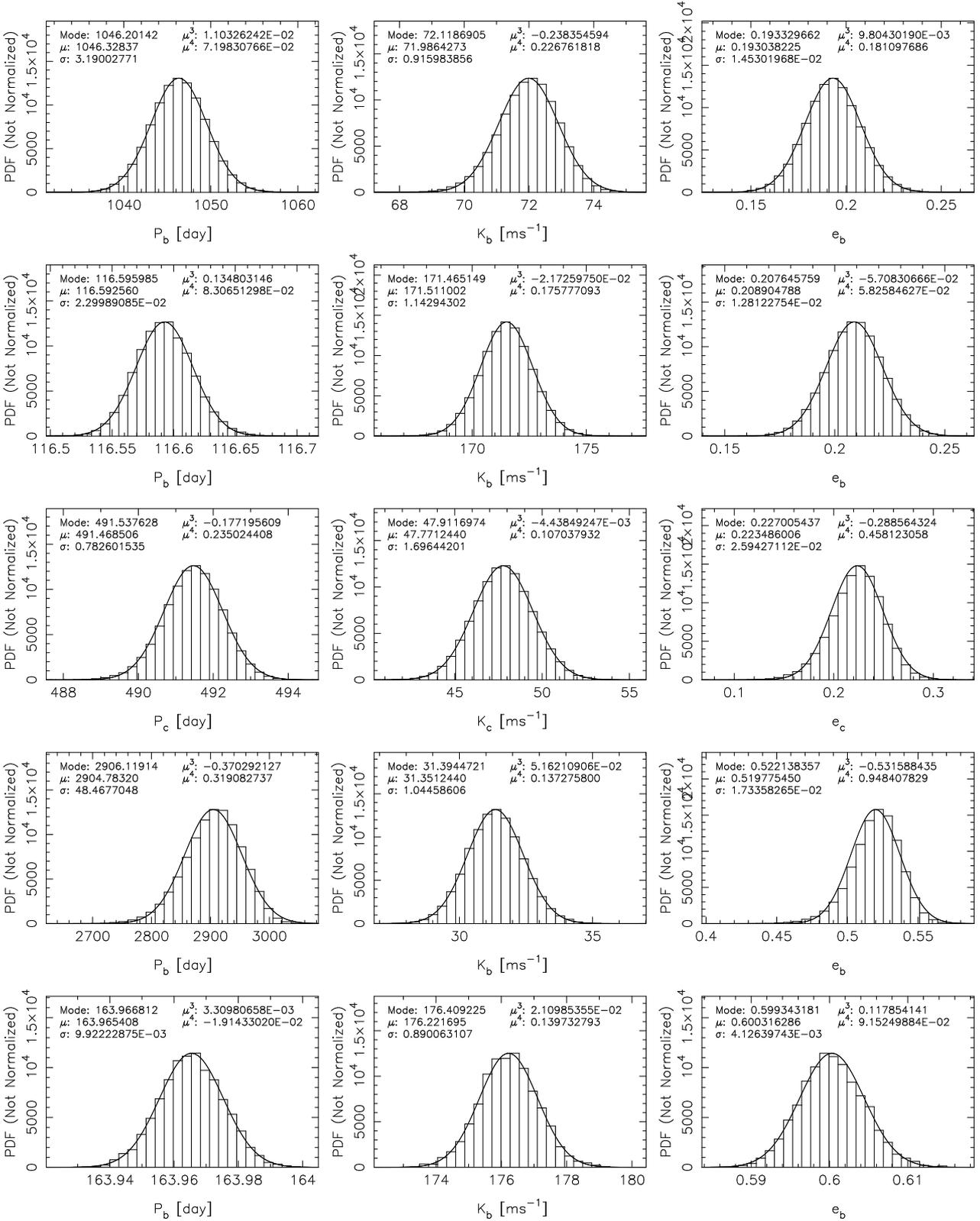

\vspace{5.5cm}
\hspace{-4.0cm}

\includegraphics{Fig55.ps}
\includegraphics{Fig56.ps}
\includegraphics{Fig57.ps}

\includegraphics{Fig58.ps}
\includegraphics{Fig59.ps}
\includegraphics{Fig60.ps}

\includegraphics{Fig61.ps}
\includegraphics{Fig62.ps}
\includegraphics{Fig63.ps}

\includegraphics{Fig64.ps}
\includegraphics{Fig65.ps}
\includegraphics{Fig66.ps}

\includegraphics{Fig67.ps}
\includegraphics{Fig68.ps}
\includegraphics{Fig69.ps}
\vspace{15.2cm}
\caption{Posterior densities for the signals HD143361$b$, HD147873$b$, HD147873$c$, HD152079$b$, and HD154672$b$.}
\label{appendix2}
\end{figure*}

\begin{figure*}
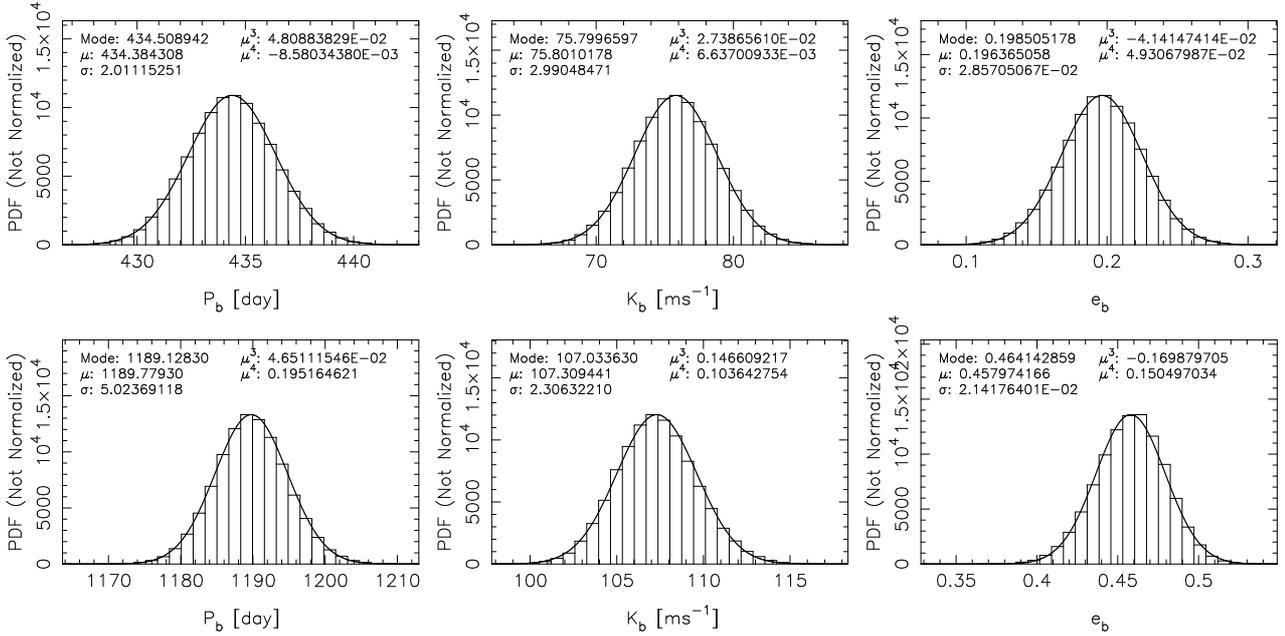

\vspace{5.5cm}
\hspace{-4.0cm}

\includegraphics{Fig70.ps}
\includegraphics{Fig71.ps}
\includegraphics{Fig72.ps}

\includegraphics{Fig76.ps}
\includegraphics{Fig77.ps}
\includegraphics{Fig78.ps}
\vspace{2.5cm}
\caption{Posterior densities for the signals HD165155$b$ and HD224538$b$.}
\label{appendix3}
\end{figure*}

\section{CORALIE Chromospheric Activity Indices}\label{appendix_act}

\begin{figure*}
\vspace{6.0cm}
\hspace{-4.0cm}
\includegraphics{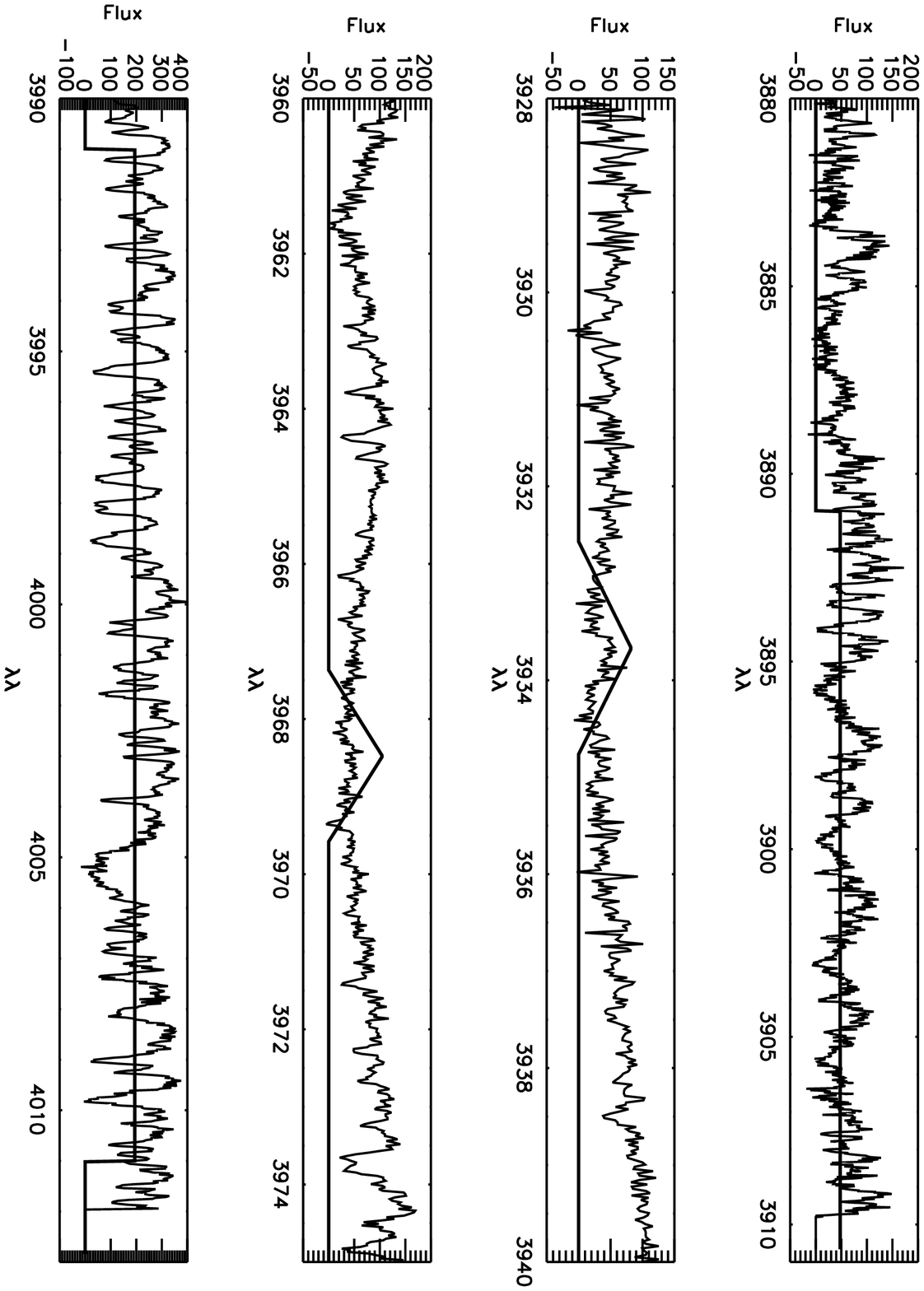}
\vspace{3.8cm}
\caption{CORALIE echelle orders used to extract the $S$ activity indices for HD128356.  The square continuum bandpasses regions V and R 
are shown by the thick solid lines in the top and bottom plots, respectively.  The thick triangular solid lines in the upper and lower center 
plots show the K and H bandpasses, respectively. }
\label{cor_passes}
\end{figure*}

We measure the CORALIE activities using only four echelle orders, even though the regions we require for the $S_{\rm{MW}}$ passbands are found across five orders.  
We drop one of the orders (order 4) due to an excess of noise at the blue end, which is due to the position of the echellogram where the V passband is found 
and therefore including this order enhances the uncertainty in the $S$-index and, in general, artificially increases the activity value making each star appear more 
active than it really is.  We note that this could be taken out by calibration to other chromospheric indexes.  We show the CORALIE extraction regions in Fig.~\ref{cor_passes}. 

We compute the activities by integrating the square continuum V (3891-3911\AA) and R (3991-4011\AA) bandpasses and taking the 
ratio of these against the integrated flux in the triangular core bandpasses, described in the following series of equations:

\begin{equation}
\label{eq:activity}
f_{j,i} = \Im_{j,i} * B_{j,i} * \delta\lambda
\end{equation}

\begin{equation}
S_{cont} = \frac{ \sum{(f_{V,i} + f_{R,i})} }{ \sum{(B_{V,i} + B_{R,i}) * \delta\lambda} } 
\end{equation}

\begin{equation}
S_{core} = \frac{ \sum{(f_{K,i} + f_{H,i})} }{ \sum{(B_{K,i} + B_{H,i}) * \delta\lambda} } 
\end{equation}

\begin{equation}
\sigma_{S_{cont}} = \frac{ \sqrt{ \sum{ \left( \sqrt{(\sigma_{f_{V,i}}^2 + \sigma_{f_{R,i}}^2)} \right) }^2 } }{ \sum{(B_{V,i} + B_{R,i}) * \delta\lambda} }
\end{equation}

\begin{equation}
\sigma_{S_{core}} = \frac{ \sqrt{ \sum{ \left( \sqrt{(\sigma_{f_{K,i}}^2 + \sigma_{f_{H,i}}^2)} \right) }^2 } }{ \sum{(B_{K,i} + B_{H,i}) * \delta\lambda} }
\end{equation}

The cont and core subscripts represent the continuum and core regions of the spectrum respectively, $\Im$ is the flux measured in each wavelength domain ($i$), 
$j$ here denotes either the V,R,K, or H bandpass regions, and $\delta\lambda$ is the wavelength step (dispersion), which at the resolution of CORALIE is $\sim$0.023\AA.

\section{Activity Indicator Periodograms}

Lomb-Scargle periodograms of the activity indicators measured from the CORALIE and HARPS timeseries spectra.  

\begin{figure*}
\vspace{5.5cm}
\hspace{-4.0cm}
\includegraphics{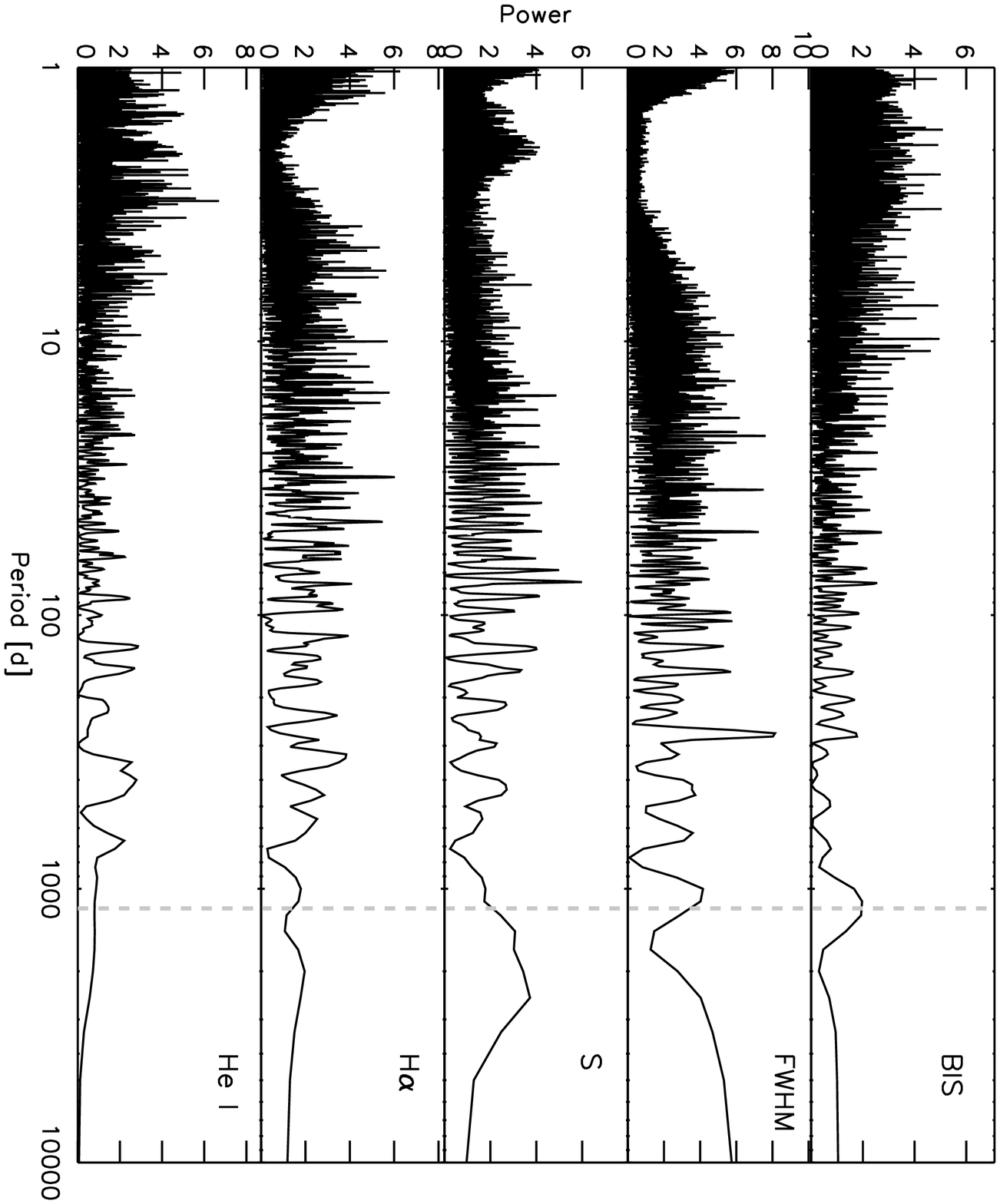}
\includegraphics{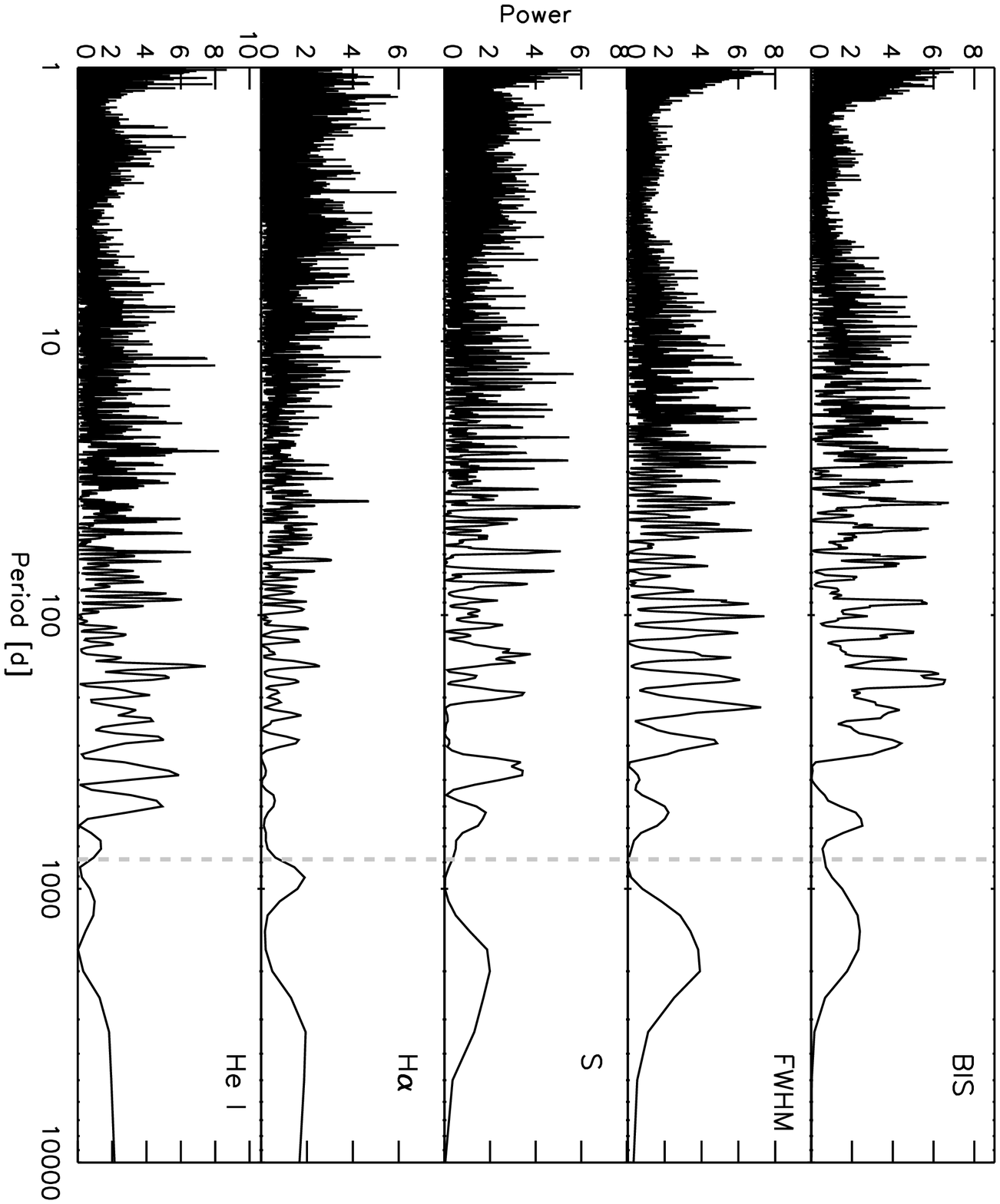}
\includegraphics{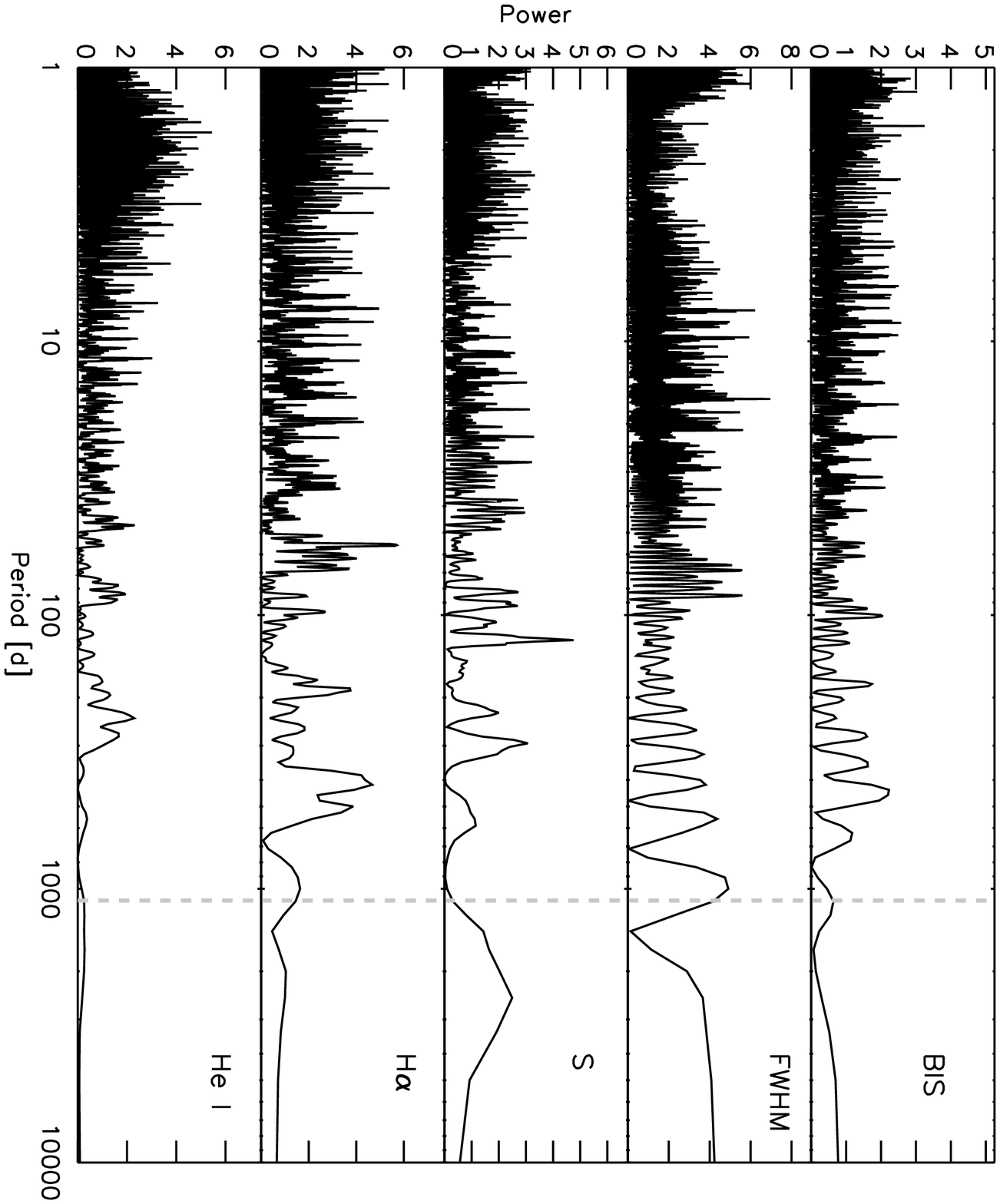}

\includegraphics{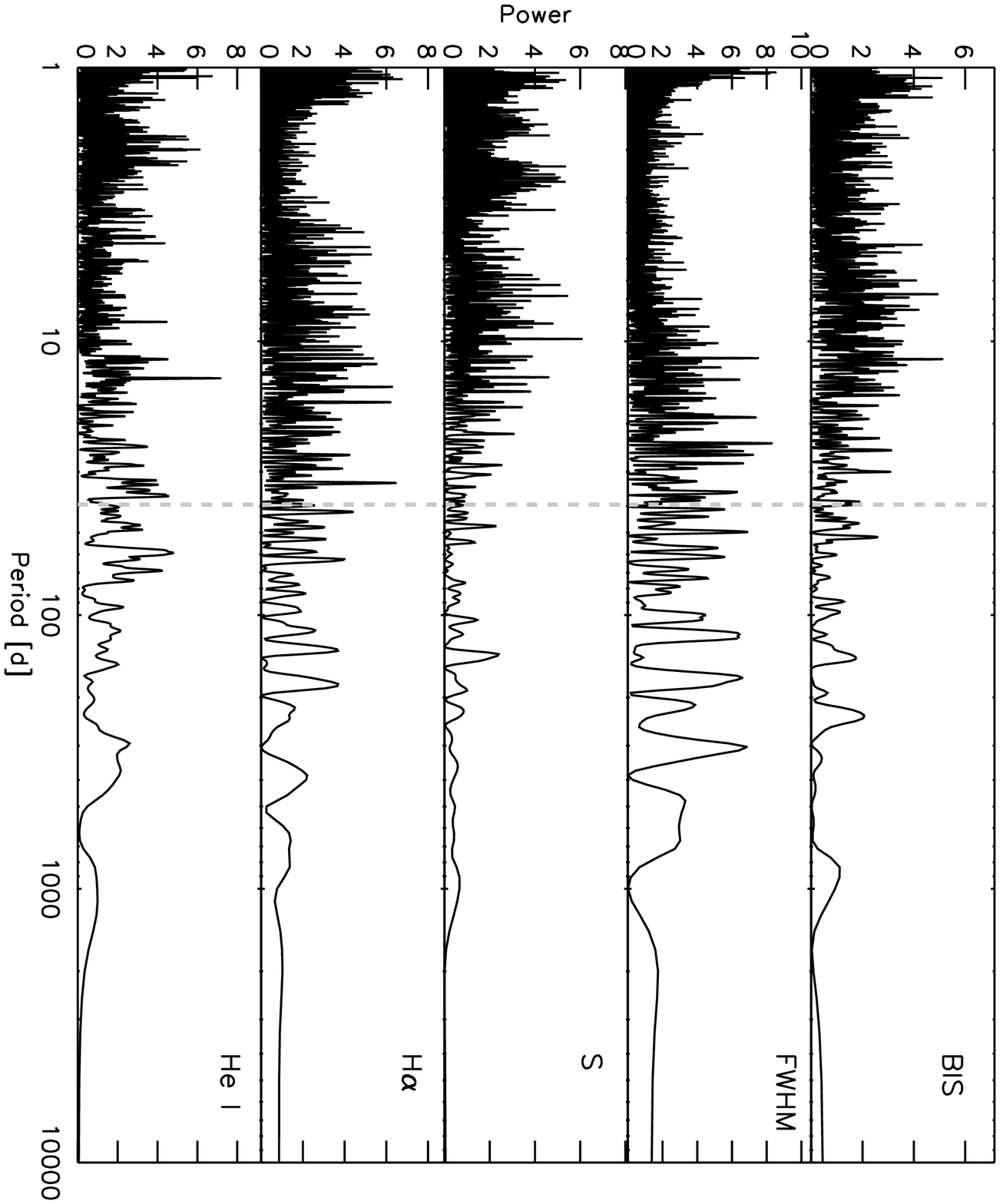}
\includegraphics{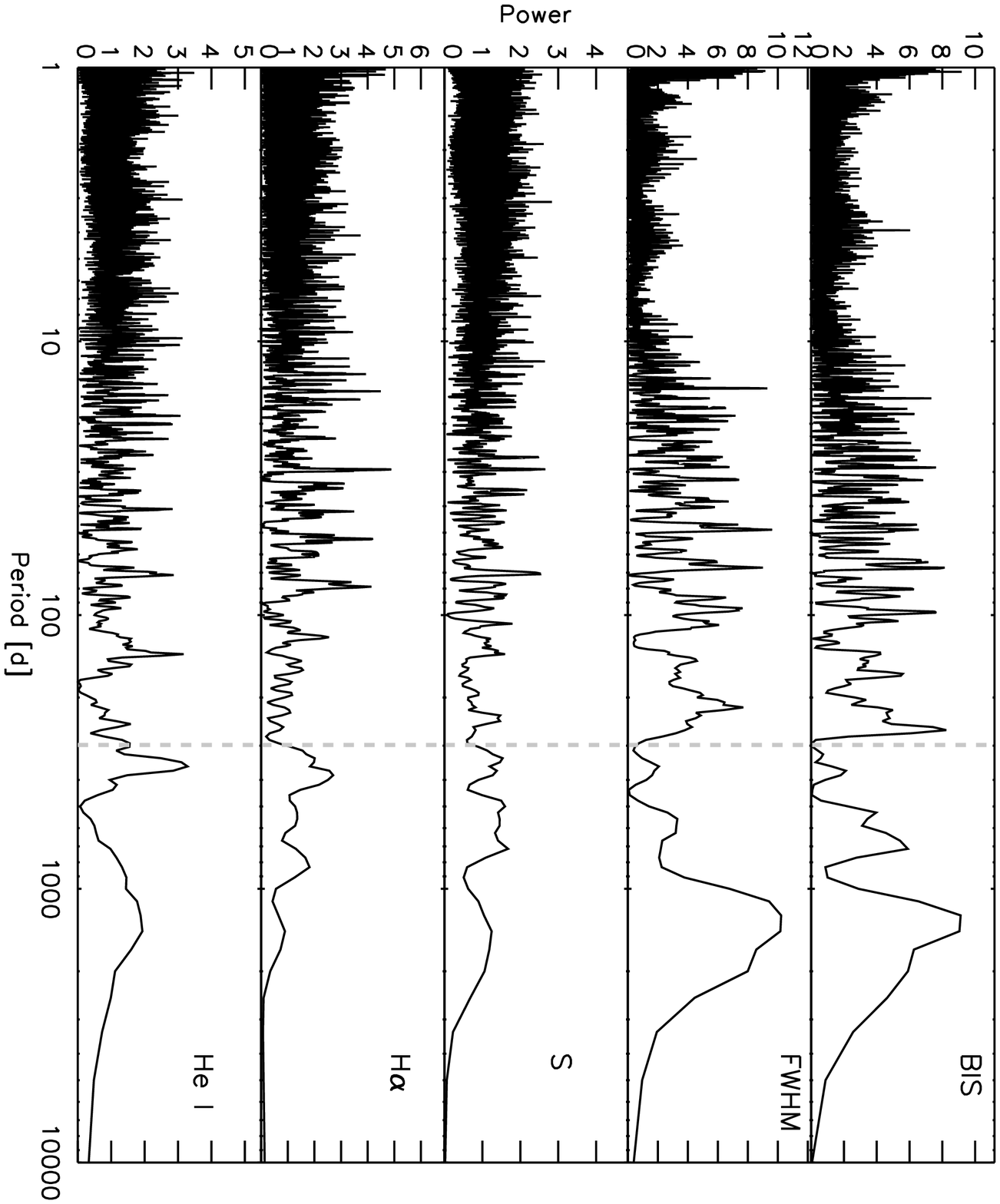}
\includegraphics{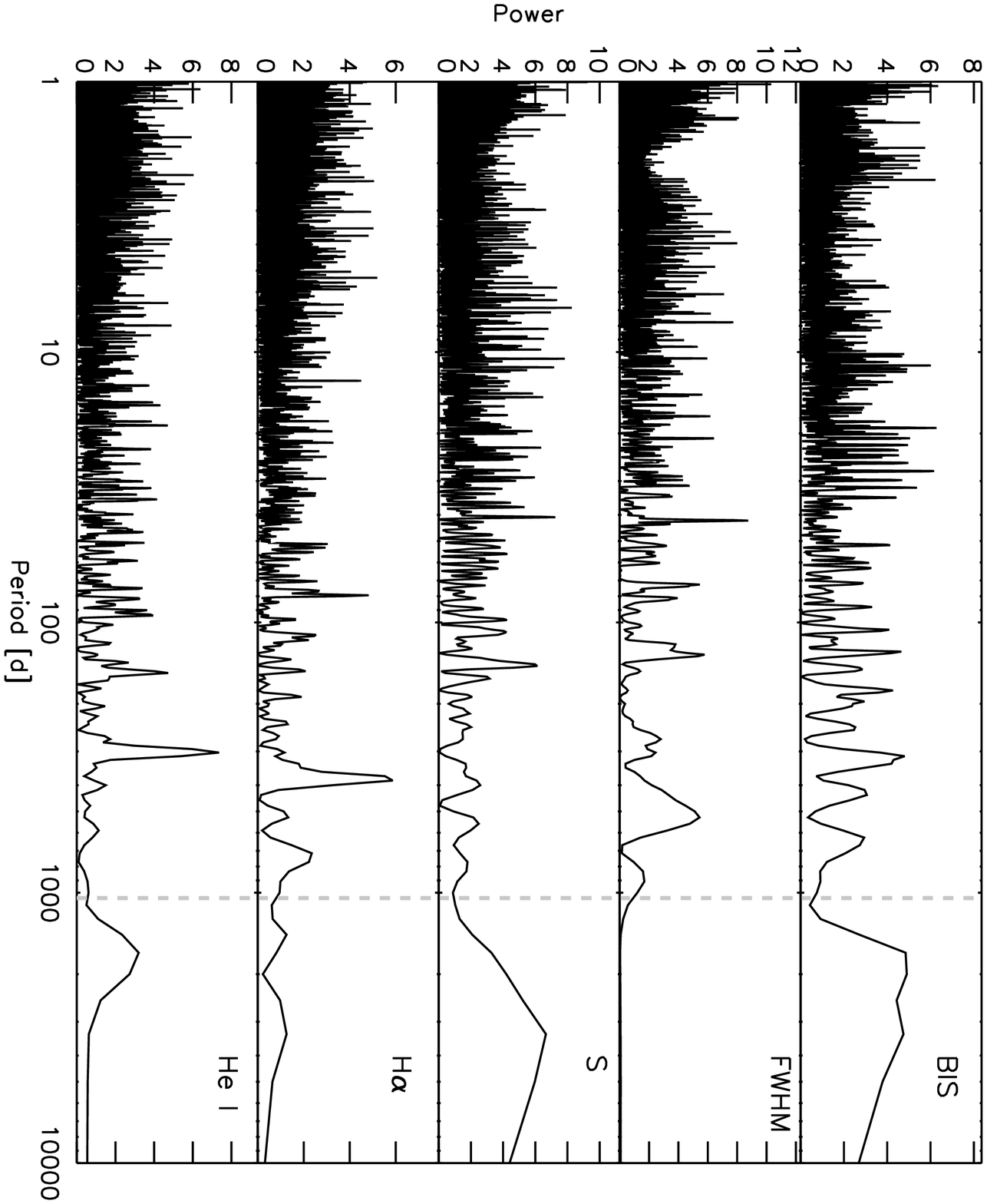}

\includegraphics{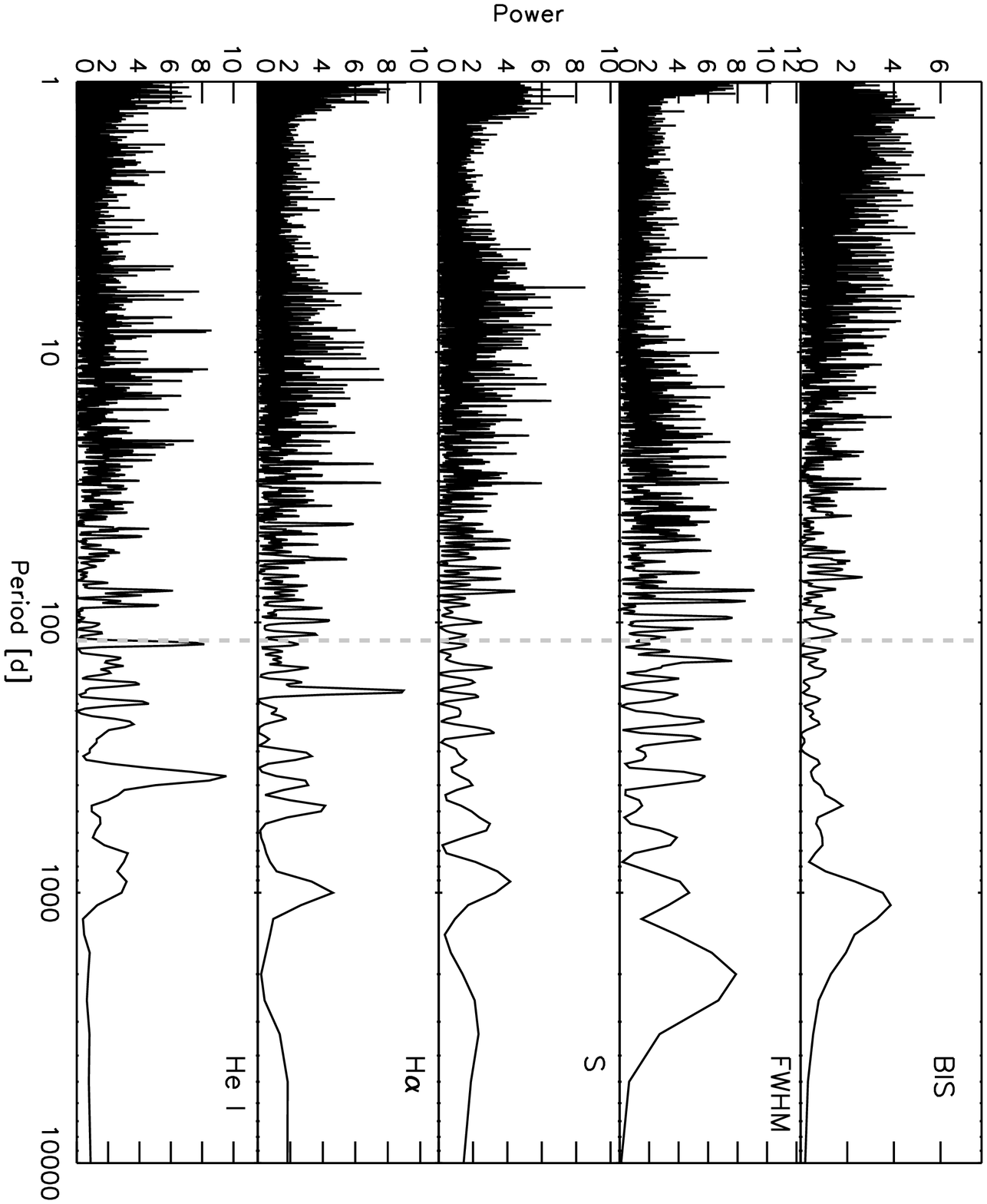}
\includegraphics{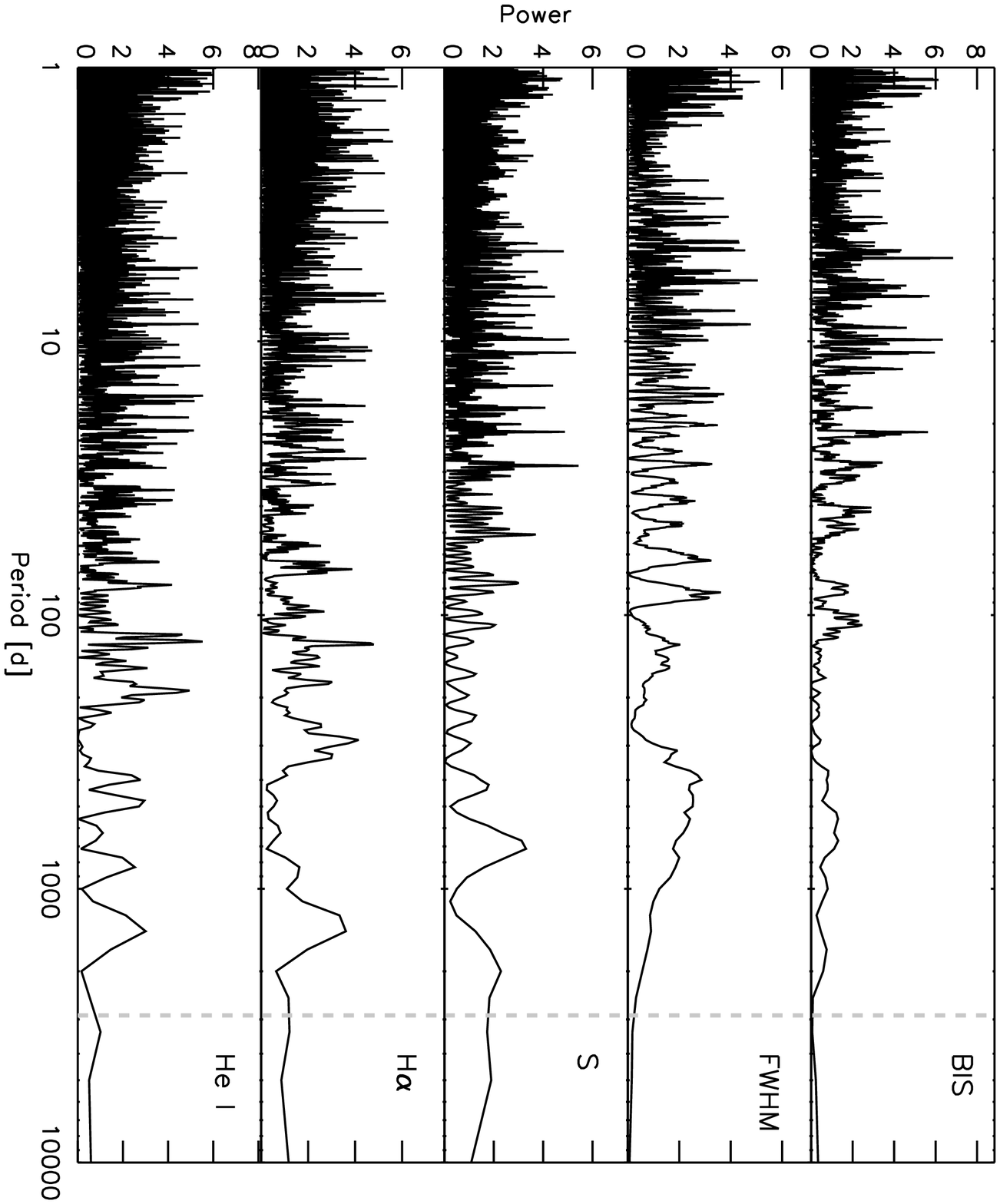}
\includegraphics{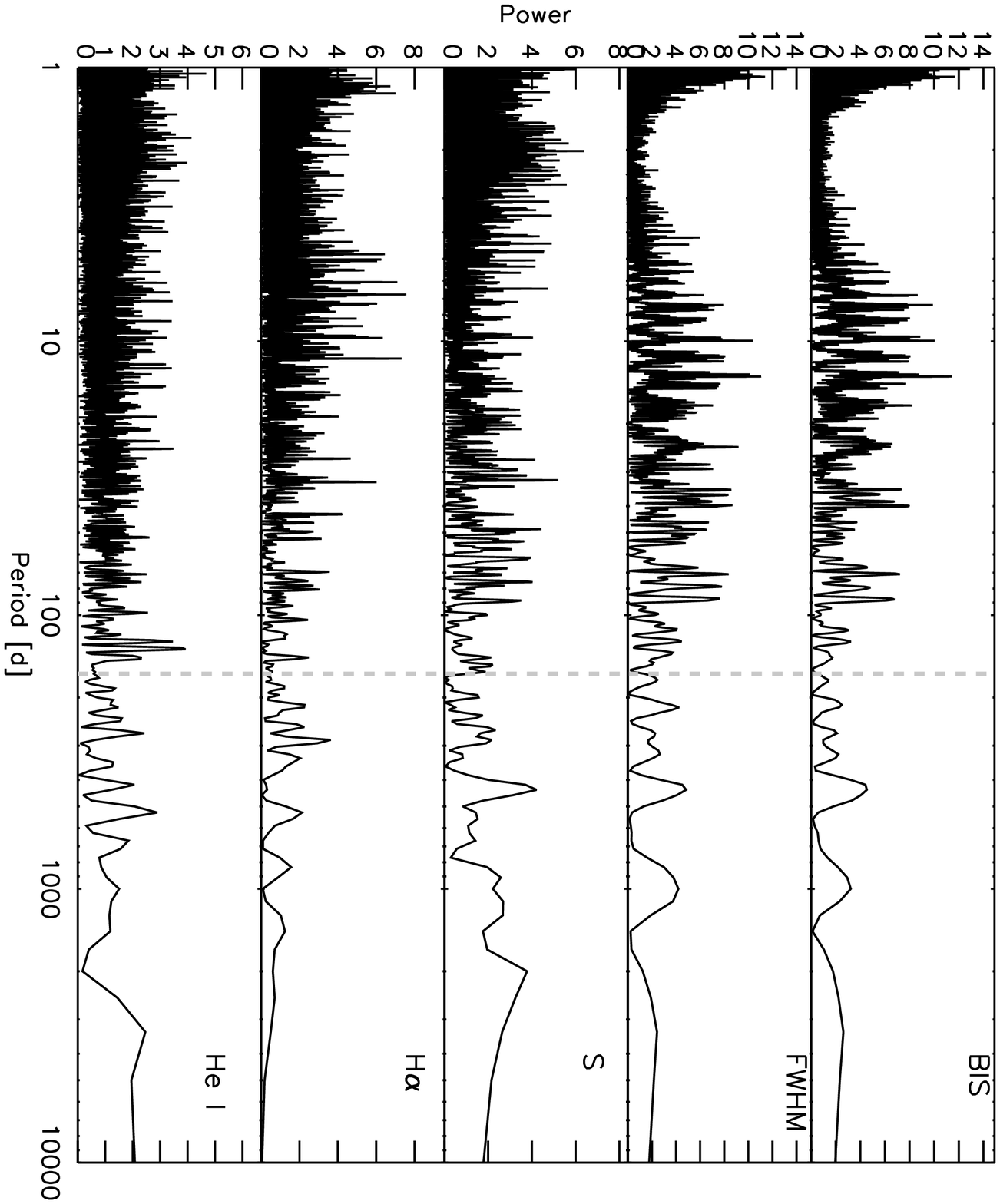}

\includegraphics{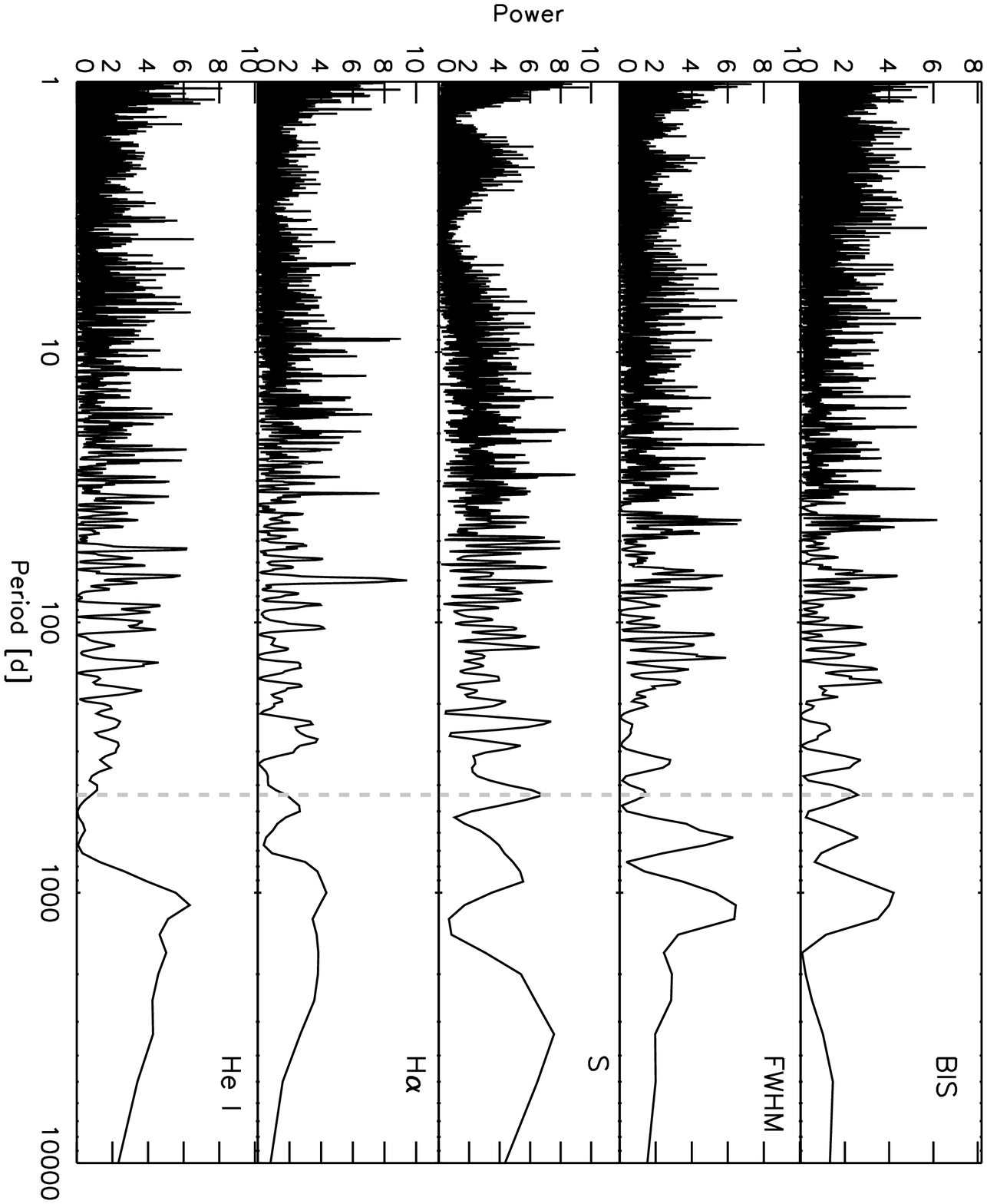}
\includegraphics{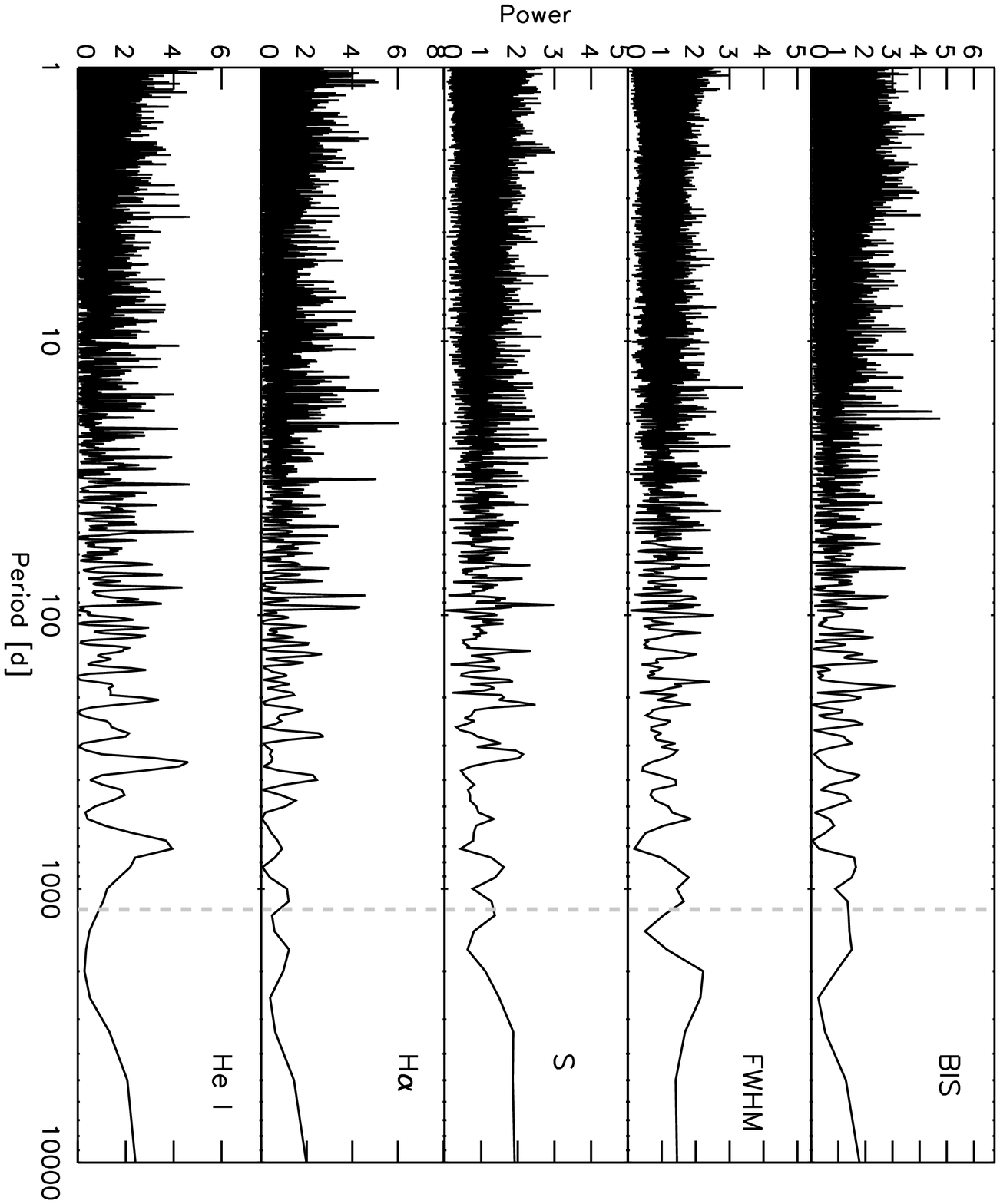}
\vspace{14.2cm}
\caption{In each plot from top to bottom we show the periodograms for the BIS, FWHM, $S$, H$\alpha$, and He~I indices.  From top left to bottom right 
we show the stars HD9174, HD48265, HD68402, HD72892, HD128356, HD143361, HD147873, HD152079, HD154672, HD165155, and HD224538, respectively.}
\label{appendix5}
\end{figure*}

\section{ASAS Periodograms}

Lomb-Scargle periodograms of the ASAS timeseries $V$-band photometric data for the six stars that show peaks that could be related to magnetic activity on the 
surface of the star.  

\begin{figure*}
\vspace{5.5cm}
\hspace{-4.0cm}
\includegraphics{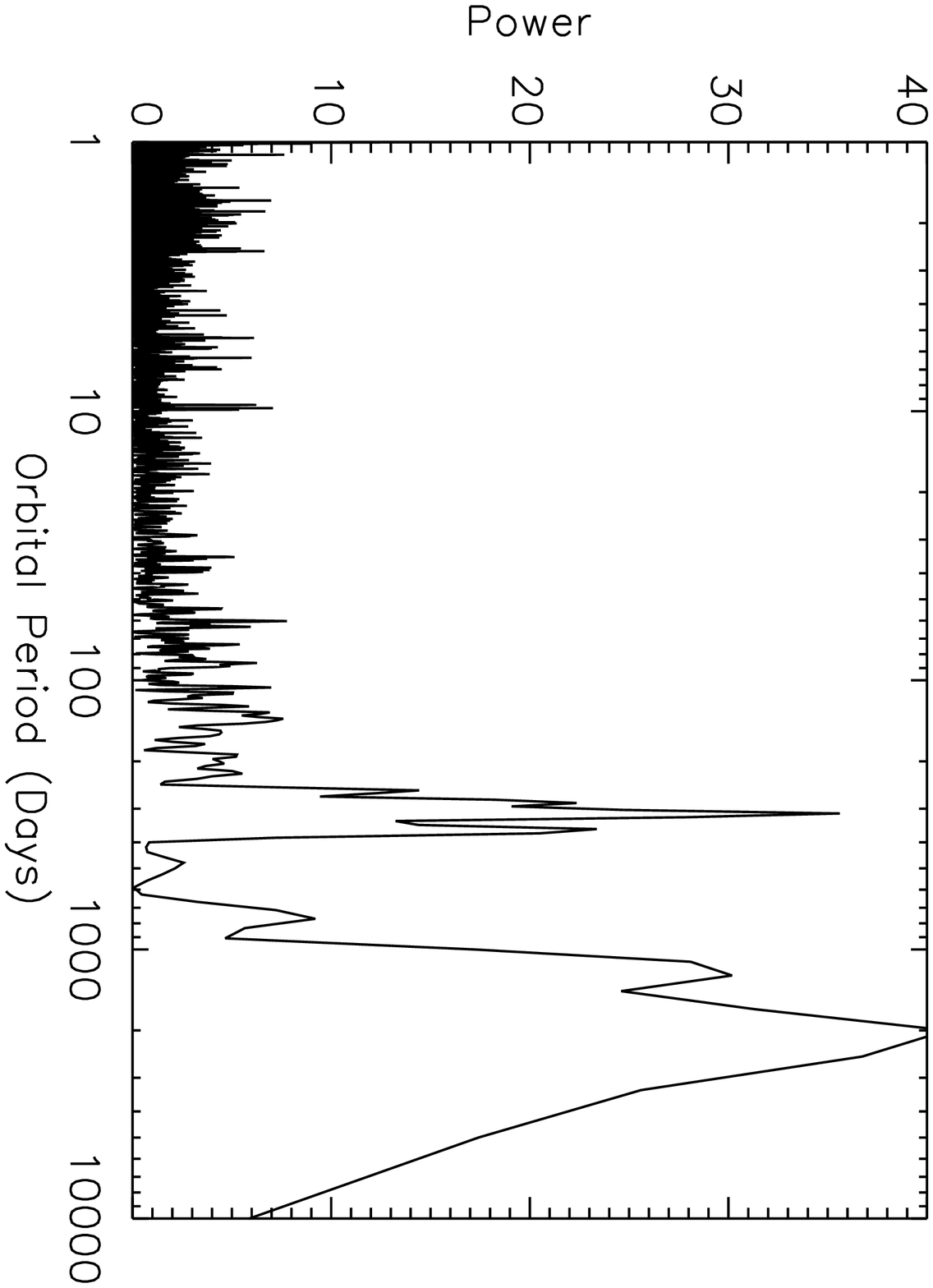}
\includegraphics{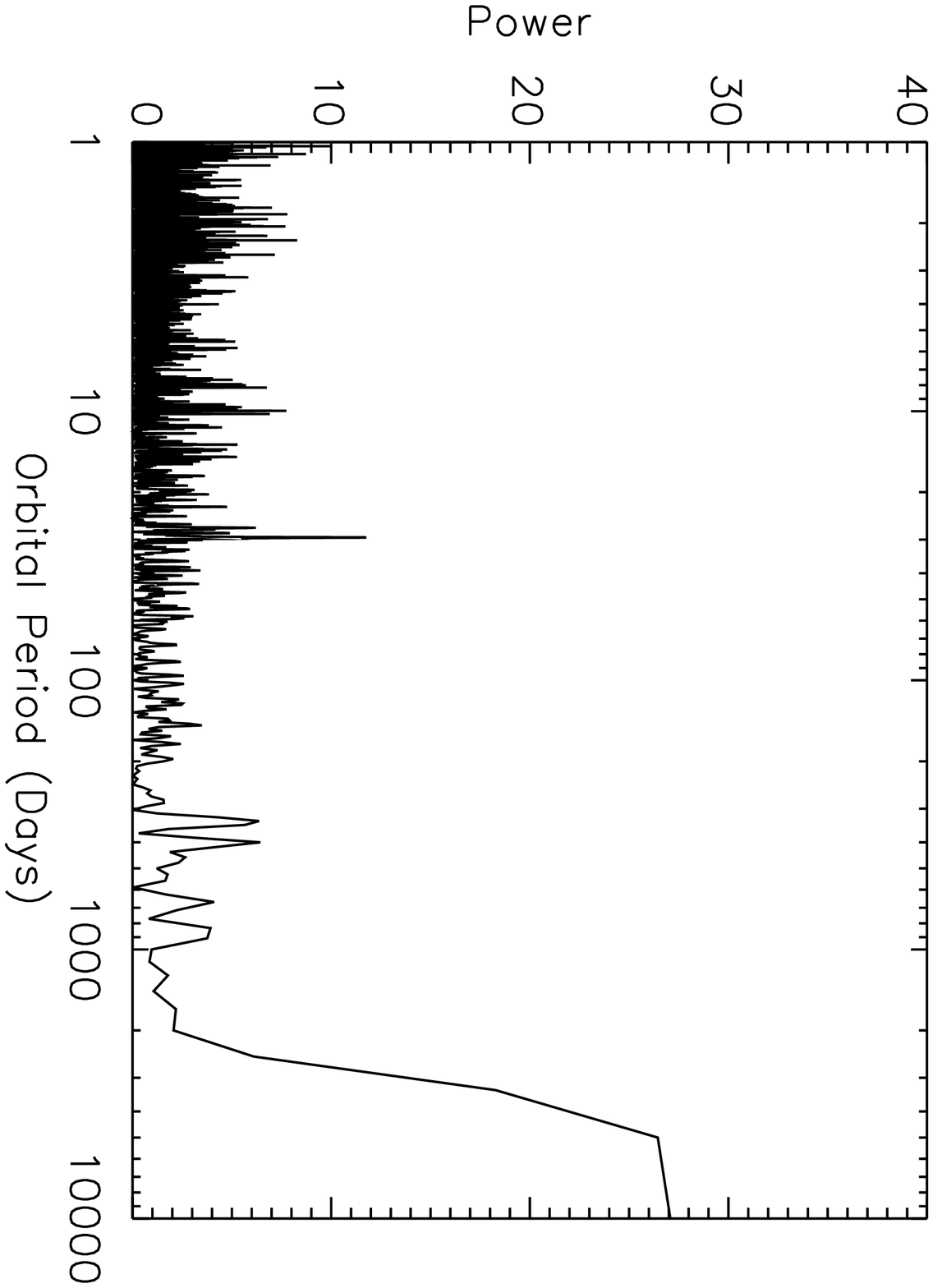}
\includegraphics{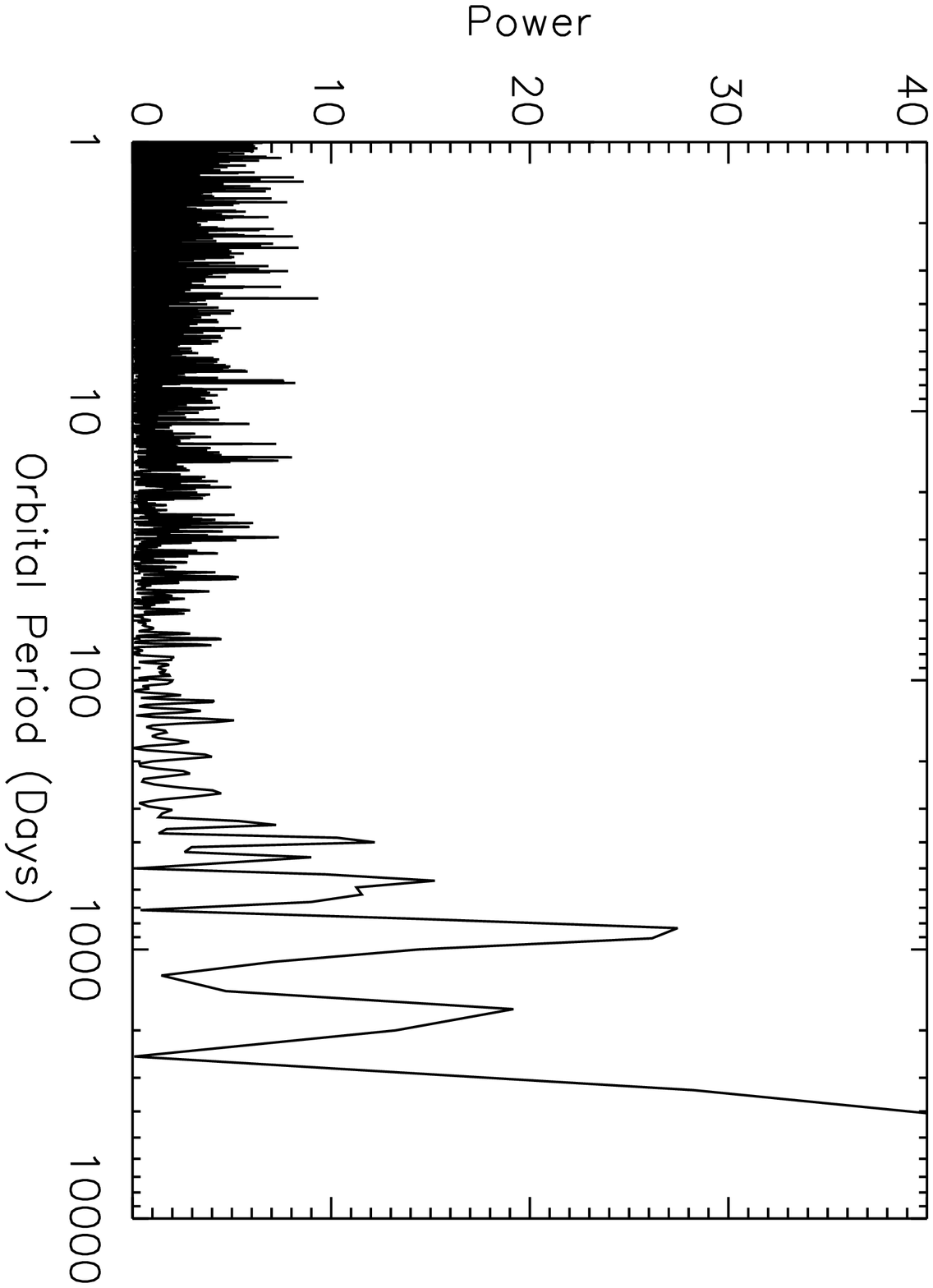}

\includegraphics{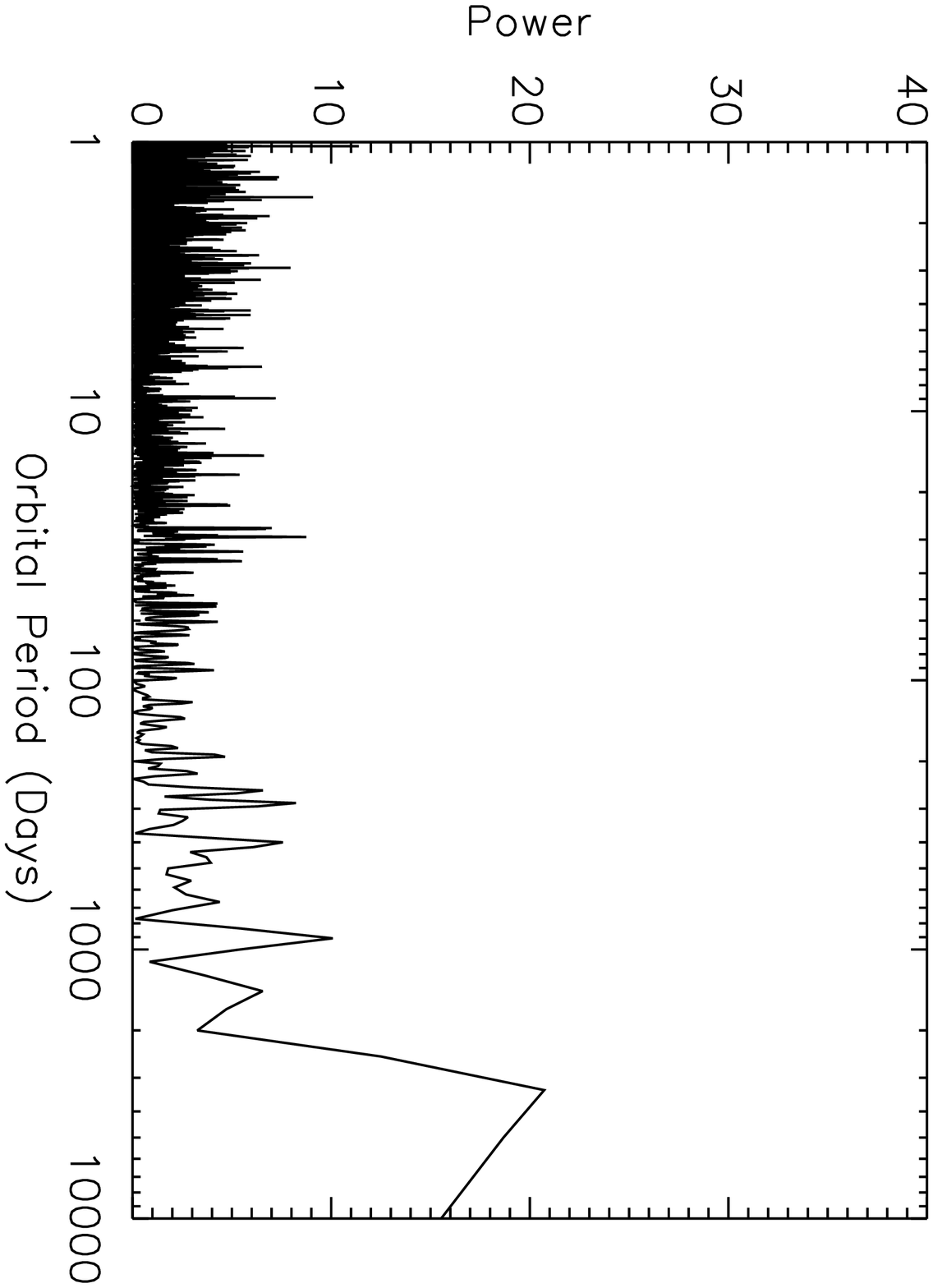}
\includegraphics{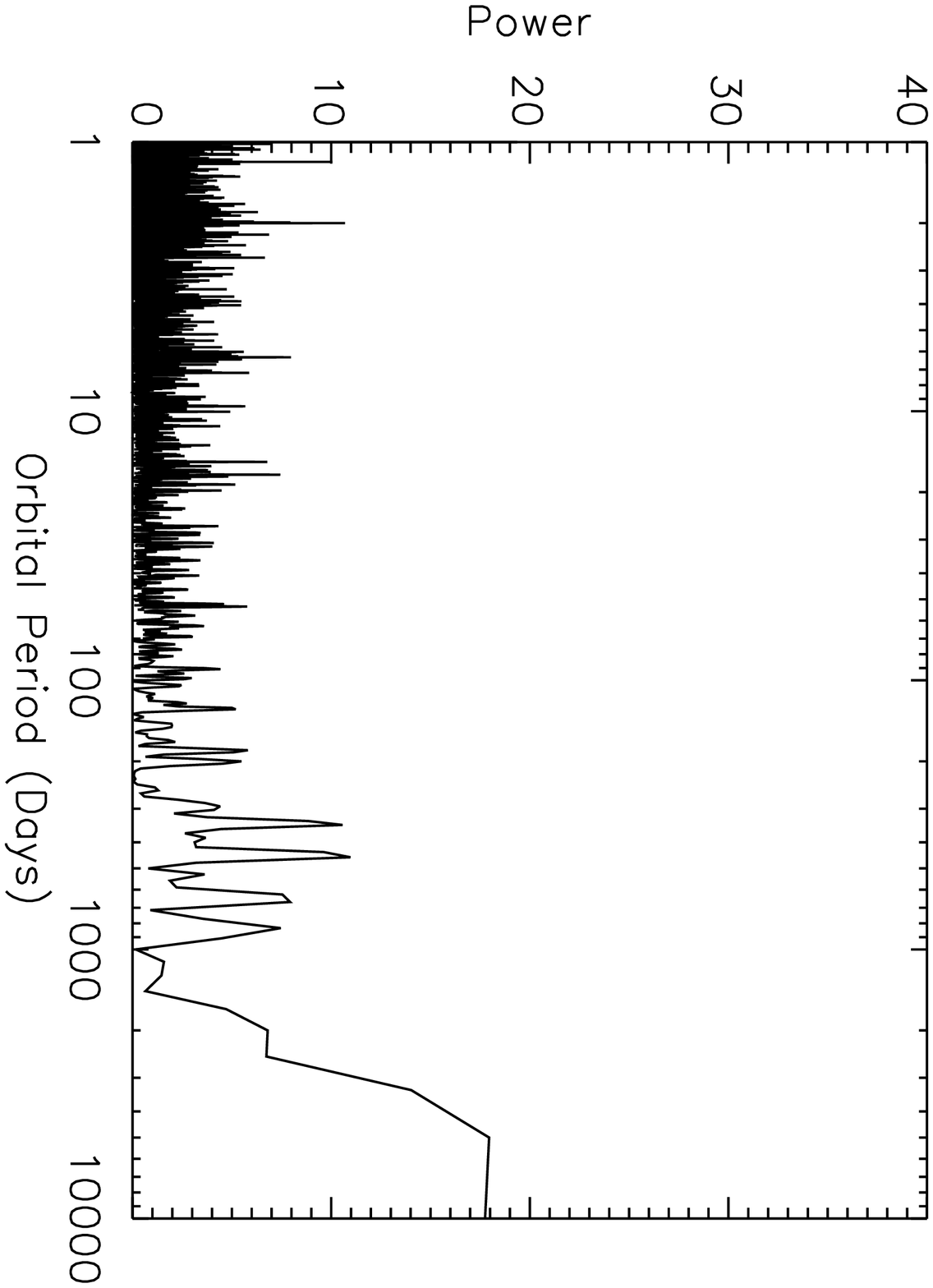}
\includegraphics{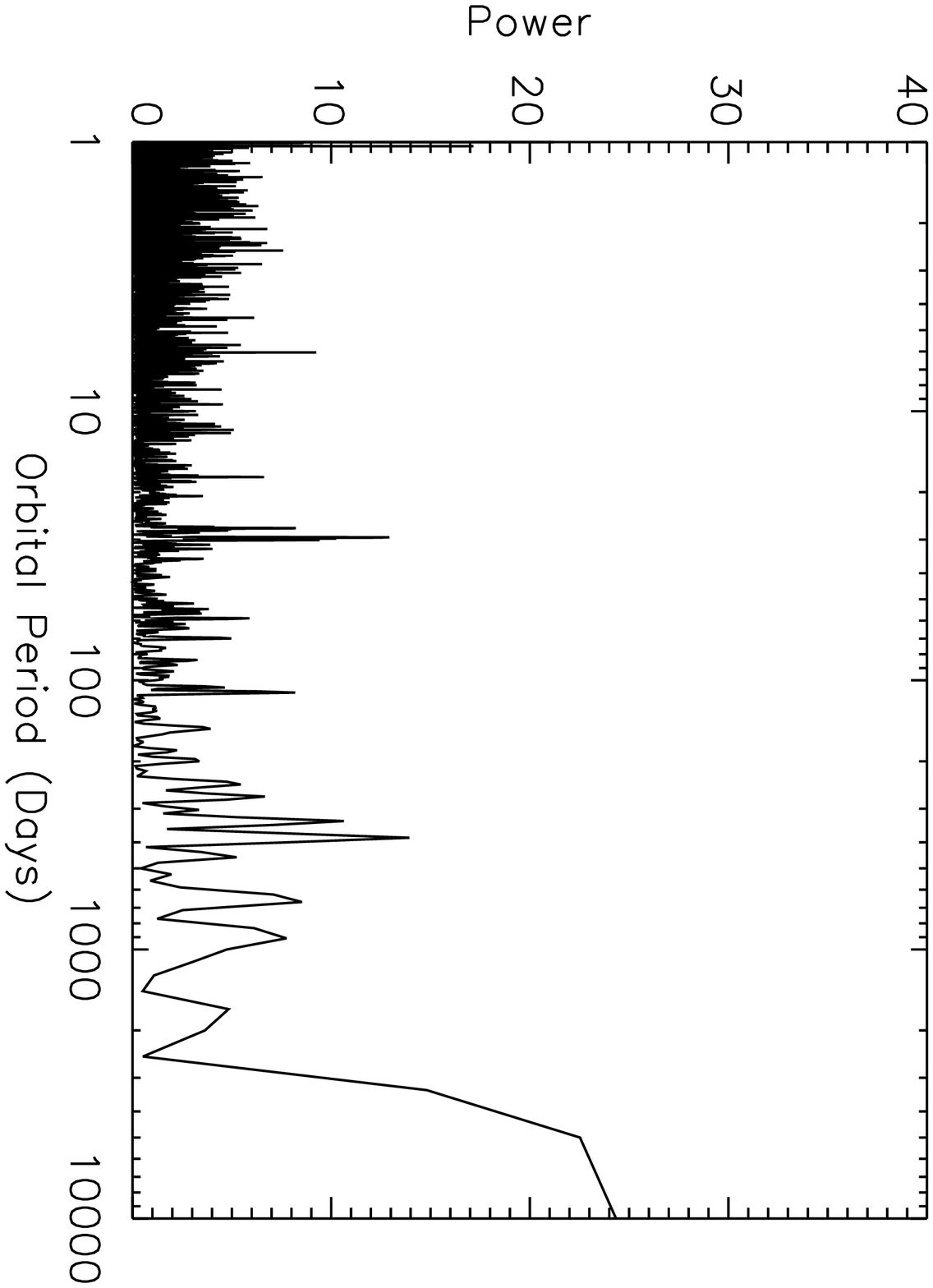}

\vspace{2.5cm}
\caption{From top left to bottom right we show the periodograms from ASAS data for the stars HD68402, HD147873, HD152079, 
HD154672, HD165155, and HD224538, respectively.}
\label{appendix4}
\end{figure*}

\section{Radial Velocities}

Here we provide all radial velocities that are discussed in this work.

\begin{table*}
\tiny
\caption{HD9174}
\begin{tabular}{cccccc}
\hline 
 & CORALIE & & & HARPS & \\
\multicolumn{1}{c}{BJD} & \multicolumn{1}{c}{RV} & \multicolumn{1}{c}{Error} & \multicolumn{1}{c}{BJD} &
\multicolumn{1}{c}{RV} & \multicolumn{1}{c}{Error} \\
\hline
2455160.5362269       &       7.2   & 9.0   & 2455188.6187049        &     10.77   & 0.55  \\
2455161.5372730       &     -11.8 &   9.0 &   2455883.6145165      &      -10.66 &    0.86 \\
2455162.5377306       &      30.2  &  9.0  &  2455885.5904695       &     -12.54  &  1.03  \\
2455467.6819298       &     -35.8 &   9.0 &   2456183.7872355      &       12.43  &  0.48 \\
2455468.7578863       &     -25.8 &   9.0 &   2456184.6697746      &        9.39   & 0.79  \\
2455878.6618959       &      -8.9  &  9.0  &  2456184.8402196       &       8.10    &0.92   \\
2455879.6756525       &      -5.9  & 15.0 &   2456185.7970768      &       13.90  &  0.55  \\
2456160.7988170       &       6.0   & 9.0   & 2456442.9352036        &      3.98    &0.69    \\
2456164.7903294       &      38.0  &  9.0  &  2456461.9331703       &       0.26   & 0.74   \\
2456307.5567717       &      23.0  &  9.0  &  2456463.8950719       &       1.48   & 0.77   \\
2456308.5698681       &      20.0  &  9.0  &  2456561.7814364       &     -13.46 &   0.50   \\
2456675.6083797       &     -22.0 &   9.0 &   2456562.5748725      &      -12.24&    0.86  \\
2456676.6025593       &     -16.0 &   9.0 &   2456563.6078357      &      -13.76&    0.78  \\
2456881.8675908       &    -20.2  & 12.0 \\ 
2456882.8265879       &    -17.2  & 12.0  \\
2456883.7721760       &    -13.2  & 12.0  \\
\hline
\end{tabular}
\end{table*}

\begin{table*}
\tiny
\caption{HD48265}
\begin{tabular}{ccccccccc}
\hline 
 & CORALIE & & & HARPS & & & MIKE & \\
\multicolumn{1}{c}{BJD} & \multicolumn{1}{c}{RV} & \multicolumn{1}{c}{Error} & \multicolumn{1}{c}{BJD} &
\multicolumn{1}{c}{RV} & \multicolumn{1}{c}{Error} & \multicolumn{1}{c}{BJD} & \multicolumn{1}{c}{RV} & 
\multicolumn{1}{c}{Error}\\
\hline
2455208.6992489       &       1.8   & 9.0 &    2454365.8248037        &    -32.25   & 0.77   &  2452920.8629    &  20.6   & 6.9  \\
2455209.7264368       &      10.8  &  9.0&     2454366.8740710       &     -25.18  &  0.99  &   2453431.6073   &  -30.2 &   2.5\\
2455210.7165285       &     -29.2 &  9.0&      2454367.8521212      &      -29.87 &   0.62 &    2453455.5573  &   -35.0&    2.6\\
2455465.8329696       &      16.8  &  9.0&     2454580.5231386       &      22.43   & 0.64   &  2453685.8138    &  15.0   & 2.9\\
2455468.8765488       &      27.8  &  9.0&     2454581.5537085       &      23.60   & 0.56   &  2453774.6744    &  25.0   & 3.3 \\
2455877.7963300       &      -4.2  & 15.0&     2454724.8337096      &      -14.14 &   0.59 &    2453775.6763  &    12.4 &   5.2\\
2455878.7843441       &      13.8  &  9.0&     2454725.8177884       &     -13.78  &  0.53  &   2453784.6887   &   25.4  &  2.6\\
2455879.7984724       &      10.8  &  9.0&     2454726.7789038       &     -14.84  &  0.80  &   2453811.5943   &   24.3  &  2.5 \\
2455969.6052648       &     -19.2 &   9.0&     2455651.5366350      &      -15.41 &   0.56 &    2453987.9168  &    -7.7 &   2.9 \\
2455970.7499814       &     -17.2 &   9.0&     2455883.7630397      &       -7.88  &  0.70  &   2454078.7826   &  -17.9 &   3.0   \\
2455971.7266404       &     -21.2 &   9.0&     2455885.7426863      &       -1.10  &  0.92  &   2454081.7133   &  -23.4 &   2.8  \\
2456034.5026492       &      13.8  &  9.0&     2455992.5816638       &      18.09   & 0.56   &  2454137.6436    & -20.7  &  2.5   \\
2456037.5096202       &      37.8  &  9.0&     2455994.5576882       &      23.81   & 0.61   &  2454138.6670    & -28.7  &  2.6   \\
2456164.8925279       &      15.8  &  9.0&     2456183.8678319       &      52.75   & 0.59   &  2454189.5669    & -29.5  &  4.4    \\
2456378.6064455       &     -15.2 &   9.0&     2456185.8516258      &       57.16  &  0.66  &   2454483.6148   &   28.5  &  3.0   \\
2456379.5444957       &     -21.2 &   9.0&     2456442.4578668      &      -16.22 &   0.91 &    2454501.6725  &    21.5 &   2.4   \\
2456380.6001445       &      -6.2  &  9.0&     2456444.4478133       &     -14.86  &  1.07  &   2454522.6219   &   25.4  &  2.7    \\
2456381.5979512       &     -15.2 &   9.0&     2456562.8313800      &      -12.48 &   0.73 &       \\
2456733.6079032       &       2.9   &  9.0 \\     
2456736.5950580      &  19.9       & 10.0 \\
2456753.5245324      & 3.9          & 10.0   \\
2456754.5405659      & 7.9          & 10.0  \\
\hline
\end{tabular}
\end{table*}

\begin{table*}
\tiny
\caption{HD68402}
\begin{tabular}{cccccc}
\hline 
 & CORALIE & & & HARPS & \\
\multicolumn{1}{c}{BJD} & \multicolumn{1}{c}{RV} & \multicolumn{1}{c}{Error} & \multicolumn{1}{c}{BJD} &
\multicolumn{1}{c}{RV} & \multicolumn{1}{c}{Error} \\
\hline
2455268.5536332       &      11.4  &  9.0 & 2456442.5125503      &      -21.04 &   1.51  \\
2455269.5471815       &       0.4   & 9.0  &2456444.4892384       &     -17.13  &  1.66  \\
2455270.5544347       &       4.4   & 9.0  &2456562.9026514       &      12.99   & 1.46  \\
2456034.5316938       &     -45.6 &   9.0&  2456563.8260976     &        13.44 &   1.51\\
2456037.5405384       &     -31.6 &   9.0&  2456563.9098232     &        11.72 &   2.07 \\
2456307.7270563       &     -32.6 &   9.0   \\
2456308.7455901       &     -18.6 &   9.0   \\
2456675.7326419       &      47.4  & 16.0     \\
2456675.7405950       &      64.4  & 17.0    \\
2456735.5968024       &   45.0     & 12.0  \\
2456736.6337750       &  33.0      & 14.0 \\  
2456752.5562350      &  38.0       &  13.0 \\  
2456754.5693098      &  53.0      & 15.0  \\  
2456823.4935737      & 33.0       & 15.0  \\ 
2456825.4656166      & 8.0         & 12.0  \\
2457075.6564372      & -43.1     & 14.0  \\ 
2457318.8514672     & -19.1      & 14.0  \\
\hline
\end{tabular}
\end{table*}

\begin{table*}
\tiny
\caption{HD72892}
\begin{tabular}{cccccc}
\hline 
 & CORALIE & & & HARPS & \\
\multicolumn{1}{c}{BJD} & \multicolumn{1}{c}{RV} & \multicolumn{1}{c}{Error} & \multicolumn{1}{c}{BJD} &
\multicolumn{1}{c}{RV} & \multicolumn{1}{c}{Error} \\
\hline
2455698.5075694      &      223.8  &  9.0  & 2456442.5223438      &     -195.55  &  1.12\\
2455699.5368477      &      337.8  &  9.0&   2456443.5281580      &     -167.05  &  1.77\\
2455700.5250521      &      444.8  &  9.0&   2456444.5101783      &     -134.83  &  1.36\\
2455968.7197064      &     -111.2 &  9.0&    2456448.4605785     &       144.30  &  1.29\\
2455969.7048152      &      -77.2  &  9.0&   2456449.4624615      &      248.19   & 0.82\\
2455970.6308925      &      -55.2  &  9.0&   2456450.4614518      &      357.16   & 1.21\\
2455972.6288495      &       61.8   &  9.0&  2456462.4520834       &    -114.51   & 1.13\\
2456034.5646797      &     -129.2 &  9.0&    2456463.4499736     &      -137.72 &   1.69\\
2456037.5717726      &     -156.2 &  9.0\\
2456307.7713660      &     -110.3 &  9.0\\
2456308.7833076      &     -126.3 &  9.0\\
2456378.6598031      &      159.7  &  9.0\\
2456379.5959537      &      128.7  &  9.0\\
2456463.4576754      &      -57.3  &  9.0\\
2456464.4530818      &      -80.3  &14.0\\
2456465.4567122      &     -111.3 &  9.0\\
2456467.4802779      &     -147.3 &16.0\\
2456675.7562736      &     -196.3 &  9.0\\
2456734.6035167      &       134.1 & 11.0\\
2456752.6156272      &    -176.9  & 13.0\\
2456754.5855485      &    -154.9  & 13.0\\
2456824.4911691      &    -138.9  & 15.0\\
2457184.4724902      &    -161.0  & 13.0\\
\hline
\end{tabular}
\end{table*}

\begin{table*}
\tiny
\caption{HD128356}
\begin{tabular}{cccccc}
\hline 
 & CORALIE & & & HARPS & \\
\multicolumn{1}{c}{BJD} & \multicolumn{1}{c}{RV} & \multicolumn{1}{c}{Error} & \multicolumn{1}{c}{BJD} &
\multicolumn{1}{c}{RV} & \multicolumn{1}{c}{Error} \\
\hline
2455268.9086205      &      -22.1  &  9.0 & 2454248.6023117     &       16.75  &  1.66\\
2455269.8196218      &      -34.1  &  9.0&  2454248.6108994     &        21.79 &   1.69\\
2455349.6784180      &      -16.1  &  9.0&  2454248.6176931     &        17.99 &   0.69\\
2455350.6814856      &       14.9   &  9.0& 2454367.5031968      &      -31.24 &   0.66\\
2455351.6612024      &       -4.1   &  9.0& 2454369.5091063      &      -37.97 &   2.49\\
2455352.6667444      &       -6.1   &  9.0& 2454577.6158842      &       35.96  &  0.48\\
2455609.8365887      &      -27.1  &14.0&  2454578.5680249     &        35.50 &   0.52\\
2455611.8334577      &      -13.1  &  9.0&  2454581.7357981     &        41.67 &   0.47\\
2455699.7015878      &        9.9    & 9.0& 2455271.8690903       &     -23.11  &  0.36\\
2455967.8551252      &       27.9   & 9.0&  2455649.7321937      &       -6.46  &  0.46\\
2455968.8240257      &       39.9   & 9.0&  2455650.7148680      &       -6.51  &  0.49\\
2455970.8305273      &       20.9   & 9.0&  2455651.7295683      &       -7.26  &  0.50\\
2455971.7742759      &       31.9   & 9.0&  2455786.4827996      &       50.87  &  0.51\\
2455972.8507044      &       46.9   & 9.0&  2455787.4566297      &       49.97  &  0.36\\
2456034.8259987      &       11.9   & 9.0&  2455992.7730262      &        9.21   & 0.39\\
2456037.8192588      &       27.9   & 9.0&  2455993.7048066      &        9.29   & 0.52\\
2456160.4758000      &       -6.1   & 9.0&  2456063.6469346      &       34.67  &  0.78\\
2456164.4757457      &       -9.1   & 9.0&  2456064.6145915      &       35.94  &  0.45\\
2456381.7122851      &       15.9   & 9.0&  2456065.6564584      &       36.26  &  0.43\\
2456463.5427541      &      -23.1  & 9.0&  2456183.4873901      &      -31.26 &   0.60\\
2456464.6240954      &      -35.1  & 9.0&  2456442.6772101      &      -46.63 &   0.46\\
2456465.6419917      &      -45.1  & 9.0&  2456443.6828424      &      -46.21 &   0.77\\
2456467.5931893      &      -26.1 &16.0&  2456444.6629996      &      -45.18 &   0.91\\
2456467.6621264      &      -24.1  & 9.0&  2456450.4938930      &      -40.30 &   0.62\\
2456676.8044908      &       19.8   & 9.0&  2456462.6290393      &      -36.01 &   0.54\\
2456734.7225188      &     -35.2   & 10.0& 2456463.5862333      &      -35.50 &   0.80\\
2456752.8173417      &     -30.2   & 12.0& 2456562.4755850      &       -3.37  &  0.56\\
2456823.7074373      &       1.8     & 14.0\\
2456825.7157702     &       6.8      & 12.0\\
2456881.5522933     &       4.8      & 12.0\\
2456882.5702439     &     20.8      & 15.0\\
\hline
\end{tabular}
\end{table*}

\begin{table*}
\tiny
\caption{HD143361}
\begin{tabular}{ccccccccc}
\hline 
 & CORALIE & & & HARPS & & & MIKE & \\
\multicolumn{1}{c}{BJD} & \multicolumn{1}{c}{RV} & \multicolumn{1}{c}{Error} & \multicolumn{1}{c}{BJD} &
\multicolumn{1}{c}{RV} & \multicolumn{1}{c}{Error} & \multicolumn{1}{c}{BJD} & \multicolumn{1}{c}{RV} & 
\multicolumn{1}{c}{Error}\\
\hline
2455269.8328453  &  31.9  &  12.0             &  2454253.7711993       &      10.50  &  3.17 &  2452864.57934  &   43.03  & 7.14    \\ 
2455269.8477334  &  30.9  &  12.0             &  2454253.7786414       &       9.06   & 2.22  & 2453130.83712   &  -6.28   &2.65     \\
2455349.6897288  &  9.9   & 14.0              &  2454367.5565354       &     -23.81 &   1.09&   2453872.73696 &    28.04 &  2.62  \\
2455349.7090115  &  6.9   & 15.0              &  2454368.5376348       &     -23.64 &   1.13&   2453987.50506 &    38.49 &  3.45  \\
2455350.6957245  &  -5.2  &  17.0             &  2454578.7885789       &     -83.52 &   0.83&   2453988.49460 &    25.06 &  2.81  \\
2455350.7085995  &  -11.2  &  22.0            & 2454581.8076461        &    -81.90  &  0.92 &  2454190.80550  &    0.00   &3.43  \\
2455351.6772114  &  3.9   & 13.0              & 2455271.8809609        &     19.62   & 0.77  & 2454217.84734   &  -7.02   &3.38  \\
2455351.6921046  &  -0.2  &  14.0             & 2455649.7540214        &    -83.51  &  0.94 &  2454277.65819  &  -29.11 &  3.43  \\
2455352.6804150  &  -8.2  &  13.0             &  2455786.5132754       &     -30.99 &   0.89&   2454299.55951 &   -37.73&   3.04  \\
2455352.6925204  &  2.9   & 12.0              & 2455992.7979663        &     55.69   & 1.23  & 2454300.58038   & -36.03  & 2.51  \\
2455465.4819439  &  -37.2  &  19.0            &   2455993.7269629      &       65.78 &   0.89&   2454339.50049 &   -52.16&   3.39  \\
2455465.4921698  &  -41.2  &  18.0            &   2456063.6718947      &       68.27 &   2.14&   2454501.87197 &   -94.15&   3.01  \\
2455466.4810091  &  -44.2  &  20.0            &  2456064.6371793       &      65.24  &  0.87 &  2454650.68383  & -105.06&   6.82  \\
2455467.4908225  &  -36.2  &  13.0            &  2456065.6805773       &      64.11  &  0.80 &  2454925.87115  &   21.04  & 2.76  \\
2455786.5181546  &  -28.2  &  13.0            &   2456184.5078021      &       53.36 &   0.84&   2454963.75707 &    34.93 &  3.05  \\
2455787.5128697  &  -32.2  &  12.0            &   2456442.6901584      &      -22.67&    0.90&  2454965.78744 &    31.39 &  2.82  \\
2455788.5005189  &  -18.2  &  12.0            &   2456443.6959898      &      -19.26&    1.96&  2455019.67861 &    27.65 &  2.46  \\
2455967.8637123  &  48.8  &  14.0             &  2456444.7117386       &     -20.46 &   1.79     \\
2455969.8564628  &  55.8  &  14.0             &   2456462.6564547      &      -25.82&    0.97     \\
2455970.8440395  &  49.8  &  14.0   \\         
2455971.8236578  &  61.8  &  14.0    \\        
2455972.8379866  &  52.8  &  14.0     \\       
2456034.8111668  &  73.8  &  14.0    \\      
2456037.8046599  &  73.7  &  12.0      \\     
2456160.4914594  &  49.7  &  11.0     \\    
2456161.4914636  &  58.7  &  12.0     \\   
2456162.5761051  &  53.7  &  18.0       \\  
2456164.4909267  &  46.7  &  12.0       \\  
2456381.7216801  &  -14.3  &  15.0    \\  
2456381.7325023  &  -13.3  &  14.0     \\  
2456463.6056985  &  -23.3  &  12.0     \\  
2456465.6944440  &  -22.3  &  12.0     \\  
2456467.5744277  &  -75.3  &  23.0    \\  
2456554.5120400  &  -28.3  &  16.0      \\  
2456554.5219817  &  -49.3  &  20.0    \\  
2456555.5573278  &  -33.3  &  13.0       \\  
2456676.8679860  &  -68.4  &  14.0         \\  
2456734.7382463  &  -61.4  &  12.0      \\  
2456752.8000194  &  -39.4  &  13.0         \\  
2456754.7548507  &  -51.4  &  13.0         \\  
2456823.7338512  &  -33.4  &  13.0         \\  
2456825.7551623  &  -36.4  &  13.0         \\  
2456881.6125148  &  5.6   & 12.0           \\  
2456882.6036411  &  17.6  &  15.0          \\  
2457281.5366030  &  71.5  &  19.0          \\  
\hline
\end{tabular}
\end{table*}

\begin{table*}
\tiny
\caption{HD147873}
\begin{tabular}{ccccccccc}
\hline 
 & CORALIE & & & HARPS & & & MIKE & \\
\multicolumn{1}{c}{BJD} & \multicolumn{1}{c}{RV} & \multicolumn{1}{c}{Error} & \multicolumn{1}{c}{BJD} &
\multicolumn{1}{c}{RV} & \multicolumn{1}{c}{Error} & \multicolumn{1}{c}{BJD} & \multicolumn{1}{c}{RV} & 
\multicolumn{1}{c}{Error}\\
\hline
2455268.8296124      &       66.7  &  9.0 & 2454365.5676419     &      -165.09 & 4.83&    2453189.66944  &    0.00  & 3.38\\
2455269.8954986      &       58.7  &  9.0&  2454580.8338639     &        86.26   & 2.48&  2453190.64093    & -9.92   &3.36\\
2455270.8623072      &       41.7  &  9.0&  2454581.8459239     &        77.96   & 1.73&  2453191.65473    & -5.91   &3.56\\
2455349.7478270      &        4.7   &  9.0& 2454724.4802846      &     -117.74  & 1.27&   2453551.62308   & -72.79 &  5.00\\
2455351.7263025      &       -9.3  &  9.0&  2454725.4983255     &      -121.37 & 1.56&    2454339.53949  &  134.00&   4.93\\
2455352.7243194      &        3.7   &  9.0& 2454726.4832450      &     -120.54  & 1.57&   2455001.70315   & 156.04 &  4.48\\
2455465.5203302      &       38.7  &  9.0&  2455271.8407105     &       124.30  & 1.42\\
2455466.5103888      &       25.7  &  9.0&  2455649.7654997     &       -73.82  & 2.01\\
2455467.5077317      &       39.7  &  9.0&  2455786.5492317     &      -102.21 & 2.08\\
2455468.4926389      &       32.7  &  9.0&  2455787.4838577     &       -93.49  & 1.66\\
2455786.5547221      &     -195.3&  9.0&    2455787.6993011   &         -92.88& 2.53\\
2455787.5512296      &     -190.3&  9.0&    2455788.4587689   &         -86.93& 1.92\\
2455788.5404119      &     -170.3&  9.0&    2455788.6109888   &         -91.50& 2.10\\
2455968.8355111      &      100.7 &  9.0&   2455788.7074614    &        -83.90 & 3.01\\
2455970.8644298      &       74.7  &  9.0&  2455992.8338522     &       -77.92  & 1.67\\
2456034.8536087      &      -19.3 &  9.0&   2455993.7871396    &        -80.38 & 2.00\\
2456037.8611090      &        8.7   &  9.0& 2455994.7989047      &      -90.35   & 2.11\\
2456160.5035894      &       15.7  &  9.0&  2456063.7089752     &       250.12  & 3.48\\
2456161.5099351      &       22.7  &  9.0&  2456064.7231089     &       258.36  & 1.87\\
2456164.5890537      &       55.7  &  9.0&  2456183.4686863     &       223.49  & 3.03\\
2456307.8687079      &      112.7 &  9.0&   2456184.4750146    &        219.57 & 3.41\\
2456308.8690886      &      117.7 &  9.0&   2456442.7882798    &         95.12  & 2.19\\
2456381.7577061      &     -119.3&  9.0&    2456444.7375998   &          62.78 & 3.16\\
2456463.6629499      &     -189.3&  9.0\\
2456464.6620073      &     -190.3&  9.0\\
2456554.5635006      &      144.7 &  9.0\\
2456555.4998324      &      118.7 &  9.0\\
2456676.8378428      &       -0.3  &  9.0\\
2456733.8400326      &     -56.8  &  12.0\\
2456734.7541033      &     -50.8  &  11.0\\
2456735.7646576      &     -55.8  &  11.0\\
2456736.8330743      &     -47.8  &  12.0\\
2456752.8898318      &     50.2   & 14.0\\
2456754.7704681     &    76.2  &  13.0\\
2456823.7610291     &    -155.8  & 14.0\\
2456824.7569383     &     -181.8  & 14.0\\
2456825.8288093     &    -168.8  & 14.0\\
2456881.6254682     &    130.2  &  14.0\\
2456882.6396241     &   151.2  &  13.0\\
2456883.5912828     &   149.2  &  13.0\\
2457179.6595698     &   -126.8  & 16.0\\
2457180.6006537     &   -127.8  & 15.0\\
2457181.5978747     &   -142.8  & 13.0\\
2457182.5980141     &   -120.8  & 13.0\\
2457183.7046858     &  -119.8   & 14.0\\
2457184.6231976     &  -111.8  & 13.0\\
\hline
\end{tabular}
\end{table*}

\begin{table*}
\tiny
\caption{HD152079}
\begin{tabular}{ccccccccc}
\hline 
 & CORALIE & & & HARPS & & & MIKE & \\
\multicolumn{1}{c}{BJD} & \multicolumn{1}{c}{RV} & \multicolumn{1}{c}{Error} & \multicolumn{1}{c}{BJD} &
\multicolumn{1}{c}{RV} & \multicolumn{1}{c}{Error} & \multicolumn{1}{c}{BJD} & \multicolumn{1}{c}{RV} & 
\multicolumn{1}{c}{Error}\\
\hline
2455786.6489688       &     -41.7 &   9.0&  2454253.8065363      &      -20.75  & 1.75&     2452917.4972   &  -24.3  &  6.2  \\   
2456463.6745808       &      14.3  &  9.0&   2454367.5811986      &      -22.95  & 1.00&     2453542.6649   &   22.5   & 3.3 \\
2456464.7049704       &      17.3  &  9.0&   2454579.8264105      &      -29.21  & 0.72&     2453872.8022   &   -8.5   & 2.5 \\
2456465.6766024       &       8.3   & 9.0&    2455649.8008277      &      -30.70  & 1.01&     2453987.5436   &  -10.3  &  2.8 \\
2456555.5745883       &      -0.8  &  9.0&   2455650.7595904      &      -29.43  & 0.96&     2453988.5202   &  -12.6  &  2.7 \\
2456676.8549118       &       2.2   &15.0&   2455651.8003619      &      -29.35  & 0.84&     2454190.8274   &  -13.7  &  2.9 \\
2456733.8533258       & -1.2   & 13.0&               2455786.5261504      &      -22.51  & 0.96&     2454277.6950   &  -19.7  &  3.4\\
2456734.8495354  & -17.2 & 11.0&              2455787.5966780      &      -22.46  & 0.85&     2454299.6134   &  -19.6  &  3.3\\
2456735.8130233 & -10.2 & 12.0&              2455992.9178275      &        1.25    & 1.01&     2454725.5353   &  -35.1  &  2.6\\
2456736.8653294 & 7.8     & 15.0&               2455993.8444865      &        3.09    & 0.93&     2454925.9161   &  -29.2  &  2.4\\
2456752.8688188 & 8.8     & 12.0&               2456064.7106335      &       19.86   & 0.89&     2454963.7753   &  -22.6  &  2.7\\
2456754.7851172 & 0.8     & 14.0&               2456184.5322797      &       44.23   & 0.93&     2454993.7093   &  -27.5  &  2.4\\
2456823.7916933 & 16.8   & 13.0&              2456442.7259868      &       24.77   & 0.90&     2455001.7291   &  -25.3  &  2.9\\
2456824.7696220 & -3.2   & 13.0&               2456443.7319729      &       27.81   & 1.62&     2455017.6624   &  -28.5  &  2.4\\
2456825.8154610 & -8.2  &  13.0&               2456444.7714764      &       25.93   & 1.28&     2455019.6938   &  -22.5  &  2.2\\
2456881.6859346 & 10.8   & 13.0&               2456462.7752093      &       19.99   & 1.30\\
2456882.6527789 & -10.2 & 12.0&               2456561.5772596      &       18.97   & 1.25\\
                                     &&&                        2456563.5044073      &       18.41   & 1.18\\
\hline
\end{tabular}
\end{table*}

\begin{table*}
\tiny
\caption{HD154672}
\begin{tabular}{ccccccccc}
\hline 
 & CORALIE & & & HARPS & & & MIKE & \\
\multicolumn{1}{c}{BJD} & \multicolumn{1}{c}{RV} & \multicolumn{1}{c}{Error} & \multicolumn{1}{c}{BJD} &
\multicolumn{1}{c}{RV} & \multicolumn{1}{c}{Error} & \multicolumn{1}{c}{BJD} & \multicolumn{1}{c}{RV} & 
\multicolumn{1}{c}{Error}\\
\hline
2455268.8925184 & -51.4 & 9.0  &            2454367.5951401       &     225.90  & 0.69& 2453189.71323 &  -141.83 &  2.89\\
2455286.7126041 & -133.4&  12.0  &          2454368.6185060       &     227.85  & 1.10& 2453190.70833 &  -143.62 &  2.73\\
2455286.7482435 & -137.4&  12.0  &          2454578.8112608       &      82.73   & 0.61& 2453191.72040 &  -147.14 &  3.33\\
2455287.7447603 & -130.4&  12.0  &          2454581.8703510       &      71.32   & 0.52& 2453254.50616 &   175.30  & 2.60\\
2455287.7803973 & -136.4&  12.0  &          2455271.8920649       &     -18.52  & 0.63& 2453596.68926 &   130.45  & 3.10\\
2455288.7337184 & -139.4&  11.0  &          2455649.8248173       &    -195.00 & 0.66& 2453810.90968 &    -5.00   &2.56\\
2455288.7776633 & -138.4&  12.0  &          2455786.5972307       &     -87.04  & 0.66& 2453872.81362 &   260.75  & 2.45\\
2455349.7906916 & 161.6 & 10.0   &          2455993.8674482       &    -126.00 & 0.78& 2454189.87086 &   -80.09  & 3.62\\
2455350.8687367 & 177.6 &  10.0  &          2455994.8361631       &     -98.02  & 0.83& 2454189.87898 &   -90.28  & 3.52\\
2455351.7067623 & 173.6 &  10.0  &          2456063.7424430       &      53.14   & 1.25& 2454190.84022 &   -57.35  & 2.84\\
2455352.7571340 & 172.6 &  9.0   &          2456064.7465191       &      50.37   & 0.66& 2454215.86050 &   242.59  & 2.52\\
2455353.6007870 & 135.6 &  10.0  &          2456065.7448182       &      48.70   & 0.74& 2454216.78927 &   243.77  & 2.54\\
2455354.6329905 & 151.6 &  10.0  &          2456183.5270912       &     188.65  & 0.85& 2454217.87254 &   250.94  & 2.76\\
2455354.7244421 & 136.6 &  11.0  &          2456442.7630531       &     -86.65  & 0.69& 2454277.70250 &    67.40   & 2.74\\
2455355.6194464 & 135.6 &  11.0  &          2456443.7800510       &     -86.90  & 1.29& 2454299.62096 &     0.00    & 2.64\\
2455355.7115550 & 148.6 &  11.0  &          2456462.7983801       &    -163.39 & 0.91& 2454339.55738 &  -145.72 &  4.15\\
2455433.5945428 & -58.4 &  11.0  &          2456462.8102434       &    -164.91 & 1.02& 2454501.89596 &  -143.22 &  2.71\\
2455434.5939320 & -68.4 &  10.0  &          2456561.4986867       &      39.66   & 0.85& 2455018.68932 &   213.92  & 2.17\\
2455467.5419737 & -152.4 &  11.0  &         2456563.5280947       &      36.30   & 0.89\\
2455468.5360292  &-140.4 &  11.0  \\      
2455786.6672674  &-128.5 &  10.0  \\     
2456034.8785575 & 94.5  & 9.0  \\         
2456037.8864541 & 86.5   &  9.0  \\       
2456160.5115180 & -99.5 &  11.0  \\       
2456161.5235072 & -68.5 &  12.0  \\      
2456164.5556060 & 60.5   &  13.0  \\     
2456381.7967325 & 47.5  & 10.0 \\        
2456463.6867223 & -212.5 &  10.0  \\      
2456464.7194041 & -226.5 &  10.0  \\     
2456734.8670344 & -17.5 &  9.0  \\       
2456735.8301098 & -20.5 &  9.0 \\         
2456823.8058978 & 153.5 &  11.0  \\       
2456825.8537984 & 178.5 &  11.0  \\       
2456882.6789128 & 44.5  & 11.0  \\       
\hline
\end{tabular}
\end{table*}

\begin{table*}
\tiny
\caption{HD165155}
\begin{tabular}{cccccc}
\hline 
 & CORALIE & & & HARPS & \\
\multicolumn{1}{c}{BJD} & \multicolumn{1}{c}{RV} & \multicolumn{1}{c}{Error} & \multicolumn{1}{c}{BJD} &
\multicolumn{1}{c}{RV} & \multicolumn{1}{c}{Error} \\
\hline
2455698.7379179      &      -70.4 &   9.0&    2454577.8573518       &      70.07  &  0.70\\
2455786.6994564      &       36.6  & 13.0&    2454578.8259992       &      67.18  &  0.80\\
2455786.7108277      &       75.6  & 13.0&    2454579.8512477       &      67.69  &  0.74\\
2455787.6641024      &       35.6  & 12.0&    2455650.8171652       &     -83.60 &   1.14\\
2455787.6754805      &       35.6  & 12.0&    2455786.6216504       &      -5.87  &  1.04\\
2455788.6765746      &       48.6  & 13.0&    2455787.4979049       &      -3.44  &  0.96\\
2455788.6875372      &       67.6  & 13.0&    2455788.6355855       &      -1.89  &  1.07\\
2456034.9332917      &       20.6  & 12.0&    2455993.8895322       &      -6.64  &  1.23\\
2456381.8066463      &       54.6  & 12.0&    2455994.8596255       &      -7.98  &  1.57\\
2456381.8190355      &       47.6  & 12.0&    2456063.7543293       &     -35.54 &   1.82\\
2456463.7613374      &       59.6  &   9.0&    2456064.7595490       &     -35.45 &   0.96\\
2456464.7345595      &       11.6  &   9.0&    2456065.7693045       &     -33.41 &   2.68\\
2456555.5907771      &      -42.4 & 11.0&    2456183.5412473       &     -15.93 &   1.13\\
2456555.6041047      &      -39.4 &   9.0&    2456442.8006921       &      14.89  &  1.18\\
2456733.8878823 & 145.2  &  10.0    &         2456443.8064819       &      19.51  &  2.63\\
2456735.8673459 & 111.2  &  10.0    &         2456444.8081381       &      17.67  &  1.57\\
2456754.8071800 & 115.2  &  13.0    &         2456461.7756434       &      13.11  &  1.08\\
2456823.8481613 & 112.2  &  11.0    &         2456462.8238912       &       9.33   & 1.42\\
2456825.8695333 & 89.2   & 11.0     &         2456463.7557595       &      11.89  &  1.71\\
2456881.7083646 & 99.2   & 11.0     &         2456561.4867891       &     -62.19 &   1.12\\
2456883.6490320 & 85.2   & 10.0  \\
2457077.8839217 & 120.2  &  13.0 \\
2457183.7828036 & -250.8  &  16.0\\
2457184.6886555 & -250.8  &  15.0\\
2457312.5069791 & -254.8  &  17.0\\
2457318.4944200 & -279.8  &  16.0\\
\hline
\end{tabular}
\end{table*}

\begin{table*}
\tiny
\caption{HD224538}
\begin{tabular}{ccccccccc}
\hline 
 & CORALIE & & & HARPS & & & MIKE & \\
\multicolumn{1}{c}{BJD} & \multicolumn{1}{c}{RV} & \multicolumn{1}{c}{Error} & \multicolumn{1}{c}{BJD} &
\multicolumn{1}{c}{RV} & \multicolumn{1}{c}{Error} & \multicolumn{1}{c}{BJD} & \multicolumn{1}{c}{RV} & 
\multicolumn{1}{c}{Error}\\
\hline
2455196.5457833       &     -63.9  &  9.0&      2454365.6728323      &      -72.92 &  0.84&  2453189.91295 &    -2.66 &  2.86\\
2455197.5258897       &     -62.9  &  9.0&      2454724.7162997      &       87.00  &  0.93& 2453190.91771  &    0.00   &3.27\\
2455198.5268449       &     -54.9  &  9.0&      2454725.6845580      &       87.50  &  1.02& 2453191.91314  &    7.12   &2.85\\
2455352.9431841       &     -89.9  &  9.0&      2454726.6664725      &       89.44  &  0.65& 2453254.73446  &  -13.98 &  3.23\\
2455465.6728602       &     -55.9  &  9.0&      2454727.6026469      &       87.24  &  0.83& 2454338.83927  &    3.92   &3.86\\
2455468.6569334       &     -51.9  &  9.0&      2455786.8072676      &       -5.84  &  0.80& 2454339.75448  &   -7.75  & 3.09\\
2455497.5904524       &     -52.9  &  9.0&      2455787.6269341      &      -12.49 &  0.96\\
2455514.6207088       &     -42.9  &  9.0&      2455787.8105224      &      -16.34 &  0.73\\
2455786.8585090       &      37.1   &  9.0&     2455787.9313791       &     -14.21  &  0.82\\
2455787.8646502       &      25.1   &  9.0&     2455788.5991810       &     -14.30  &  1.26\\
2455788.8218648       &      53.1   &19.0&      2455788.7826534      &      -14.01 &  1.33\\
2455877.6116650       &      93.1   &22.0&      2455883.5865245      &       46.83  &  1.19\\
2455878.5958784       &      92.1   &  9.0&     2455885.5121047       &      54.40   &  1.30\\
2455879.6063410       &     108.1  &  9.0&      2455885.5744505      &       52.39  &  1.53\\
2455969.5203878       &     158.1  &16.0&       2455885.7033086     &        51.81 &  1.43\\
2455970.5204393       &     167.1  &16.0&       2456064.9384446     &        67.21 &  0.85\\
2456160.6541736       &     -25.9  &16.0&       2456183.7248653     &       -67.99&  0.86\\
2456164.7802416       &      -4.9   &16.0&      2456185.6754003      &      -70.76 &  1.02\\
2456554.6541123       &     -68.9  &  9.0&      2456442.9267855      &     -116.37&  1.14\\
2456554.6624176       &     -54.9  &  9.0&      2456461.8887446      &     -110.55&  1.03\\
2456555.7132051       &     -76.9  &  9.0&      2456561.7050322      &     -109.01&  0.97\\
2456675.5534417       &     -27.9  &16.0\\
2456824.9242966       &      5.9     &  13.0\\
2457317.6711775       &     -3.1    &  36.0\\
\hline
\end{tabular}
\end{table*}

\label{lastpage}

\end{document}